\documentclass{article}

\usepackage{amsmath,amssymb,amsfonts,amsthm,makeidx,graphicx,colortbl,url,booktabs,authblk}
\usepackage{txfonts}
\usepackage{helvet}
\usepackage[T1]{fontenc}

\usepackage{dsfont}
\usepackage{verbatim}
\usepackage{paralist}
\usepackage{subfigure}
\usepackage{tikz}
\usepackage[european resistor, european voltage, european current]{circuitikz}
\usetikzlibrary{arrows,shapes,positioning,snakes, decorations.markings,decorations.pathmorphing,decorations.pathreplacing,calc,patterns,shapes.geometric,topaths}
\usepackage{pgfplots}
\usepackage{xparse} 
\usepackage{scrextend}
\usepackage{nth}

\tikzset{
    block/.style={rectangle, draw, line width=0.5mm, black, text width=8em, text centered,
                 minimum height=2em},
    line/.style={draw, -latex}}

\newtheorem{theorem}            {Theorem}[section]

\newtheorem{definition}         [theorem]{Definition}
\newtheorem{example}            [theorem]{Example}
\newtheorem{lemma}              [theorem]{Lemma}

\newtheorem{remark}             [theorem]{Remark}

\newcommand{\metalevel}{\text{meta-level }}
\newcommand{\viewcount}{\text{view count }}
\newcommand{\viewcounts}{\text{view count }}

\newcommand{\Y}{\mathbb{Y}} \newcommand{\X}{\mathbb{X}}
\def\S{{\mathbf S}}

\newcommand{\obs}{y}    
\newcommand{\A}{\mathcal{A}}

\newcommand{\beq}{\begin{equation}}
\newcommand{\eeq}{\end{equation}}
\newcommand{\E}{\mathbf{E}}

\newcommand{\history}{\mathcal{H}}
\newcommand{\full}{\mathcal{F}}

\newcommand{\argmin}{\operatorname{argmin}}

\newcommand{\Bs}{R^\pi} 
\newcommand{\ta}{\tilde{a}}
\newcommand{\sigs}{\sigma}
  \def\1{{\mathbf 1}}
\newcommand{\reals} {\Bbb{R}}
\newcommand{\nn}{\nonumber}

\newcommand{\lbelief}{l}

    \newcommand{\Ts}{T}
    \newcommand{\ca}{c_a}
\newcommand{\p}{\prime}
    \def\S{{S}} 
  
\newcommand{\norm}[1]{\lVert#1\rVert}
\newcommand{\tindx}{t}
\newcommand{\Tindxter}{T}
\newcommand{\nindx}{n}

\newcommand{\probe}{p}
\newcommand{\response}{x}

\newcommand{\utility}{u}
\newcommand{\budget}{I}
\newcommand{\dataset}{\mathcal{D}}
\newcommand{\setresponse}{X}
\newcommand{\potfun}{V}
\newcommand{\estpotfun}{\hat{V}}
\newcommand{\tweet}{\tau}

\newcommand{\tp}{P}

\newcommand{\belief}{{\pi}}
\newcommand{\tbelief}{\pi^0}

\newcommand{\priv}{\eta}

  \def\1{{\mathbf 1}}

\def\lbr{\left\lbrace}
\def\rbr{\right\rbrace}

\renewcommand{\P}                 {\Bbb{P}}

\newcommand{\minc}{w}

\def\lb{\left[}
\def\rb{\right]}

\newcommand{\failprob}{p_F}

\NewDocumentCommand{\weight}{gg}
  {\IfNoValueTF{#2}
     {\IfNoValueTF{#1}
        {W}
        {W{(#1)}}%
     }
     {W{(#1, #2)}}%
  }

\NewDocumentCommand{\oprob}{gg}
  {\IfNoValueTF{#2}
     {\IfNoValueTF{#1}
        {B}
        {B_{#1}}%
     }
     {B_{#1 #2}}%
  }

\newcommand{\sigmaf}{\mathcal{F}}

\newcommand{\aindex}{a}

\newcommand{\ole}{\stackrel{\text{defn}}{=}}

\NewDocumentCommand{\tpdiff}{gggg}
     {\IfNoValueTF{#4}
     {\IfNoValueTF{#3}
         {\IfNoValueTF{#2}
        {{p}_{#1} }
        {{p}_{#1}{(#2)} } }%
     {{p}_{#1}(#2,#3) }}%
     {p_{#1}(#2,#3,#4)}
  }

\NewDocumentCommand{\atp}{gggg}
     {\IfNoValueTF{#4}
     {\IfNoValueTF{#3}
         {\IfNoValueTF{#2}
          {\IfNoValueTF{#1}
          {\bar{p}}
        {{\bar{p}}_{#1} } }
        {{\bar{p}}_{#1}{(#2)} } }%
     {{\bar{p}}_{#1}(#2,#3) } }%
     {{\bar{p}_{#1}(#2,#3,#4) }}
  }

\NewDocumentCommand{\sentiment}{gggg}
     {\IfNoValueTF{#4}
     {\IfNoValueTF{#3}
         {\IfNoValueTF{#2}
        {{z}_{#1} }
        {{z}_{#1}{(#2)} } }%
     {{z}_{#1}(#2,#3) }}%
     {z_{#1}(#2,#3,#4)}
  }

\NewDocumentCommand{\onoise}{gggg}
     {\IfNoValueTF{#4}
     {\IfNoValueTF{#3}
         {\IfNoValueTF{#2}
        {{v}_{#1} }
        {{v}_{#1}{(#2)} } }%
     {{v}_{#1}(#2,#3) }}%
     {v_{#1}(#2,#3,#4)}
  }

\NewDocumentCommand{\var}{gggg}
     {\IfNoValueTF{#4}
     {\IfNoValueTF{#3}
         {\IfNoValueTF{#2}
        {\sigma^2{(#1)} }
        {\sigma^2{(#1, #2)}  } }%
     {\sigma^2_{#1}(#2,#3) }}%
     {\sigma^2_{#1}(#2,#3,#4)}
  }

\NewDocumentCommand{\steady}{g}
  {
     {\IfNoValueTF{#1}
        {\pi}
        {\pi{(#1)}}%
     }
     }

\NewDocumentCommand{\cost}{g}
  {
     {\IfNoValueTF{#1}
        {c}
        {c^{(#1)}}%
     }
     }

\newcommand{\reward}{r}

\newcommand{\costpdf}{P_{\cost,\target{\dtime}}}

\NewDocumentCommand{\target}{g}
  {
     {\IfNoValueTF{#1}
        {s}
        {s_{#1}}%
     }
     }

\NewDocumentCommand{\targetobs}{g}
  {
     {\IfNoValueTF{#1}
        {o}
        {o_{#1}}%
     }
     }

\NewDocumentCommand{\targetoprob}{gg}
  {\IfNoValueTF{#2}
     {\IfNoValueTF{#1}
        {\bar{B}}
        {\bar{B}_{#1}}%
     }
     {\bar{B}_{#1 #2}}%
  }

\NewDocumentCommand{\tptarget}{gg}
  {\IfNoValueTF{#2}
     {\IfNoValueTF{#1}
        {A}
        {A_{#1}}%
     }
     {A_{#1 #2}}%
  }

\newcommand{\tstate}{s}
\newcommand{\tstatep}{s'}

\newcommand{\indicator}{I}

\newcommand{\network}{G}
\newcommand{\Vertexset}{V}
\newcommand{\vertexset}{\{1,2,\ldots,\vertexnum\}}
\newcommand{\edgeset}{E}

\NewDocumentCommand{\infectdist}{gg}
  {\IfNoValueTF{#2}
     {\IfNoValueTF{#1}
        {\rho}
        {\rho_{#1}}%
     }
     {\rho_{#1}(#2)}%
  }

\NewDocumentCommand{\minfectdist}{gg}
  {\IfNoValueTF{#2}
     {\IfNoValueTF{#1}
        {\bar{\rho}}
        {\bar{\rho}_{#1}}%
     }
     {\bar{\rho}_{#1}(#2)}%
  }

\NewDocumentCommand{\tinfectdist}{gg}
  {\IfNoValueTF{#2}
     {\IfNoValueTF{#1}
        {\tilde{\rho}^\vertexnum}
        {\tilde{\rho}^{\vertexnum}_{#1}}%
     }
     {\tilde{\rho}^{\vertexnum}_{#1}(#2)}%
  }

\NewDocumentCommand{\pa}{gg}
  {\IfNoValueTF{#2}
     {\IfNoValueTF{#1}
        {\theta}
        {\theta_{#1}}%
     }
     {\theta_{#1}^{#2}}%
  }

\NewDocumentCommand{\nodeobs}{gg}
  {\IfNoValueTF{#2}
     {\IfNoValueTF{#1}
        {y}
        {y_{#1}}%
     }
     {y_{#1}^{(#2)}}%
  }

\newcommand{\nodeobsdim}{Y}

\NewDocumentCommand{\sample}{g}
  {
     {\IfNoValueTF{#1}
        {\alpha}
        {\alpha{(#1)}}%
     }
     }

\NewDocumentCommand{\obsm}{g}
  {
     {\IfNoValueTF{#1}
        {H}
        {H{(#1)}}%
     }
     }

\NewDocumentCommand{\feedforward}{g}
  {
     {\IfNoValueTF{#1}
        {D}
        {D{(#1)}}%
     }
     }

\newcommand{\degreediff}[1]{D^{(#1)}}
\newcommand{\nbhood}[1]{\mathcal{N}^{(#1)}}

\newcommand{\dtime}{k}
\newcommand{\ctime}{t}

\newcommand{\finaltime}{T}

\newcommand{\seq}{l}

\NewDocumentCommand{\nodem}{g}
  {
     {\IfNoValueTF{#1}
        {m}
        {m_{#1}}%
     }
     }

\newcommand{\noden}{n}
\renewcommand{\deg}{d}
\newcommand{\degmax}{\bar{D}}

\NewDocumentCommand{\state}{gg}
  {\IfNoValueTF{#2}
     {\IfNoValueTF{#1}
        {x}
        {x_{#1}}%
     }
     {x_{#1}^{(#2)}}%
  }

\NewDocumentCommand{\vertexnum}{g}
  {
     {\IfNoValueTF{#1}
        {N}
        {N{(#1)}}%
     }
     }

\NewDocumentCommand{\degdist}{g}
  {
     {\IfNoValueTF{#1}
        {P}
        {P{(#1)}}%
     }
     }

\newcommand{\statev}{x}

\newcommand{\neighbor}[1]{N^{(#1)}}

\newcommand{\nactive}[1]{A^{(#1)}}

\newcommand{\statea}{i}
\newcommand{\stateb}{j}

\newcommand{\Prb}{\mathbb{P}}

\newcommand{\cvr}{\text{CVaR}_{\alpha}(c(x_{k},a))}

\begin{document}

\title{Dynamics of Information Diffusion and Social Sensing}

\author[1]{Vikram Krishnamurthy}
\affil[1]{ School of Electrical \& Computer Engineering,  Cornell Tech,  Cornell University, New York email: vikramk@cornell.edu}

\author[2]{William Hoiles}
\affil[2]{Department of Electrical \& Computer Engineering, University of British Columbia, Vancouver, Canada  email: whoiles@ece.ubc.ca}
\maketitle

\begin{abstract}
Statistical inference using social sensors is an area that has witnessed remarkable progress in the last decade. It is relevant in a variety of applications including localizing  events for targeted advertising, marketing, localization of natural disasters and predicting sentiment of investors in financial markets. This chapter presents a tutorial description of four  important aspects of sensing-based information diffusion in social networks from a communications/signal processing perspective. First,   diffusion models for  information exchange in large scale social networks  together with social sensing via social media networks such as Twitter is considered. Second, Bayesian social learning models and risk averse social learning is considered with applications in finance and online reputation  systems.  Third, the principle of revealed preferences arising in  micro-economics theory is used to parse datasets to determine if social sensors are utility maximizers and then determine their utility functions. Finally, the interaction of social sensors with YouTube channel owners is studied using time series analysis methods.
All four topics are explained in the context of actual experimental datasets from health networks, social media and psychological experiments. Also, algorithms are given that exploit the above models to infer underlying events based on social sensing. The overview, insights, models and algorithms presented in this chapter stem from recent developments in network science, economics  and signal processing. At a deeper level, this chapter considers mean field dynamics of networks, risk averse Bayesian social learning filtering and quickest change detection, data incest in decision making over a directed acyclic graph of social sensors, inverse optimization problems for utility function estimation  (revealed preferences) and statistical modeling of interacting social sensors in YouTube social networks.
\end{abstract}


\section{Introduction and Motivation}
\label{sec:introduction}

Humans can be viewed as
social sensors that interact over a social network  to  provide information about their environment.  Examples of information produced by such social sensors include Twitter posts, Facebook status updates, and ratings on online reputation systems like Yelp and Tripadvisor.  
Social sensors go beyond physical sensors -- for example, user opinions/ratings (such as the quality of a restaurant) are available on Tripadvisor
but  are difficult to measure via  physical sensors.  Similarly, future situations revealed by the Facebook status of a user  are impossible to predict using  physical sensors \cite{RMZ11}.

Statistical inference using social sensors is an area that has witnessed remarkable progress in the last decade. It is relevant in a variety of applications including localizing special events for targeted advertising \cite{LS10,CCL10}, marketing~ \cite{TBB10,LAH07}, localization of natural disasters~\cite{SOM10}, and predicting sentiment of investors in financial markets~\cite{BMZ11,PL08}. 
For example, \cite{AH10} reports sthat models built from the rate of tweets for particular products can outperform  market-based predictors.

\subsection{Context: Why social sensors?}
Social sensors  present unique challenges from a statistical estimation point of view.
First,  social sensors  interact with and influence other social sensors. For example, ratings posted on online reputation systems strongly influence the behavior of  individuals.\footnote{It is reported
 in \cite{IMS11} that  81\% of hotel managers  regularly check Tripadvisor reviews.  \cite{Luc11} reports that a one-star increase in the Yelp rating maps to 5-9 \% revenue increase. \label{foot}}
Such interacting sensing can result in  non-standard information patterns due to correlations introduced by the structure of the underlying social network.
Thus
 certain events  ``go viral" \cite{LAH07,GWG12}.
Second, due to privacy
concerns and time-constraints,
social sensors typically do not reveal  raw observations of the underlying state of nature.
Instead, they reveal their decisions 
(ratings, recommendations, votes) which can be viewed as a low resolution (quantized)   function of their raw measurements and interactions with other social sensors.
This can result
in misinformation propagation,  herding and information cascades.
Third, the response of a social sensors  may not be consistent with that of an utility maximizer; social sensors are typically risk averse.
%

Social sensors are enabled by technological networks. Indeed,
social media sites that  support interpersonal communication and
collaboration using Internet-based social network platforms, are growing rapidly. 
 McKinsey estimates that the economic impact of social media on business is potentially greater than \$1 trillion since social media facilitates  efficient communication and collaboration within and across organizations.

\subsection{Main Results and Organization}
There is strong motivation to construct  models that  facilitate understanding the dynamics of information
flow in social networks. This chapter  presents a tutorial description of  four mportant aspects of  sensing-based information diffusion in social networks from a signal processing perspective:

 \subsubsection{Information Diffusion in Large Scale Social Networks}
 The first topic considered in this chapter (Sec.\ref{sec:diffusion})
is diffusion of information in   social networks comprised of a  population of interacting social sensors.
The  states of sensors evolve  over time as a probabilistic function of the states of their neighbors and an underlying
target process.
Several recent  papers   investigate such information diffusion in real-world social networks. Motivated by marketing applications, \cite{SRM09} studies  diffusion (contagion) behavior in Facebook. Using data from 260,000 Facebook \textit{pages} (which advertise  products, services  and  celebrities), \cite{SRM09} analyzes information diffusion.   In  \cite{RMK11}, the spread of  \textit{hashtags} on Twitter is studied.
There is a wide range of social phenomena such as diffusion of technological innovations, sentiment, cultural fads, and economic conventions~\cite{Cha04,Pin08} where individual decisions are influenced by the decisions of others.  

We consider the so called  Susceptible-Infected-Susceptible (SIS) model \cite{PV01} for information diffusion in a social network.
It is shown for social networks comprised of a large number of agents how the dynamics of  degree distribution can  
 be approximated by the mean field dynamics. Mean field dynamics have been studied in~\cite{BW03} and applied to social networks in~\cite{Pin08} and
 leads to a tractable model for the dynamics social sensors.

 We demonstrate using  influenza datasets from the U.S Centers for Disease Control and Prevention (CDC) how
Twitter can be used as a real time social sensor for tracking the spread of influenza.
That is, a health  network (namely, Influenza-like Illness Surveillance Network (ILInet))
 is sensed by a real time microblogging social media network (namely, Twitter).

 We  also  review two recent methods for sampling social networks, namely, social sampling and respondent-driven sampling.
Respondent-driven sampling is now used  by the  U.S. Centers for Disease Control and Prevention (CDC)  as part of the National HIV Behavioral Surveillance System in health networks.

\subsubsection{Bayesian Social Learning in Online Reputation Systems}  The second topic of this chapter
(Sec.\ref{sec:socialmain}) considers
 online reputation  systems where individuals make recommendations based on their private observations and recommendations of friends.
 Such  interaction of individuals and their social  influence  is modelled as Bayesian
  social learning \cite{Ban92,BHW92,Cha04} on a directed acyclic graph.  We consider two important classes of such problems; risk averse social learning in financial systems,
  and data incest in reputation systems. The risk averse social learning and associated quickest change detection is important in detecting market shocks in high frequency
  trading.
  Data incest (misinformation propagation) arises as a result of correlations in recommendations due to the intersection of multiple paths in the information exchange graph.  Necessary and sufficient conditions are given on the structure of information exchange graph  to mitigate data incest. 
Experimental results on human subjects are presented to illustrate the effect of social influence  and data incest on decision making.

 The setup differs from classical signal processing  where sensors use  noisy observations to compute estimates - in social learning agents
use  noisy observations together with decisions made by previous agents, to estimate the underlying state of nature.
Social learning has been used widely in economics, marketing, political science and sociology  to model the behavior of financial markets, 
crowds, social groups and social networks; see \cite{Ban92,BHW92,AO11,Cha04,LADO07,ADLO08} and numerous references therein. Related models have been studied in the context of sequential decision making in
information theory \cite{CH70,HC70}  and statistical signal processing \cite{CSL13,KP13} in the electrical engineering literature. 
 Social learning can result in   unusual behavior such as herding  \cite{BHW92} where agents eventually  choosing the same action irrespective of their private observations. As a result, the actions
contain no information about the private observations and so the Bayesian estimate of the underlying random variable freezes. Such behavior
 can be undesirable, particularly if individuals
 herd and make incorrect decisions.

\subsubsection{Revealed Preferences and Detection of Utility Maximizers}
The third topic considered in this chapter (Sec.\ref{sec:revealed})  is 
 the principle of revealed preferences arising in microeconomics.
 It  is used as  a constructive test to determine: Are social sensors utility optimizers in their response to external influence?
 The key question considered is as follows:
  Given a time-series of data $\dataset=\{(\probe_\tindx,\response_\tindx), \tindx\in\{1,2,\dots,\Tindxter\}\}$ where $\probe_\tindx\in\mathds{R}^m$ denotes the external influence, $\response_\tindx$ denotes the response of an agent,  is it possible to detect if the
   agent is a {\it utility maximizer}?

These issues are  fundamentally different to the {\em model-centric} theme  used in the signal processing literature where
one postulates an objective function (typically convex) and then proposes  optimization algorithms.  In contrast the revealed
  preference approach is {\em data centric} - given a dataset,  we wish to determine if is consistent with utility maximization.

   We present a remarkable result called Afriat's theorem \cite{Afr67,Var82} which provides a necessary and sufficient
  condition for a finite dataset $\dataset$ to have originated from a utility maximizer.
  Also a multi-agent version of Afriat's theorem is  presented to determine if the  dataset generated by multiple agents 
  is consistent with playing from the equilibrium of a potential game. 

 Unlike model centric applications of game theory in
signal processing, the  revealed preferences approach is  data centric: 1) Given a time series dataset of probe and response signals, how can one detect if the response signals are consistent with a Nash equilibrium generated by players in a concave potential game? 2) If consistent with a concave potential game, how can the utility function of the players be estimated?
  
  We present three  datasets involving social sensors to illustrate Afriat's theorem of revealed preferences.
These datasets are: (i) an auction conducted by undergraduate students at Princeton University,
(ii)  aggregate power consumption in the
electricity market of Ontario province and (iii) Twitter dataset for specific hashtags.

Varian   has written several influential papers on Afriat's theorem  in the economics
literature. These include
measuring the welfare effect of price discrimination~\cite{Var85}, analysing the relationship between prices of broadband Internet access and time of use service~\cite{Var12}, and ad auctions for advertisement position placement on page search results from Google~\cite{Var12,Var09}. 
Despite  widespread use in economics, revealed preference theory is relatively unknown in the electrical engineering literature.

\subsubsection{Social Interaction of YouTube Consumers}  
The fourth topic considered in this chapter  (Sec.\ref{sec:SocialInteractionofChannelOwnersandYouTubeConsumers}) is the engagement dynamics of social sensors to online video content. Specifically, we consider how users interact with video content created on the YouTube social network. YouTube is the largest user-driven video content provider in the world and has become a major platform for disseminating multimedia information. YouTube contains over 1 billion users who collectively watch millions of hours of YouTube videos
and generate billions of views every day (e.g. 150 years of video are watched every day). Additionally, users upload over 300 hours of video content every minute. YouTube generates billions in revenue through advertising and also shares the revenue with the popular users that upload videos through the Partner program. YouTube is clearly a social media site, however is YouTube also a social networking site? In classical online social networks the interaction is directly between users--that is, user-user interactions. However YouTube is unique in that the interaction between users includes video content--that is, the interaction follows users-content-users. In fact the interaction between users is incentivized using the posted videos. In this way it is not merely the interest preferences between users that promote user-user interaction, but also the content of the videos that governs the social interactions between users. 

Using real-world data consisting of over $6$ million videos spread over $25$ thousand channels, we empirically examine the sensitivity of YouTube \metalevel features on the engagement dynamics of users in YouTube.
Insight into the dynamics of social sensors in YouTube can be used to predict how users will interact with posted video content. These results are important for designing methods for optimizing user engagement and for improving the efficiency of content distribution networks~\cite{HAV17,SHV17,HGKDZ15}. Estimating the popularity of YouTube videos based on \metalevel features is a challenging problem given the diversity of users and content providers. Time-series methods for modeling YouTube the engagement dynamics of videos over time include ARMA time series models~\cite{GCM11}, multivariate linear regression models~\cite{PAG13}, and Gompertz models~\cite{RAEJLP14,REJAL15}. These methods do not utilize any of the meta-level features of the YouTube video to estimate the users engagement dynamics. In~\cite{Zha15} a bag-of-words Bernoulli Naive Bayes classifier is applied to perform a binary classification (popular/unpopular) of YouTube videos based on the title alone. The classifier was able to achieve a classification accuracy of 66\%. In~\cite{HAV17} visual perception and extreme learning machines were applied to the meta-level features of videos and found to be able to accurately estimate the (popular/unpopular) videos with an accuracy of 80\%. It was determined that the main meta-level features that impact video engagement include: first day \viewcount, number of subscribers, and contrast of the video thumbnail. 

The above methods focus on how to estimate user engagement to specific videos; however, they do not consider the social learning dynamics that are present between users and channel owners. The key topics focused on in Sec.\ref{sec:SocialInteractionofChannelOwnersandYouTubeConsumers} are: (i) how user engagement is affected by changes in meta-level (title, thumbnail, tag) features of the videos, (ii) the causal relationship between channel subscribers and user engagement, (iii) the engagement dynamics of videos over time with exogenous social media events, and (iv) the engagement of users to videos in a channel's video playlist. The insight provided can be used by channel owners to design policies for maximizing user engagement by adjusting video meta-level features, promoting on external social media venues, and periodically adjusting the uploading schedule of videos.

\subsection{Perspective}

The  unifying theme  that underpins  the  four topics in this chapter  stems from statistical signal processing and controlled sensing. These are used
to predict global behavior given local behavior:  individual social sensors 
interact with other sensors 
 and we are interested in understanding the behavior of the entire network.
Information  diffusion, social learning  and revealed preferences
 are important issues for  social sensors. 
We  treat these issues in a highly stylized manner so as to provide easy accessibility to a signal processing audience.
The underlying  tools
used in this chapter are widely used in signal processing,  economics  and network science.

\begin{figure}[h] \centering
\begin{tikzpicture}[node distance = 5cm, auto]
    \node [block] (BLOCK1) {Network Diffusion Models};
     \node [block,right of =BLOCK1] (youtube) {User Interaction in YouTube};
    \node [block, below of=BLOCK1,node distance=1.5cm] (BLOCK2) {Bayesian Social Learning Models};
    \node[block, right of=BLOCK2] (grad) {Revealed Preferences};
 
    

\end{tikzpicture}
\caption{Four themes considered in this chapter}
\label{fouraspects}
\end{figure}
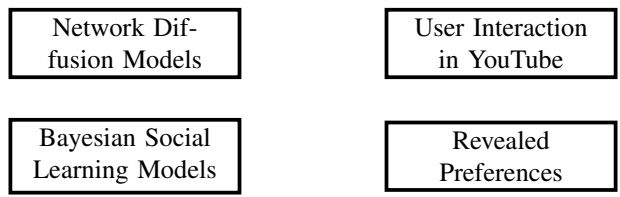

Let us briefly discuss how the four themes of this chapter interact; these four themes are depicted in Fig.\ref{fouraspects}.

The network diffusion models are non-Bayesian and describe the behavior of large numbers of social sensors.
The mean field dynamics model for the diffusion of information has the form of an averaged stochastic approximation algorithm
(which is widely used in adaptive filtering). Note however, here that the stochastic approximation type equation is a generative model, and
not an algorithm. 

The Bayesian social learning models in contrast describe highly stylized individual behavior of social sensors. At this level,
it is important to model risk-averse human decision making and the Bayesian social learning model serves as a useful 
generative model.

Underpinning both the network diffusion and Bayesian social learning models, are utility (cost) functions which the social sensors 
optimize in order to make decisions. The natural question is: Given real world data, is the behavior of agents consistent with optimizing
a utility function? If yes, can the utility function be estimated?  Revealed preferences yield a useful set of algorithms that can answer
both these questions. More generally, it can be used to detect play from the Nash equilibrium of a potential game.
Put simply, revealed preferences provide the data-driven justification for the utility function models.

Finally, the detailed analysis of the YouTube data provides for an  interesting real world study of how social sensors interact.
It is important to note that while YouTube is clearly a social media site, it  is also a social networking site. Classical online social networks (OSNs) are dominated by user-user interactions. However YouTube is unique in that the interaction between users includes video content?that is, the interaction follows users-content-users.
The interaction between users in the YouTube social network is incentivized using the posted videos.
In addition to the social incentives, YouTube also gives monetary incentives to promote users increasing their popularity. As more users view and interact with a users video or channel, YouTube will pay the user proportional to the advertisement exposure on the users channel. Therefore, users not only maximize exposure to increase their social popularity, but also for monetary gain which introduce unique dynamics in the formation of edges in the YouTube social network.

\subsubsection*{Books and Tutorials}
The literature in social learning, information diffusion and revealed preferences  is extensive. In each of the following sections, we provide  a brief review of relevant works.
Seminal books in social networks,  social learning  and network science include \cite{Veg07,Jac10,Cha04,EK10}. 
There is  a growing literature dealing with  the interplay of technological  and social networks \cite{CCP13}. Social networks overlaid on technological networks account
for a significant fraction of Internet use. As discussed in \cite{CCP13}, three key aspects  that cut across social and technological networks are the emergence of global coordination through local actions,
resource sharing models and the wisdom of crowds. These themes  are addressed in the current chapter in the context of social learning, diffusion and revealed preferences. Other tutorials include \cite{KP14,KNH14}.

\section{Information Diffusion in Large Scale Social Networks} \label{sec:diffusion}
This section addresses the first topic of the chapter, namely, information diffusion models and their mean field dynamics in social
networks.  The setting is as follows: The  states of individual nodes in the social network evolve  over time as a probabilistic function of the states of their neighbors {\em and} an underlying
target process (state of nature). The underlying target  process can be viewed as the market conditions or competing  technologies that evolve with time and affect the information diffusion.
The  nodes in the social network are sampled randomly to determine their state.
As the adoption of the new technology diffuses through the network, its effect is observed via sentiments  (such as tweets) of these selected members of the population.  
 These selected nodes act as social sensors.
In signal processing terms, the underlying target  process can be viewed as a signal, and the social network can be viewed as a sensor.
The key difference compared to classical sensing is that the sensor now is a social network with diffusion dynamics and noisy measurements (due to sampling nodes).

As described  in Sec.\ref{sec:introduction},  
a wide range of social phenomena such as diffusion of technological innovations, cultural fads, ideas, behaviors, trends and economic conventions~\cite{Gra78,MR07,Che09,Cha04} can be modelled by diffusion  in social networks.
Another important application is sentiment analysis (opinion mining) where the spread of opinions amongst people is monitored via
social media.

Motivated by the above setting, this section  proceeds as follows:
\begin{compactenum}
\item We describe  the  Susceptible-Infected-Susceptible (SIS) model  for  diffusion of  information in social networks 
which has been extensively studied in~\cite{Jac10,Pin06,Pin08,PV01,Veg07}.
\item Next, it is shown how the dynamics of the infected degree distribution of the social network can be approximated by
the mean 
field dynamics. The mean field dynamics state that as the number of agents 
in the social network goes to infinity, the dynamics of the infected degree distribution converges to that of  an 
ordinary differential (or difference) equation. Such averaging theory results are widely used to analyze adaptive filters. For social networks, they yield a useful tractable model for the diffusion dynamics.

\item 
We illustrate the diffusion model by using data sets from 3 related social networks to track the  spread of influenza during the period  September 1 to December 31, 2009.
The friendship network of 744 undergraduate students at Harvard college is used together with the  U.S. outpatient Influenza-like Illness Surveillance Network (ILInet) to monitor the spread of influenza. Then it is shown that  Twitter posts related to influenza during this period 
are correlated with the spread of influenza.
Thus in this example, influenza diffuses in  a human health network (Harvard friendship network at a local level and ILInet at a global level) and  
Twitter  is used as a social sensor to monitor the spread of the influenza.
\item 
Finally, this section also  describes how 
social networks can be sampled.
We  review two recent methods for sampling social networks, namely, social sampling and respondent-driven sampling; the latter being used in health  networks.
\end{compactenum}

 \begin{figure}
  \centering



\subfigure[Dynamics of Health Network impact the \# of tweets.]{\label{fig:ARMAXintroa}
\begin{tikzpicture}[font = \small, scale =0.8,transform shape, american voltages]
\node[inner sep=0pt] at (-2.85,1.25) {\includegraphics[width=3.7cm,height=2.5cm]{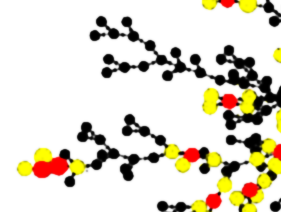}};
\draw [fill= none] (-5,0) coordinate (topleft) rectangle (-1,3) coordinate (bottomright);
\node at (-3,2.75) {\normalsize Influenza Health Network};
\node[inner sep=0pt] at (-3,-2.5) {\includegraphics[width=3.7cm,height=2.5cm]{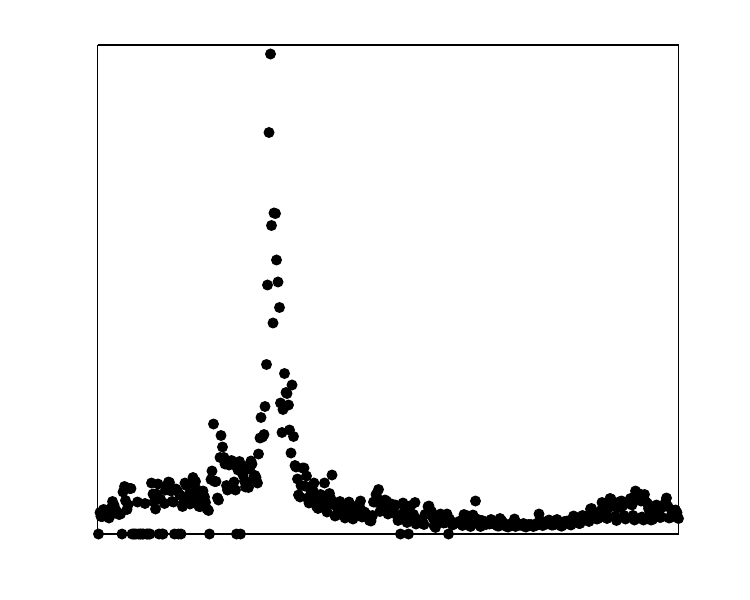}};
\draw [fill= none] (-5,-4) coordinate (topleft) rectangle (-1,-1) coordinate (bottomright);
\node at (-3,-1.25) {\normalsize Twitter (Social Sensor)};
\node[rotate=90] at (-4.75,-2.5) {\# tweets $\tau_\dtime$};
\node at (-3,-3.75) {Days};
\draw[thick, black, ->] (-3,0) -- (-3,-1) node[midway,xshift=-18pt] {Infected} node[midway,xshift=15pt] {Nodes};
\end{tikzpicture}
 }
\subfigure[Model for the number of tweets resulting from the Influenza Health Network.]{\label{fig:fig:ARMAXintrob}
\begin{tikzpicture}[font = \small, scale =0.8,transform shape, american voltages]
\node[draw,thick,rectangle,minimum width=3cm,minimum height=1.0cm,align=center] (R1) at (-3,1.5) {Susceptible-Infected-Susceptible \\ Diffusion Model};
\draw [fill= none,draw=none] (-5,-4) coordinate (topleft) rectangle (-1,-1) coordinate (bottomright)node[pos=0.5] {};
\node[draw,thick,rectangle,minimum width=4.5cm,minimum height=0.75cm] (R2) at (-3,-2.5) {Time-series model for tweets};
\draw[thick, black, ->] (R1.south) -- (R2.north) node[midway,xshift=-23pt] {Number of} node[midway,xshift=30pt] {Infected Nodes};
\end{tikzpicture}
 }
  \caption{Dynamics of Health Network are modelled using the SIS model and a Linear and Nonlinear Autoregressive with Exogenous input time series models, refer to Sec.\ref{sec:diffusion} for details.} 
\label{fig:schematicsocialnetwork}
\end{figure}
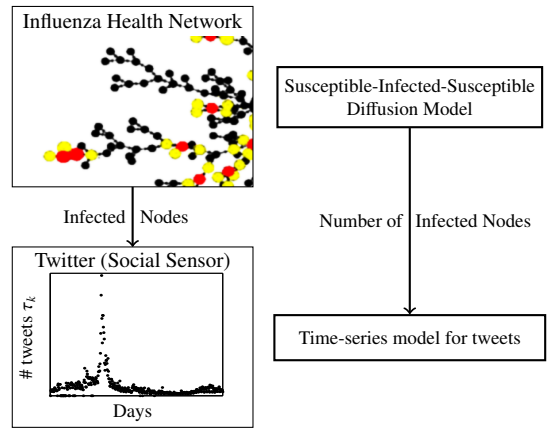

The  aim is to estimate the underlying target state that is being sensed by the social network
and also and the state probabilities of the nodes by sampling  measurements at nodes in the social network.
In a Bayesian estimation context, this is equivalent to a  filtering problem involving estimation of the state of a prohibitively large scale Markov chain in noise. The  mean field dynamics yields
a tractable  approximation with provable bounds
for the information diffusion.
Such mean field dynamics have been studied in~\cite{BW03} and applied to social networks in~\cite{Pin06,Pin08,Veg07}.
For an excellent recent  exposition of interacting particle systems comprising of agents each with a finite state space, see~\cite{Ald13}, where the more apt term
``Finite Markov Information Exchange (FMIE) process'' is used. 

Regarding real datasets, in addition to the case study presented below,
for other examples of diffusion datasets and their analysis see \cite{SRM09,RMK11}. A repository of social network datasets can be obtained at \cite{SNAP}.

\subsection{Social Network Model}
\label{subsec:SocialNetworkModel}

A social network is modelled as a  graph with $N$ vertices:
\begin{equation}
\network = (\Vertexset,\edgeset),  \text{ where } \Vertexset= \vertexset, \text{ and } \edgeset  \subseteq \Vertexset \times \Vertexset.
\end{equation}
Here, $\Vertexset$ denotes the finite set of vertices, and $\edgeset $ denotes the set of edges.
 In social networks, it is customary to use the terminology {\em network}, {\em nodes} and {\em links} for {\em graph}, {\em vertices} and {\em edges}, respectively.

We use the notation $(\nodem,\noden)$ to refer to a link between node $\nodem$ and $\noden$. The network may be undirected in which case $ (\nodem,\noden) \in \edgeset$ implies $(\noden,\nodem) \in \edgeset$. In undirected graphs, to simplify notation, we use the notation $\nodem,\noden$ to denote the undirected link between node $\noden$ and $\nodem$. If the graph is directed, then $(\nodem,\noden) \in \edgeset$ does not imply that $(\noden,\nodem) \in \edgeset$. We will assume that self loops (reflexive links) of the form $i,i$  are excluded from $\edgeset$.

An important parameter  of a social network $\network = (\Vertexset,\edgeset)$ is the connectivity of its nodes.
Let $\nbhood{\nodem}$ and $\degreediff{\nodem}$ denote the neighbourhood set
and   degree (or connectivity)  of a node $\nodem \in \Vertexset$, respectively. That is,  
with $|\cdot|$ denoting cardinality,
\begin{equation}
\nbhood{\nodem} =  \{\noden \in \Vertexset :  \nodem,\noden \in \edgeset \},
\quad \degreediff{\nodem} =  \big| \nbhood{\nodem}\big|.
\end{equation}
For convenience, we assume that the maximum degree of the network is uniformly bounded by some fixed integer $\degmax$.

Let $\vertexnum{\deg}$ denote the number of nodes with degree $\deg$, and let the degree distribution
 $ \degdist{\deg}$ specify the fraction of nodes with degree $\deg$. That is, for $\deg=0,1,\ldots,\degmax$,
$$
\vertexnum{\deg} =  \sum_{\nodem \in \Vertexset}  \indicator\lbr\degreediff{\nodem} = \deg\rbr , \quad
\degdist{\deg} =  \frac{ \vertexnum{\deg} }{\vertexnum} .
 $$
Here, $\indicator\lbr\cdot\rbr$ denotes the indicator function. Note that
 $\sum_\deg \degdist{\deg}=1$. The degree distribution  can be viewed as the probability that a node selected  randomly with uniform distribution on $\Vertexset$ has a connectivity $\deg$.

 Random graphs generated to have a degree distribution $\degdist$ that is Poisson were formulated by Erd\"{o}s and Renyi~\cite{ER59}.
 Several recent works show that large scale social networks are characterized by connectivity distributions that are different to Poisson distributions.
 For example, the internet, www have a power law connectivity distribution $\degdist(\deg) \propto \deg^{-\gamma}$, where $\gamma$ ranges between 2 and 3.
Such scale free networks are studied in~\cite{BR99}. In the rest of this chapter, we assume that the degree distribution of the social network is arbitrary but known---allowing an arbitrary degree distribution facilities modelling complex networks.

Let $\dtime = 0,1,\ldots$ denote discrete time.
Assume the target process $\target$ is a finite state Markov chain with transition probability
\begin{equation}
\tptarget{\tstate}{\tstatep} = \P\left(\target{\dtime+1}=\tstatep  |  \target{\dtime} = \tstate \right).
\end{equation}
In the example of technology diffusion, the target process can denote the availability of competition or market forces that determine whether a node
adopts the technology.  In the model below, the target state will affect the  probability that an agent adopts the new technology.

\subsection{SIS Diffusion Model for Information in Social Network}
The model we present below for the diffusion of information in the social network is called the 
{\em Susceptible-Infected-Susceptible (SIS)} model \cite{PV01,Veg07}.
The diffusion of information  is modelled by the time evolution of the state of individual nodes in the network.
Let  $\state{\dtime}{\nodem} \in \{0,1\}$ denote the state at time $\dtime$ of each node $\nodem$ in the social network.
Here, $\state{\dtime}{\nodem}= 0$ if the agent at time $\dtime$ is susceptible and $\state{\dtime}{\nodem} = 1$  if the agent is infected.
At time $\dtime$, the state vector of the $\vertexnum$ nodes is
\begin{equation}
\state{\dtime} = \lb\state{\dtime}{1} , \ldots, \state{\dtime}{\vertexnum}\rb^\p \in \{0,1\}^{\vertexnum}.
\end{equation}

Assume that the process $\statev$  evolves as a discrete time Markov process with transition
law depending on the target state $\target$.
If node $\nodem$ has degree $\degreediff{\nodem}=\deg$, then the probability of node $\nodem$ switching from state $\statea$ to $\stateb$ is
\begin{equation}
\P\left(\state{\dtime+1}{\nodem} =\stateb | \state{\dtime}{\nodem}= \statea, \state{\dtime}{i-},\target{\dtime}=\tstate\right) = \tpdiff{\statea\stateb}{\deg}{\nactive{\nodem}_\dtime}{\tstate},\; \statea,\stateb \in \{0,1\}.
\label{eq:tp}
\end{equation}
Here, $\nactive{\nodem}_\dtime$ denotes the number of infected neighbors of node $\nodem$ at time $\dtime$. That is,
\begin{equation}
\nactive{\nodem}_\dtime = \sum_{\noden \in \neighbor{\nodem} } I\lbr\noden:  \state{\dtime}{\nodem} = 1\rbr.
\end{equation}
In words, the transition probability of an  agent depends on its degree distribution and the number of active neighbors.

With the above probabilistic model, we are interested in modelling the evolution of infected agents over time.
Let   $\infectdist{\dtime}{\deg}$  denote the fraction of infected nodes at each time $\dtime$ with degree $\deg$.
We call $\infectdist$ the {\em infected   node distribution}.
So with $\deg=0,1,\ldots,\degmax$, 
\begin{equation}
\infectdist{\dtime}{\deg} =  \frac{1}{\vertexnum(\deg)} \sum_{\nodem \in \Vertexset} \indicator\lbr\degreediff{\nodem} = \deg,  \state{\dtime}{\nodem} = 1\rbr  .
\label{eq:ind}
\end{equation}
The SIS model assumes that the infection spreads according to the following dynamics:
\begin{enumerate}
\item
At each time instant $\dtime$, a single agent, denoted by $\nodem$,  amongst the $\vertexnum$ agents is chosen uniformly.
%
Therefore, the probability that the chosen agent $\nodem$ is infected and of degree $\deg$  is $\infectdist{\dtime}{\deg}\, \degdist{\deg}$. The probability that the chosen agent $\nodem$ is  susceptible and of degree $\deg$ is $(1-\infectdist{\dtime}{\deg})\, \degdist{\deg}$.
\item
 Depending on whether its state $\state{\dtime}{\nodem}$ is infected
or susceptible, the
 state of agent $\nodem$ evolves according
to the transition probabilities  specified in~(\ref{eq:tp}).
\end{enumerate}

With the Markov chain transition dynamics of individual agents specified above, it is clear that the infected  distribution
$\infectdist{\dtime} =
 \big(\infectdist{\dtime}{1},\ldots, \infectdist{\dtime}{\degmax}\big)$  is an $\prod_{\deg=1}^{\degmax} \vertexnum{\deg}$ state Markov  chain. Indeed,
 given $\infectdist{\dtime}{\deg}$, due to the infection dynamics specified above
\begin{equation}
\infectdist{\dtime+1}{\deg} \in \left\{ \infectdist{\dtime}{\deg} - \frac{1}{\vertexnum{\deg} }, \;
 \infectdist{\dtime}{\deg} + \frac{1}{\vertexnum{\deg} }
\right\}.
\end{equation}
Our aim below is to specify the transition probabilities of the Markov chain $\infectdist$.
Let us start with the following statistic that forms a convenient parametrization of the transition probabilities. Given the infected node distribution $\infectdist{\dtime}$ at time $\dtime$,
define $\pa(\infectdist{\dtime})$ as the probability that at time $\dtime$ a uniformly sampled   link in the network points to an infected node. We call $\pa(\infectdist{\dtime})$ as the
{\em infected
link probability}. Clearly 
{\small \begin{align}
 \pa(\infectdist{\dtime})
&= \frac{ \sum_{\deg=1}^{\degmax}\text{(\# of links from infected node of degree $\deg$)}}
{\sum_{\deg=1}^{\degmax}\text{(\# of links  of degree $\deg$)}}  \nonumber\\
&=
 \frac{ \sum_{\deg=1}^{\degmax} \deg \, \degdist{\deg}\, \infectdist{\dtime}{\deg} } {\sum_{\deg}^{\degmax} \deg \, \degdist{\deg} }.  \label{eq:infectasymp}
\end{align}}
In terms of the infected link probabilities, the scaled transition probabilities\footnote{The transition probabilities are scaled by the degree distribution $\degdist{\deg}$ for notational convenience. Indeed, since $\vertexnum{\deg} = \vertexnum \degdist{\deg}$,
by using these scaled probabilities we can express the dynamics of the process $\infectdist$ in terms of the same-step size $1/\vertexnum$ as described
in Theorem~\ref{thm:minc}.
Throughout this chapter, we
assume that  the degree distribution $\degdist(\deg)$, $\deg \in \{1,2,\ldots,\degmax\}$, is uniformly bounded away from zero.  That is, $\min_\deg \degdist{\deg} > \epsilon$ for some positive
constant $\epsilon$.
} of the process
$\infectdist$ are:
{\small \begin{align}
 \atp{01}{\deg}{\pa{\dtime}}{\tstate} 
&\ole \frac{1}{\degdist{\deg}}\,
\P\left(\infectdist{\dtime+1}{\deg} =  \infectdist{\dtime}{\deg} + \frac{1}{\vertexnum{\deg} } \Big\vert  \target{\dtime}=\tstate \right) \nonumber \\
&\hspace{-1.5cm}= (1-\infectdist{\dtime}{\deg})  \,  \sum_{\aindex = 0}^\deg \tpdiff{01}{\deg}{\aindex}{\tstate} \,  \P
(\text{$\aindex$ out of $l$ neighbours infected})
\nonumber\\
& \hspace{-1cm}= (1-\infectdist{\dtime}{\deg})  \,  \sum_{\aindex = 0}^\deg \tpdiff{01}{\deg}{\aindex}{\tstate} \binom{\deg}{\aindex}
\pa{\dtime}{\aindex}  (1 - \pa{\dtime})^{\deg -\aindex}, \nn
\\
 \atp{10}{\deg}{\pa{\dtime}}{\tstate}  &\ole \frac{1}{\degdist{\deg}}\,
\P\left(\infectdist{\dtime+1}{\deg} =  \infectdist{\dtime}{\deg} - \frac{1}{\vertexnum{\deg} } \Big\vert \target{\dtime} = \tstate\right) \nonumber \\ &
\hspace{-1cm}=
\infectdist{\dtime}{\deg} \,\sum_{\aindex = 0}^\deg \tpdiff{10}{\deg}{\aindex}{\tstate}\,  \binom{\deg}{\aindex}
\pa{\dtime}{\aindex}  (1 - \pa{\dtime})^{\deg -\aindex}.
\end{align}}
In the above, the notation  $\pa{\dtime}$ is the short form for $\pa(\infectdist{\dtime})$.
The  transition probabilities  $\atp{01}$ and $\atp{10}$ defined above model the diffusion of information about the target state $\target$ over the social network.
We have the following martingale representation theorem for the evolution of Markov process $\infectdist$.

Let $\sigmaf_\dtime$ denote the sigma algebra generated by $\{\infectdist_{0},\dots,\infectdist_{\dtime}, \target{0},\ldots\target{\dtime} \}$.

\begin{theorem} For $\deg=1,2,\ldots, \degmax$, the infected distributions evolve as
\begin{multline} \label{eq:minc}
\infectdist{\dtime+1}{\deg} = \infectdist{\dtime}{\deg} + \frac{1}{\vertexnum}
\bigl[
  \atp{01}{\deg}{\pa(\infectdist{\dtime})}{\target{\dtime}}  -
   \atp{10}{\deg}{\pa(\infectdist{\dtime})}{\target{\dtime}}  
    + \minc_{\dtime+1}  \bigr]
\end{multline}
where $\minc$ is a martingale increment process, that is $\E\{\minc_{\dtime+1}| \sigmaf_\dtime \} = 0$. Recall $\target$ is the finite state Markov chain that models  the target process. \label{thm:minc} \qed
\end{theorem}

The above theorem is a well-known martingale representation of a Markov chain \cite{EAM95}---it says that a discrete time Markov process can be obtained by discrete
time filtering of a martingale increment process.
The theorem implies that the infected distribution dynamics resemble what is commonly called a stochastic approximation (adaptive filtering) algorithm in statistical signal
processing: the new estimate is the old estimate plus a noisy update (the ``noise'' being a martingale increment)  that is weighed by a small step size $1/\vertexnum$ when $\vertexnum$ is large. Subsequently, we will exploit the structure in Theorem \ref{thm:minc} to devise a mean field dynamics  \index{mean field dynamics} model which has a state of dimension~$\degmax$.
This is to be compared with the intractable state dimension $\prod_{\deg=1}^{\degmax} \vertexnum{\deg}$ of the Markov chain $\infectdist$.

\subsection{Mean Field Dynamics of Information Diffusion}
\label{subsec:MeanFieldDynamicsofInformationDiffusion}


The mean field dynamics state
that as the number of agents $\vertexnum$ grows to infinity, the dynamics of the infected distribution  $\infectdist$, described by (\ref{eq:minc}),
in the social
network evolves according to the following
 deterministic difference equation that is modulated by a Markov chain that depends on the target state evolution $\target$:

For $\deg=1,2,\ldots, \degmax$,
\begin{align}
 \minfectdist{\dtime+1}{\deg}   &=  \minfectdist{\dtime}{\deg} +  \frac{1}{\vertexnum} \bigl[ \atp{01}{\deg}{\pa(\minfectdist{\dtime})}{\target{\dtime}}  
 -   \atp{10}{\deg}{\pa(\minfectdist{\dtime})}{\target{\dtime}}\bigr]  \nonumber  \\
%
%
\atp{01}{\deg}{\pa}{\target{\dtime}} & = (1-\minfectdist{\dtime}{\deg}) 
 \sum_{\aindex = 0}^\deg \tpdiff{01}{\deg}{\aindex}{\target{\dtime}}
\binom{\deg}{\aindex}
\pa^{\aindex}  (1 - \pa)^{\deg-\aindex}  \nonumber\\
 \atp{10}{\deg}{\pa}{\target{\dtime}} &= \minfectdist{\dtime}{\deg}  \,  \sum_{\aindex = 0}^\deg \tpdiff{10}{\deg}{\aindex}{\target{\dtime}}
\binom{\deg}{\aindex}
\pa^{\aindex}  (1 - \pa)^{\deg-\aindex} \nonumber  \\
%
\pa(\minfectdist{\dtime}) & = \frac{ \sum_{\deg=1}^{\degmax} \deg \, \degdist{\deg}\, \minfectdist{\dtime}{\deg} } {\sum_{\deg}^{\degmax} \deg \, \degdist{\deg} }
   \label{eq:system}
 \end{align}


That the above mean field dynamics follow from~(\ref{eq:minc}) is intuitive. Such averaging results are well known in the adaptive filtering community where they are deployed
to analyze the convergence of adaptive filters. The difference here is that the limit mean field dynamics are not deterministic but Markov modulated. Moreover, the
mean field dynamics here constitute a model for information diffusion, rather than the asymptotic behavior of an adaptive filtering algorithm.
As mentioned earlier, from an engineering point of view, the mean field dynamics yield a tractable model for estimation.

We then have the following exponential bound  result for the error of the mean field dynamics approximation.

\begin{theorem}  \label{thm:mftmatched}
For a discrete time horizon of $\finaltime$ points,  the deviation between the mean field dynamics $\minfectdist{\dtime}$ in (\ref{eq:system}) and actual infected distribution in  $\infectdist{\dtime} $ (\ref{eq:minc}) satisfies
\begin{equation}
\P\lbr \max_{0 \leq \dtime \leq \finaltime} \left\| \infectdist{\dtime} - \minfectdist{\dtime} \right\|_\infty \geq \epsilon\rbr
\leq  C_1  \exp(-C_2 \epsilon^2 \vertexnum)
\end{equation}
where $C_1$ and $C_2$ are positive constants and $\finaltime = O(\vertexnum)$.
\qed
\end{theorem}

The proof of the above theorem follows from~\cite[Lemma 1]{BW03} and is presented in \cite{Kri16}. Actually in \cite{BW03} the mean field dynamics are presented
in continuous time as a system of ordinary differential equations. The exponential bound follows from an application of the Azuma-Hoeffding inequality.
The above theorem provides an exponential bound (in terms of the number of agents $\vertexnum$)  for the probability of  deviation of the sample path of the infected distribution from the mean field dynamics for any finite time interval~ $\finaltime$.

The stochastic approximation and adaptive filtering literature~\cite{BMP90,KY03} has several averaging analysis methods for
recursions of the form~(\ref{eq:minc}).  The well studied mean square error analysis~\cite{BMP90,KY03} computes bounds on $ \E\| \minfectdist{\dtime} - \infectdist{\dtime} \|^2$ instead of the maximum
deviation in Theorem \ref{thm:mftmatched}. A mean square error analysis of estimating a Markov modulated empirical distribution is given in~\cite{YKI04}. Such mean
square analysis assume a finite but small step size $1/\vertexnum$ in~(\ref{eq:minc}).

\subsubsection*{Related Literature}
Given the above SIS model, it is appropriate to pause briefly and review related literature.
There are several other models for studying the spread of infection and technology in complex networks including Susceptible-Alert-Infected-Susceptible~(SAIS), and Susceptible-Exposed-Infected-Vigilant~(SEIV); see \cite{Het00,EK10}.  Susceptible-Infected-Susceptible (SIS) models  have been extensively studied in~\cite{Pin08,Jac10,PV01,Veg07,KNH14} to model information/infection diffusion, for example, the adoption of a new technology in a consumer market. 

Degree-based mean field dynamics approximations for SIS models have been derived in \cite{Pin08,PG16}. Pair approximations (PA) and approximate master equations (AME) yield more general  models for the complex dynamics of large scale networks  \cite{PG16}. However, the resulting differential/difference equations that characterize the dynamics in PA and AME are no longer  polynomial functions of the state. In this more general case, however, a suboptimal filter such as a particle filter can be used to track the infection diffusion.

It is also important to note that the right hand side of the mean field difference equation (\ref{eq:system}) is a polynomial function of  the infected degree distribution $\minfectdist$.
As a result, when the graph is sampled, resulting in noisy observations of $\minfectdist$, one can construct an exact finite dimensional Bayesian filter for the conditional
mean estimate of $\minfectdist$ at each time $k$ using the filtering algorithms  in \cite{HB14}. We refer the reader to \cite{KBP17} for details and also posterior Cramer-Rao
lower bounds for estimating the infected degree distribution in the case of Erdos-R\'enyi  and also power law (scale free) networks such as Twitter.
In comparison, \cite{HVY14} provides a stochastic approximation algorithm and analysis on a Hilbert space for tracking the degree distribution of evolving random networks with a duplication-deletion model.

On networks having fixed degree distribution, \cite{Pin08}
identified conditions under which a network is susceptible
to an epidemic using a mean-field approach and provided
a closed form solution for the infection diffusion threshold.
The diffusion properties of networks was investigated using
stochastic dominance of their underlying degree distributions
like in \cite{JR07}.  We generalize these stochastic dominance results for evolving networks by considering a simple preferential attachment model as this can generate a scale-free network \cite{GCB13}. 

 Finally, \cite{PV01} studies the link between the power law exponent and the diffusion threshold. For the preferential attachment model, \cite{GCB13} studies the connection between the parameters that dictate the evolution (node and edge addition probability) and the degree distribution. \cite{KBP17} has similar results using stochastic dominance, but, the key emphasis is on providing a structured way to study such ordinal sensitivity relationships in large networks.

\subsubsection*{Numerical Example}
We simulate the diffusion of information through a network comprising of $ N=100$ nodes (with maximum degree $\degmax = 17$).
It is  assumed that at time $\dtime = 0 $, $5\%$ of nodes are infected. The mean field dynamics model is investigated in terms of the infected link probability~(\ref{eq:infectasymp}). The infected link probability $\pa(\infectdist{\dtime})$ is  computed using ~(\ref{eq:system}).

Assume each agent is a myopic optimizer  and, hence, chooses to adopt the technology only if $\cost{\nodem} \leq  \nactive{\nodem}_\dtime$; $\reward = 1$. At time $k$, the costs $\cost{\nodem}$, $m = 1,2,\ldots,100$, are i.i.d. random variables simulated from uniform distribution $U[0,C(\target{\dtime})]$. Therefore, the transition probabilities  in (\ref{eq:tp}) are
\begin{equation}
\tpdiff{01}{\deg}{\nactive{\nodem}_\dtime}{\target{\dtime}} = \P\left( \cost{\nodem} \leq   \nactive{\nodem}_\dtime\right)
=\left\{\begin{array}{l} \frac{\nactive{\nodem}_\dtime)}{C(\target{\dtime})}, \quad \nactive{\nodem}_\dtime \leq C(\target{\dtime}), \\ 1, \quad\nactive{\nodem}_\dtime > C(\target{\dtime}).  \end{array}\right. \label{eq:ex1}
\end{equation}

\begin{figure}[t]
\centering
  \includegraphics[width=0.6\linewidth]{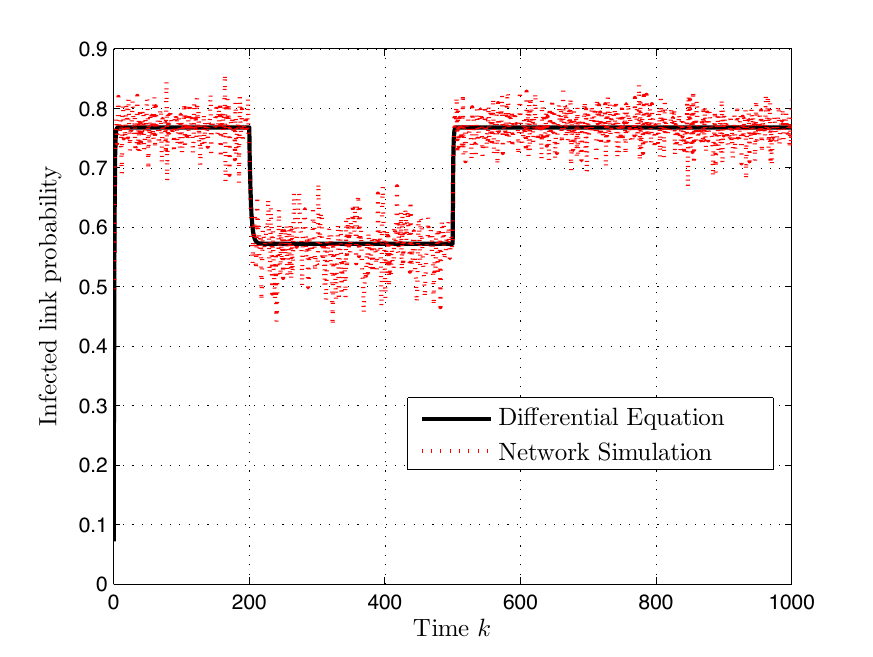}
 \caption{The infected link probability obtained from network simulation compared to the one obtained from the mean field dynamics model~(\ref{eq:system}).  The transition probabilities in (\ref{eq:ex1}) depend only on the  number of infected neighbors $\nactive{\nodem}_\dtime$ (the parameters are defined in Scenario~1).}
\label{fig:diffusion}
\end{figure}

The probability that a product fails is  $\failprob =0.3$, i.e.,  $$\tpdiff{10}{\deg}{\nactive{\nodem}_\dtime}{\target{\dtime}} = 0.3.$$ The infected link probabilities obtained from network simulation~(\ref{eq:infectasymp}) and from the discrete-time mean field dynamics model~(\ref{eq:system}) are illustrated in Figure~\ref{fig:diffusion}. To illustrate that the infected link probability computed from (\ref{eq:system}) follows the true one (obtained by network simulation), we assume that the value of $C$ jumps from $1$ to $10$ at time $\dtime = 200$, and from $10$ to $1$ at time $\dtime = 500$. As can be seen  in Figure~\ref{fig:diffusion}, the mean field dynamics provide an excellent approximation to the true infected distribution.

\subsection{Example: Social Sensing of Influenza  using Twitter}
In this section, we utilize datasets from 3 different social networks (namely, (i)
 Harvard college social network, (ii)  influenza datasets from the U.S Centers for Disease Control and Prevention (CDC) and  (iii) Twitter,   to show how
Twitter can be used as a real time social sensor for detecting outbreaks of  influenza. 

\subsubsection{Twitter as a social sensor}
 A key advantage of using social media for rapid sensing of disease outbreaks  in health networks is that it is low cost and provides rapid results compared with traditional techniques. For example, CDC must contact thousands of hospitals to query the data which causes a reporting lag of approximately one to two weeks~\cite{Cul13}. 
Using real time microblogging platforms such as Twitter for disease detection has several advantages: the tweets are publicly available, high tweet posting frequency users often provide meta-data (i.e. city, gender, age), and Twitter contains a diverse set of users~\cite{Cul13}. 

Several papers have considered using Twitter data for estimating influenza infection rates. In \cite{SSP11,ZLLWZ14} support vector regression supervised learning algorithms is used to relate the volume of Twitter posts that contain specific words (i.e. {\it flu, swine, influenza}) to the number of confirmed influenza cases in the U.S. as reported by the CDC. Multiple linear regression~\cite{Cul10,Cul13b}, and unsupervised Bayesian algorithms~\cite{LC13} have been used to relate the number of tweets of specific words to the influenza rate reported by the CDC. The detection algorithms~\cite{SSP11,Cul10,LC13} do not consider the dynamics of the disease propagation and the dynamics of information diffusion in the Twitter network. To reduce the effect of information diffusion in the network,  \cite{BPD13} proposes a support vector machine (SVM) classifier to detect: a) if the tweet indicates the users awareness of influenza or indicates the user is infected, and b) if the influenza reference is in reference to another person. The classified tweets are then used to train a multiple linear regression model. To account for the diffusion dynamics of Twitter \cite{AGLYL12,AGLYL11} utilize an Autoregressive with Exogenous input (ARX) model. The exogenous input is the number of unique Twitter users with influenza related tweets, and the output is the number of infected users as reported by the CDC. 

If the social network is known then the influenza spread can be formulated in terms of the diffusion model (\ref{eq:minc}). Given the U.S. population of several
hundred millions, it is reasonable to adopt the  mean field dynamics  (\ref{eq:system}). With the influenza infection rate modelled using (\ref{eq:system}), the results can be used as an exogenous input to an ARX or Nonlinear ARX (NARX) models to predict the volume of Twitter messages related to 
influenza  as illustrated in Fig.\ref{fig:schematicsocialnetwork}. In this framework,  the Twitter messages are used to validate the underlying propagation model of influenza of use for predicting the infection rate and outbreak detection. 

\subsubsection{Social Network Influenza Dataset}
\label{subsec:TwitterandInfluenzaExperimentalData}
We consider the dataset \cite{CF10} obtained from a social network of 744 undergraduate students from Harvard College.  The health of the 744 students was monitored from September 1, 2009 to December 31, 2009 and was reported by the university Health Services. To construct the social network, students were presented with a background questionnaire. 
 In the questionnaire students are asked: ``Please provide the contact information for 2-3 Harvard College students who you know and who you think would like to participate in this study'', and ``...provide us with the names and contact information of 2-3 of your friends...''. This information was used to construct the degree distribution and links of the social network. A movie containing the spread of the influenza in the 744 college students over the 122 day sampling period can be viewed as the Youtube video titled ``Social Network Sensors for Early Detection of Contagious Outbreaks" at
{\tt  \url{http://www.youtube.com/watch?v=0TD06g2m8qM}}. Fig.\ref{fig:harvardsocialnetwork}(a-c) display 3 illustrative snapshots from this video;
red nodes denote infected students while yellow nodes depict their neighbors in the social network.

\subsubsection{Models for Influenza Diffusion}
\label{subsubsec:ModelsforInfluenzaDiffusion}

\begin{figure}[htp]
  \centering
  \subfigure[October 10, 2009]{\label{fig:harvardsocialnetworka} \fbox{\includegraphics[angle=0,width=2.0in]{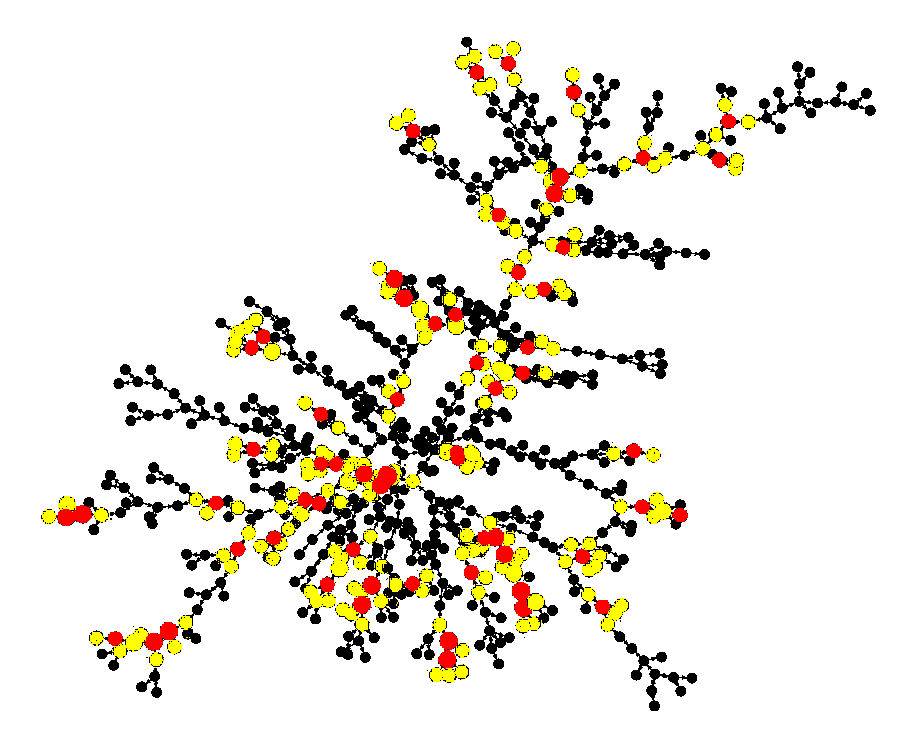}}}
   \subfigure[November 11, 2009]{\label{fig:harvardsocialnetworkb} \fbox{\includegraphics[angle=0,width=2.0in]{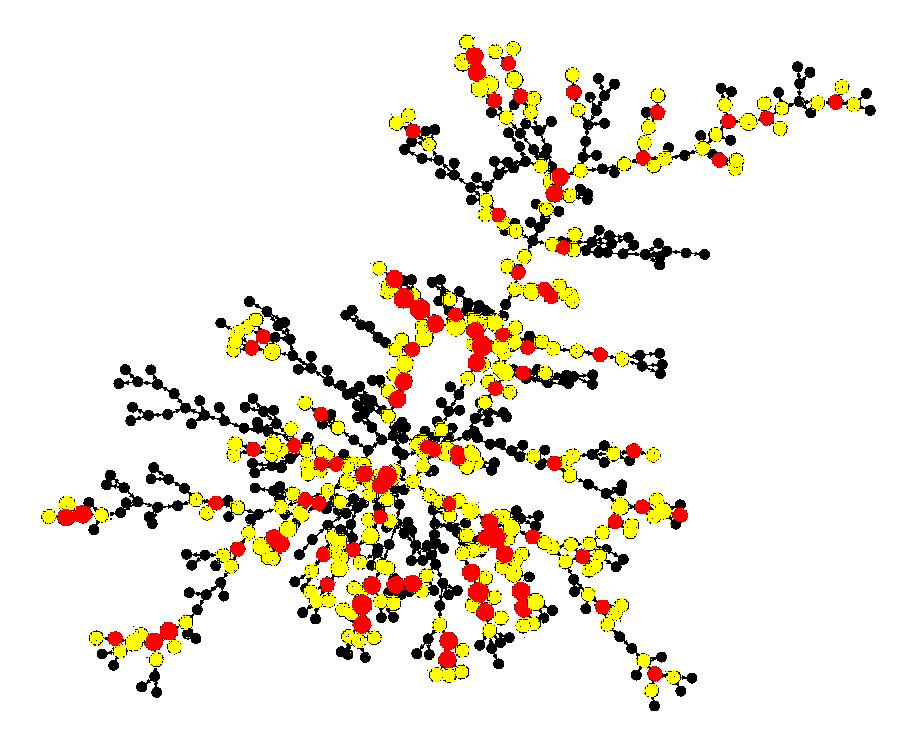}}} \\
   \subfigure[December 23, 2009]{\label{fig:harvardsocialnetworkc} \fbox{\includegraphics[angle=0,width=2.0in]{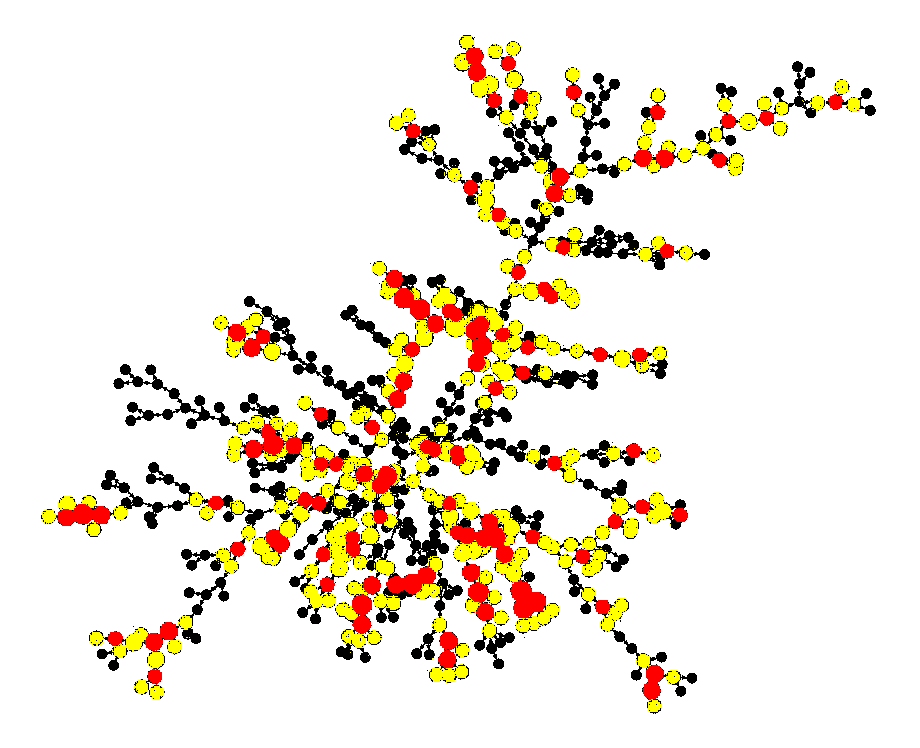}}}
\subfigure[Oct.\ 10 to Dec.\ 23, 2009]{\label{fig:harvardsocialnetworkd} \includegraphics[angle=0,width=2.3in]{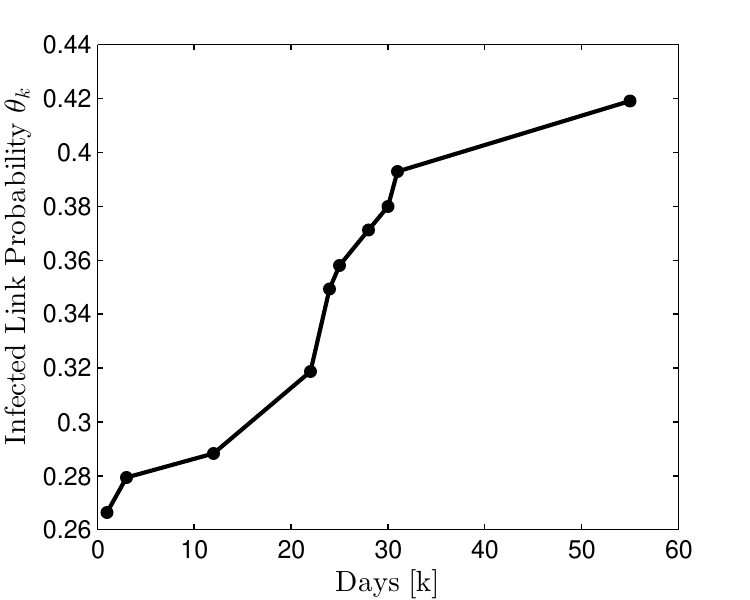}}
\vspace{-5pt}
\caption{Snapshots from youtube video of Harvard undergraduate social network propagation of influenza and the estimated infected link probability $\pa_k$ (\ref{eq:infectasymp}) for October 10 to December 23, 2009.}
\label{fig:harvardsocialnetwork} 
\end{figure}

From 
the data in  the youtube video for the Harvard students, 
 we observed the following regarding the transition probabilities $\tpdiff{\statea\stateb}{\deg}{A}{\tstate}$ defined in (\ref{eq:tp}).
As expected, students with a larger number of infected neighbors  $A$ contract influenza sooner.
The data shows  that
the transition probabilities were approximately independent of the degree of the node $d$.
Since the data provided 
was during an actual influenza outbreak we set 
the target state of the network (i.e. $\target$) constant. Therefore the transition probabilities  depend only on the number of infected neighbours and 
  were estimated as
$$\tpdiff{01}(a=1) = 0.02, \; \tpdiff{01}(a=2) = 0.15.$$ That is, the dataset reveals that the probability of getting infected given
$a=2$ infected neighbors is substantially higher than with $a=1$ infected neighbor, as expected. The estimated infected link probability  $\pa_k$ in (\ref{eq:infectasymp}) versus time (days) $k$ is displayed in Fig.\ref{fig:harvardsocialnetwork}(d). Recall from Sec.\ref{subsec:MeanFieldDynamicsofInformationDiffusion} that the infected link probability $\pa_k$ is related to the mean field dynamics equation (\ref{eq:system}). This allows the transition probabilities and $\pa_k$ to be used to predict the infection rate dynamics.

Other graph-theoretic measures also play a role in the analysis of the diffusion. Students with high $k$-coreness\footnote{$k$-coreness is the largest subnetwork comprising nodes of degree at least $k$}
  are expected to contract influenza earlier. 
  %
 Additionally, students that have high betweenness centrality (i.e. number of shortest paths from all students to all others that pass through that student) contract influenza earlier then students with low betweenness centrality. These observations show that the diffusion of  influenza in the network depends  strongly on the underlying health network structure. The dynamic model (\ref{eq:ind}) accounts for the effects of the degree of nodes, however to account for the effects from betweenness centrality and $k$-coreness would require a more sophisticated formulation then that presented in Sec.\ref{subsec:SocialNetworkModel}.

\subsubsection{Time series model for Influenza Tweets}
\label{subsec:ARMAXforPredictingInfluenzaTweets}
In Sec.\ref{subsubsec:ModelsforInfluenzaDiffusion} we illustrated how the mean field dynamics model (\ref{eq:system}) can be used to estimate the influenza infection rate with the model parameters estimated from a sampled set of the entire population. To validate the estimated parameters for the entire network requires that the infection rate be related to an observable response, in this case the number of Twitter mentions of a specific keyword. Two time series models are considered for relating the infection rate to the number of Twitter mentions. The models are validated using two real-world datasets of Twitter mentions and number of influenza cases in the U.S..

The number of influenza cases in the U.S. is obtained from the CDC\footnote{http://gis.cdc.gov/grasp/fluview/fluportaldashboard.html} which publishes weekly reports from the U.S. outpatient Influenza-like Illness Surveillance Network (ILInet). The data reported by the CDC is comprised of reports from over 3000 health providers nationwide and was obtained for the dates between September 1, 2012 to October 1, 2013. The associated Twitter data for the 122 day period was obtained using the software PeopleBrowsr\footnote{http://gr.peoplebrowsr.com/}. The pre-specified Twitter search terms used are: {\it flu, swine and influenza}. Since our focus is on monitoring influenza dynamics in the U.S., we excluded all tweets as tagged as originating from outside the U.S. The total number of mentions of a specific keyword on each is obtained using PeopleBrowsr. 

We used two time series models for the volume of tweets and compared their performance. The first time series model considered is 
 the  ARX model defined by:
\begin{equation}
\tweet_\dtime=\sum_{i=1}^{n_a}a_i\tweet_{\dtime-i}+\sum_{i=0}^{n_b-1}b_i\infectdist_{\dtime-\Delta-i}+d+v_\dtime.
\label{eqn:ARMAX}
\end{equation}
In (\ref{eqn:ARMAX}), $\tweet_\dtime$ is the number of influenza related tweets at $\dtime$, $\infectdist_\dtime$ is the exogenous input of the infected influenza patients, $n_a,n_b,a_i,b_i,\Delta$ and $d$ are model parameters with $v_\dtime$ an iid noise process. $\Delta$ models the delay between patient contraction, and the respective individual tweeting their symptoms. $d$ models the mean number of tweets related to influenza that are not related to an actual infection. 

The second time series model we used is the nonlinear autoregressive exogenous (NARX) model given by:
\begin{equation}
\tweet_\dtime=F(\tweet_{\dtime-1},\dots,\tweet_{\dtime-n_a}, \infectdist_{\dtime-\Delta},\dots,\infectdist_{\dtime-\Delta-n_b})+v_\dtime.
\label{eqn:NARX}
\end{equation}
In (\ref{eqn:NARX}) $F$ denotes a nonlinear function which relates the exogenous input and previous tweets to the current number of tweets. Here we consider $F$ as a support vector machine which can be trained using historical data. Note that if $F$ was independent of previous tweets, previous exogenous inputs, and no delay (i.e. $n_b=0$ and $\Delta=0$), then (\ref{eqn:NARX}) would be identical to the SVM classifier used in \cite{SSP11,ZLLWZ14} to relate the number of tweets to number of infected agents.  

\begin{figure}
  \centering
  \subfigure[Experimental data of \# of tweets and \# of Infected with influenza.]{\label{fig:ARMAXinfa}\includegraphics[scale=0.45]{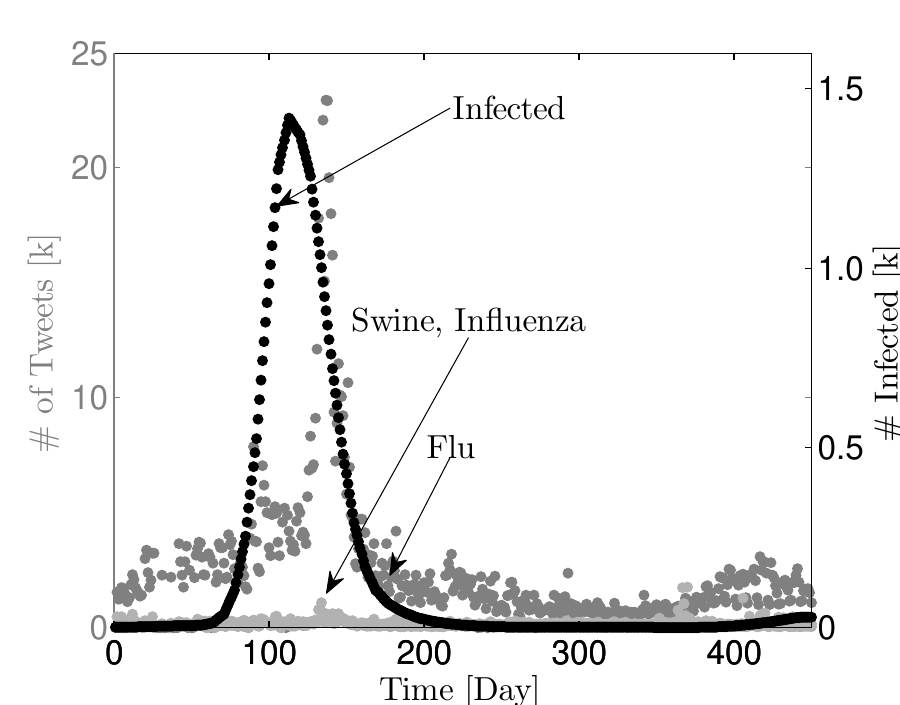}}
\subfigure[ARX model (\ref{eqn:ARMAX}) and NARX model (\ref{eqn:NARX}) for the \# of tweets given the \# of infected influenza patients. The ARX and NARX models are trained using the initial 200 days of data. The predictive accuracy of the models is illustrated for the remaining 230 days.]{\label{fig:ARMAXinfb}\includegraphics[scale=0.55]{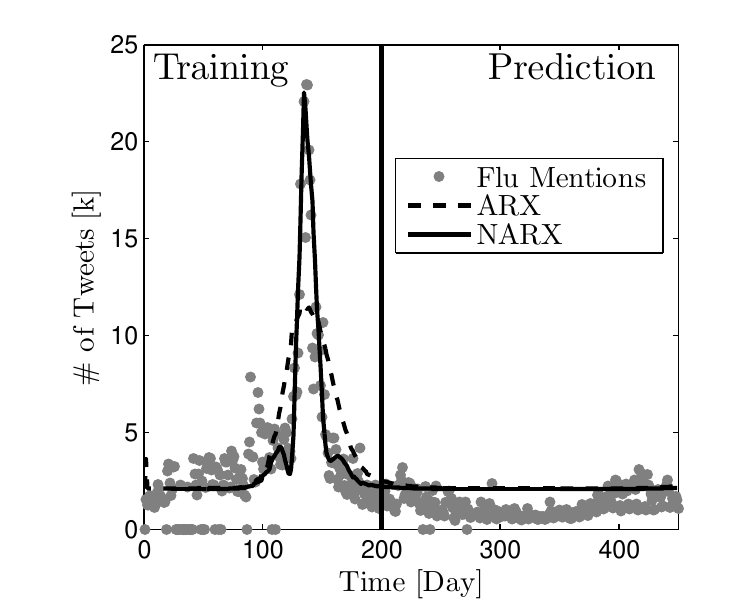}}
\vspace{-5pt}
\caption{The experimental data is obtained for September 2012 to October 2013 as described in Sec.\ref{subsec:TwitterandInfluenzaExperimentalData}. The ARX (\ref{eqn:ARMAX}) and NARX (\ref{eqn:NARX}) models are utilized to estimate the number of tweets with {\it flu} given the number of infected influenza patients.}
\label{fig:ARMAXinf} 
\end{figure}

The number of reported influenza cases, associated Twitter data, and results of the model training and prediction are displayed in Fig.\ref{fig:ARMAXinf} for the ARX (\ref{eqn:ARMAX}) and NARX (\ref{eqn:NARX}) models. As seen from Fig.\ref{fig:ARMAXinfa} ,the dominant word for indicating a possible influenza outbreak is {\it flu} as compared with {\it swine} and {\it influenza}. Notice that there is a lag between the maximum confirmed influenza cases and the \# of tweets; however, there is an increase in the number of tweets prior to the peak of infected patients. These dynamics are a result of a combination of infection propagation dynamics and the diffusion of information on Twitter. To account for these dynamics the ARX and NARX models presented in Sec.\ref{subsec:ARMAXforPredictingInfluenzaTweets} are utilized. The training and prediction accuracy of these models for $n_a=0, n_b=2$ (i.e model input parameters $\infectdist_{\dtime-\Delta}$ and $\infectdist_{\dtime-\Delta-1}$) are displayed in Fig.\ref{fig:ARMAXinfb}. As seen, the NARX (\ref{eqn:NARX}) model provides a superior estimate as compared with the ARX model (\ref{eqn:ARMAX}). Interestingly there is a $\Delta=18$ day delay between the maximum number of infected patients and the maximum number of Twitter mentions containing the word {\it flu}. This is contrast to the dynamics observed for the 2009~\cite{SSP11} and 2010-2011~\cite{AGLYL11} influenza outbreaks which show that the increase in Twitter mentions occurs earlier or at the same time as the number of infected patients increases. This also emphasizes the importance of using the mean field dynamics model for influenza propagation as compared with only using Twitter data for predicting the influenza infection rate. Here we have used the CDC data to estimate the number of infected agents, however the mean field dynamics model (\ref{eq:system}) could be used to estimate the dynamics of disease propagation and relate this to the observable number of tweets in real-time.

To summarize, 
the above datasets
illustrate  how Twitter can be used as a sensor for monitoring the spread of influenza in a heath network.
The propagation of influenza was modeled according to  the SIS model and the dynamics of tweets according to an autoregressive model.

\subsection{Sentiment-Based Sensing Mechanism}  \index{sentiment}
\label{sec:sensing}

In the above dataset, samples of influenza affected individuals  were obtained from a Harvard college social network. More generally,
it is often necessary to sample individuals in a social network to estimate an underlying state of nature such as the sentiment.
An important question regarding sensing in a social network is:
 How can one construct a small but representative sample of a social network with a large number of nodes? 
 In \cite{LF06} several scale-down and back-in-time sampling procedures are studied.
 Below we review three sampling schemes.
 The simplest possible sampling scheme  is uniform sampling.  We also briefly describe {\em social sampling} and  {\em respondent-driven sampling} which
are recent methods that have become increasingly popular.

\subsubsection{Uniform Sampling}
Consider the following sampling-based measurement strategy. At each period $\dtime$,   $\sample{\deg}$ individuals are
sampled\footnote{For large population sizes $\vertexnum$, sampling with and without replacement are equivalent.}  independently and uniformly  from
the population $\vertexnum{\deg}$ comprising of agents with connectivity  degree $\deg$.
That is, a uniform distributed i.i.d. sequence of nodes, denoted by$\{\nodem{\seq}, \seq=1: \sample{\deg}\}$, is generated from the population  $\vertexnum{\deg}$.
The messages  $\nodeobs{\dtime}{\nodem{\seq}}$
of these  $\sample{\deg}$  individuals  are recorded.
From these independent samples, the empirical sentiment distribution $\sentiment{\dtime}{\deg}$ of degree $\deg$ nodes at each time $\dtime $ is obtained as
\beq \sentiment{\dtime}{\deg}{\nodeobs} = \frac{1}{\sample{\deg} }  \sum_{
\seq=1}^{\sample{\deg}} I\lbr\nodeobs{\dtime}{\nodem{\seq}}=\nodeobs\rbr,
\quad \nodeobs =1 , \ldots, \nodeobsdim.  \label{eq:sentiment} \eeq
At each time $\dtime$, the empirical sentiment distribution $\sentiment{\dtime}$ can be viewed as noisy observations of the infected distribution $\infectdist{\dtime}$ and target state process
$\target{\dtime}$.

\subsubsection{Social Sampling} \index{social sampling}
Social sampling is an extensive area of research; see~\cite{DKS12} for recent results. In social  sampling,  participants in a poll respond with a summary of their friend's
responses. This leads to a reduction in the number of samples required.
If the average degree of nodes in the network is $\deg$, then  the savings in the number of samples is by a factor of $\deg$, since a
randomly chosen node summarizes the results form $\deg$ of its friends. However, the variance and bias of the estimate
depend strongly on the social network structure\footnote{In~\cite{DKS12}, nice intuition is provided in terms of intent polling and expectation polling. In intent polling,
individual are sampled and asked who they intend to vote for. In expectation polling,  individuals are sampled and  asked who they think would win the election.
For a given sample size, one would believe that expectation polling is more accurate than intent polling
since in expectation polling, an individual would typically consider its own intent together with the intents of its friends.}.
In~\cite{DKS12}, a social sampling method is introduced and analyzed where nodes of degree $\deg$ are sampled with probability proportional to $1/\deg$.  This is intuitive since weighing neighbors' values by the reciprocal of the degree undoes the bias introduced by large degree nodes.
It  then illustrates  this social sampling method and variants on the {\sc LiveJournal} network (livejournal.com)  comprising of more than 5 million nodes and 160 million
 directed edges.

\subsubsection{MCMC Based Respondent-Driven Sampling (RDS)} \index{respondent-driven sampling}
Respondent-driven sampling~(RDS) was   introduced by Heckathorn~\cite{Hec97,Hec02,Lee09} as an approach for sampling from hidden populations
in social networks and  has gained
enormous popularity in recent years.
There are more than 120 RDS studies worldwide involving sex workers and  injection drug users~\cite{MJK08}.
%
As mentioned in~\cite{GS09}, the U.S. Centers for Disease Control and Prevention~(CDC) recently selected RDS for a 25-city study of injection drug users that is part of the National HIV Behavioral Surveillance System~\cite{LACH07}. 

RDS  is a variant of the well known method of snowball sampling where current sample members recruit future sample members. The RDS procedure is as follows:  A small number of people in the target
population serve as seeds. After participating in the study, the seeds recruit other people they know through the social network in the target population. The sampling continues according to this procedure  with current sample members recruiting
the next wave of sample members until the desired sampling size is reached. Typically,  monetary compensations are provided for participating in the data collection and recruitment.

RDS can be viewed as a form of Markov Chain Monte Carlo~(MCMC) sampling (see~\cite{GS09} for an excellent exposition).
 Let $\{\nodem{\seq},\seq = 1:\sample{\deg}\}$  be the realization of an aperiodic irreducible Markov chain with
state space  $\vertexnum{\deg}$ comprising of nodes
of degree $\deg$. This Markov chain models the individuals of degree $\deg$ that are snowball sampled, namely, the first individual $\nodem{1}$ is sampled and then recruits the second
individual $\nodem{2}$ to be sampled, who then recruits $\nodem{3}$ and so on.
Instead  of the independent sample estimator~(\ref{eq:sentiment}),
an asymptotically unbiased MCMC estimate is then generated as
\beq \frac{ \sum_{\seq = 1}^{\sample{\deg}}  \frac{I(\nodeobs{\dtime}{\nodem{\seq}}=\nodeobs)}{\steady{\nodem{\seq}}} }{ \sum_{\seq=1}^{\sample{\deg}}  \frac{1}
{\steady{\nodem{\seq}}}
}
\label{eq:mcmcrds}
\eeq
where  $\steady(\nodem)$, $\nodem \in \vertexnum{\deg}$, denotes the stationary distribution of the Markov chain.  For example, a reversible Markov chain  with
prescribed stationary distribution  is straightforwardly generated by the Metropolis Hastings algorithm.

In RDS, the transition matrix  and, hence, the stationary distribution $\steady$
in  the estimator~(\ref{eq:mcmcrds})
 is specified as follows: Assume that  edges between any two nodes $\nodem$ and $\noden$ have symmetric weights $\weight{\nodem}{\noden}$
(i.e.,
 $\weight{\nodem}{\noden} = \weight{\noden}{\nodem}$, equivalently, the network is undirected). In RDS,
node  $\nodem$ recruits node $\noden$ with transition probability
  $\weight{\nodem}{\noden}/ \sum_{\noden} \weight{\nodem}{\noden}$. Then, it can be easily seen that
the stationary distribution is
$\pi(\nodem) = \sum_{\noden \in \Vertexset} \weight{\nodem}{\noden}/ \sum_{\nodem \in \Vertexset, \noden \in \Vertexset} \weight{\nodem}{\noden}$. Using this stationary
distribution, along with the above transition probabilities for sampling agents in~(\ref{eq:mcmcrds}), yields the RDS algorithm.

It is well known that a Markov chain over a non-bipartite connected undirected network $\network$ is aperiodic. Then, the initial seed for the RDS
algorithm can be picked arbitrarily, and the above estimator is an asymptotically unbiased estimator.

Note the difference between RDS and social sampling: RDS uses the network to recruit the next respondent, whereas social sampling seeks
to reduce the number of samples by using people's knowledge of their friends' (neighbors') opinions.

Finally, the reader may be familiar with the DARPA network challenge in 2009 where the locations of 10 red balloons in the continental US
were to be determined using social networking. In this case, the winning MIT Red Balloon Challenge Team used a recruitment based sampling method.
The strategy can also be viewed as a variant of the Query Incentive Network model of~\cite{KR05}.

\subsection{Summary and Extensions}

This section has discussed the diffusion of information in social networks. Mean field dynamics were used to approximate the asymptotic infected
degree distribution. An illustrative example of the spread of influenza was provided. Finally, methods for sampling the population in a social networks were 
reviews. Below we discuss some related concepts and extensions.

\subsubsection*{Bayesian Filtering Problem} 
Given the  sentiment observations described above, how can the infected degree distribution $\infectdist_\dtime$ and target state $\target_\dtime$ be estimated at each time instant?
 The partially observed state space model with dynamics~(\ref{eq:system}) and
discrete time observations from sampling the network   can be use to obtain Bayesian filtering estimates of the underlying state
of nature. 
Computing the conditional mean estimate  $\target{\dtime}, \infectdist{\dtime}$  given the sentiment observation  sequence
 is a Bayesian filtering problem. In fact, filtering of such jump Markov linear systems have been studied extensively in the signal processing literature~\cite{DGK01,LK99}
 and can be solved via the use of sequential Markov chain Monte-Carlo methods. For example, \cite{SOM10} reports on how a particle filter is used to localize
 earthquake events using Twitter as a social sensor.

\subsubsection*{Reactive Information Diffusion}
A key difference between social sensors and conventional sensors in statistical signal processing is that social sensors are reactive:  A social sensor uses  additional information gained to modify
its behavior.  Consider the case where the sentiment-based observation process is made available in a public blog. Then, these observations will affect the transition dynamics of the agents and, therefore,
the mean field dynamics.   \index{sentiment}


\subsubsection*{How Does Connectivity Affect Mean Field Equilibrium?}
The papers~\cite{Pin06,Pin08} examine the structure of  fixed points of the mean field differential equation~(\ref{eq:system}) when the underlying target process $\target$ is not present (equivalently, $\target$ is a one state process).  They consider the case where the agent transition probabilities are parametrized by $\tpdiff{01}{\deg}{\aindex}  = \mu F(\deg,\aindex)$ and $\tpdiff{10} = \failprob$.
Then, defining $\lambda = \mu/\failprob$,
 they study how the following two thresholds behave with
the degree distribution and diffusion mechanism:
\begin{enumerate}
\item {\em Critical threshold $\lambda_c$:} This is defined as the  minimum value of $\lambda$ for which there exists a fixed point of~(\ref{eq:system}) with positive fraction of infected agents, i.e.,
$\infectdist{\infty}{\deg} > 0$ for some $\deg$ and, for $\lambda \leq \lambda_c$,
such a fixed point does not exist.
\item  {\em Diffusion threshold $\lambda_d$:} Suppose the initial condition $\infectdist{0}$ for the infected distribution is infinitesimally small. Then, $\lambda_d$ is the minimum value of $\lambda$ for which
$\infectdist{\infty}{\deg} > 0$ for some $\deg$, and such that, for $\lambda \leq \lambda_d$, $\infectdist{\infty}{\deg}  = 0$  for all $\deg$.
\end{enumerate}
Determining how these thresholds vary with degree distribution and diffusion mechanism is very useful for understanding the long term behavior of agents in the social network.

\section{Bayesian Social Learning Models for Online Reputation Systems} 
\label{sec:socialmain}

\begin{figure}[h]
\centering
{\includegraphics[scale=0.3]{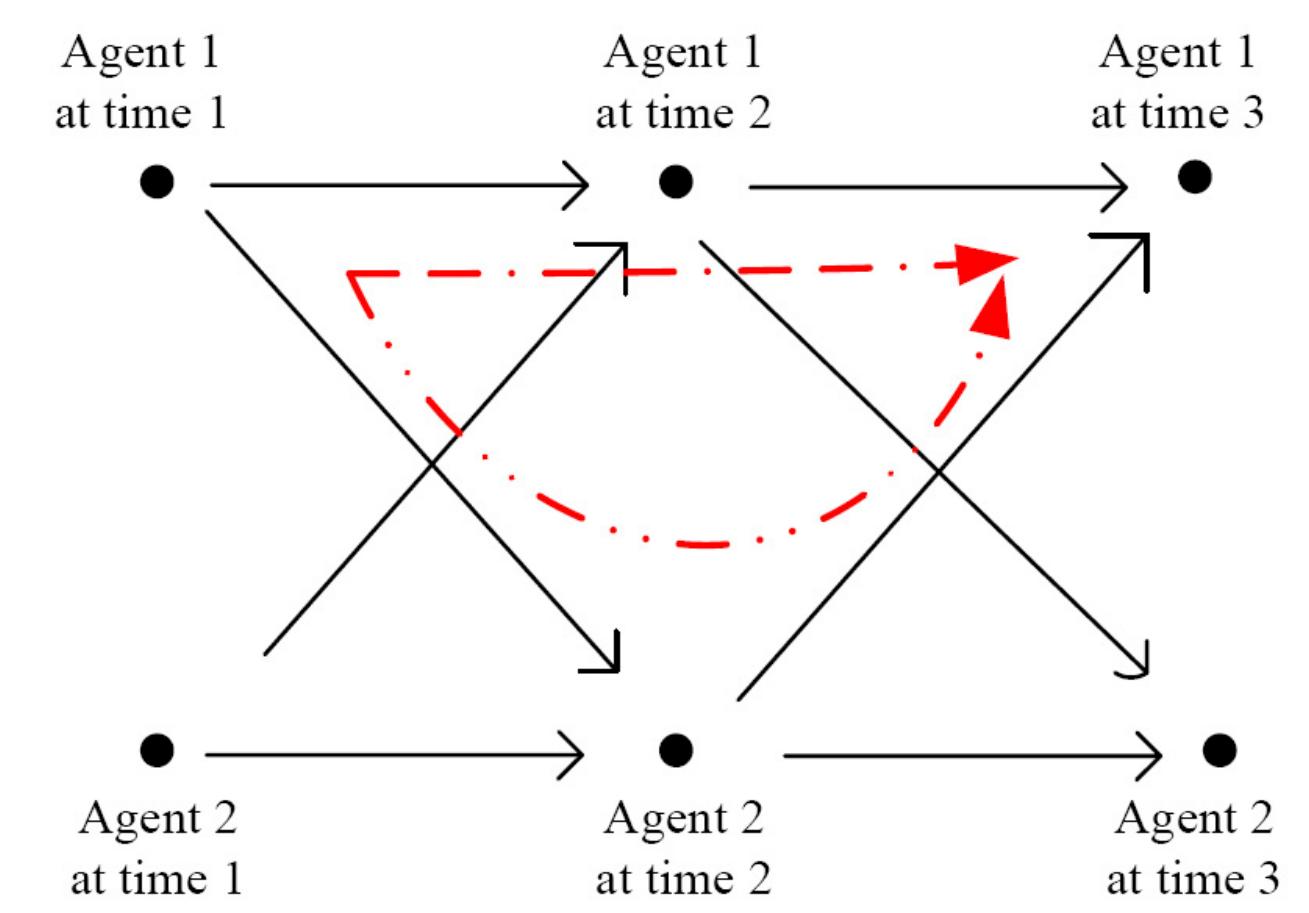}}
\caption{ Example of the information flow (communication graph) in a social network with two agents  and over three event epochs. The arrows represent exchange of information.}
\label{fig:sample}
\end{figure}

In this section we address the second topic of the chapter, namely, Bayesian social learning amongst social sensors.
The motivation  can be understood in terms of the following  social sensing example.
 Consider the following interactions in a multi-agent social network where  agents seek to estimate an underlying state of nature. Each agent visits a restaurant based on reviews on an online reputation website. The agent then obtains  a private measurement of the state  (e.g., the quality of food in a restaurant) in noise. After that, he reviews the restaurant on the same online reputation website.  The information exchange in the social network is modelled by a directed graph. 
Data incest  \cite{KH13} arises due to  loops in the information exchange graph.
 This is illustrated in the graph of Fig.\ref{fig:sample}. Agents 1 and 2 exchange beliefs (or actions) as depicted in Fig.\ref{fig:sample}. The fact
 that there are two distinct paths between Agent 1 at time 1 and Agent 1 at time 3 (these  paths are denoted in red)  implies that 
 the information of Agent 1 at time 1 is double counted leading to a data incest event.

How can data incest be removed so that agents obtain a fair (unbiased) estimate of the underlying state?
The methodology of this section  can be interpreted in terms of the recent {\em Time} article \cite{Tut13} which provides interesting rules for online reputation systems. These include: (i) review the reviewers, and  (ii) censor fake (malicious) reviewers. The data incest removal algorithm proposed in this chapter can be viewed as  ``reviewing the reviews" of other agents to see if they are associated with data incest or not.

The rest of this section is organized as follows:
\begin{compactenum}
\item  Sec.\ref{sec:herd} describes the social learning model that is used to mimic the behavior of agents in online reputation systems. 
Sec.\ref{sec:finance} describes how risk averse social learning models apply to detecting market shocks in high frequency financial systems.

\item  Sec.\ref{sec:incest} to Sec.\ref{sec:removal} deal with modelling data incest and incest removal algorithms for online reputation systems. The information exchange between agents in the social network is  formulated on a family of time dependent directed acyclic graphs.
achieves a fair rating. A necessary and sufficient condition is given on the graph structure of information exchange between agents so that a fair rating is achievable.
\item Sec.\ref{sec:ordinal} discusses conditions under which treating individual social sensors as Bayesian optimizers is a useful
idealization of their behavior. In particular, it is shown that the ordinal behavior of humans can be mimicked by Bayesian optimizers under reasonable conditions.
\item Sec.\ref{sec:expt}  presents a dataset obtained from a psychology experiment to illustrate social learning and data incest patterns.
\end{compactenum}

\paragraph*{Related work}  Collaborative recommendation systems are reviewed and studied in \cite{AT05,KSJ09}. The books \cite{Cha04,EK10}
study information cascades in social learning.  In
\cite{KT12},  
a model of Bayesian social learning is considered in which  agents receive private information about the state of nature and observe actions of their neighbors in a tree-based network.
 Another type of mis-information caused by influential agents (agents who heavily affect  actions of other agents in social networks) is investigated in 
\cite{AO11}. Mis-information in the context of this chapter is motivated by sensor networks where the term ``data incest" is used \cite{KH13}.   Data incest also arises in  Belief Propagation (BP) algorithms \cite{Pea86, MWJ99} which are used in computer vision and error-correcting coding theory. 
BP algorithms require passing local messages over the graph (Bayesian network) at each iteration. 
For graphical models with loops, BP algorithms are only approximate due to the over-counting of local messages \cite{YFW05} which is similar to data incest in  social learning.  With the algorithms presented in this section, data incest can be mitigated from Bayesian social learning over non-tree graphs that satisfy a topological constraint.
The closest work to the current chapter is \cite{KH13}. However,  in \cite{KH13}, data incest is considered in a network where agents exchange their private belief states - that is, no social learning is considered.  Simpler versions of this information exchange process and estimation were investigated in  \cite{Aum76,GP82,BV82}.  We also refer the reader to \cite{CCP13} for a discussion
of recommender systems.

\subsection{Classical Social Learning}\label{sec:herd}

 We briefly review the classical  social learning model  for  the interaction of individuals. Subsequently, we will deal with more general models over a social network.
 
Consider a multi-agent system that aims to estimate the state of an underlying finite state random variable $x \in
\X = \{1,2,\ldots,X\}$ with known prior distribution $\pi_0$.
Each agent acts once  in a predetermined sequential order indexed by $k=1,2,\ldots$   
Assume at the beginning of iteration $k$,
all agents have access to the public belief $\pi_{k-1}$ defined in  Step (iv) below.
The social learning protocol proceeds as follows
 \cite{BHW92,Cha04}:\\
 (i) {\em Private Observation}: At time $k$,
agent $k$  records a private observation $y_k\in \Y $ 
from the observation distribution $B_{iy} = P(y|x=i)$, $i \in \X$.
Throughout this section we assume that $\Y = \{1,2,\ldots,Y\}$ is finite.
\\
(ii) 
{\em Private Belief}:  Using the public belief $\pi_{k-1} $ available at time $k-1$ (Step (iv) below), agent $k$   updates its private
posterior belief  $\priv_k(i)  =  P(x_k = i| a_1,\ldots,a_{k-1},y_k)$ using Bayes formula:
\begin{align}  \label{eq:hmm} \priv_k &= 
\frac{B_{y_k} \pi}{ \mathbf{1}_X^\p B_y  \pi}, \;  
B_{y_k} = \text{diag}(P(y_k|x=i),i\in \X) . 
 \end{align}
 Here $\mathbf{1}_X$ denotes the $X$-dimensional vector of ones, $\eta_k$ is an $X$-dimensional probability mass function (pmf). \\
  (iii)   {\em Myopic Action}: Agent  $k$  takes  action $a_k\in \A = \{1,2,\ldots, A\}$ to  minimize its expected cost 
 \begin{equation}  
  a_k =  \arg\min_{a \in \A} \E\{c(x,a)|a_1,\ldots,a_{k-1},y_k\} =\arg\min_{a\in \A} \{c_a^\p\priv_k\}. 
    \label{eq:myopic}
  \end{equation}
  Here $\ca = (c(i,a), i \in \X)$ denotes an $X$ dimensional cost vector, and $c(i,a)$ denotes the cost  incurred when the underlying state is $i$ and the  agent chooses action $a$.\\
  Agent $k$ then broadcasts its  action $a_k$.\\
(iv) {\em Social Learning Filter}:   
Given the action $a_k$ of agent $k$,  and the public belief $\pi_{k-1}$, each  subsequent agent $k' > k$ 
performs social learning to
update the public belief $\pi_k$ according to the  ``social learning  filter":\ \beq \pi_k = \Ts(\pi_{k-1},a_k), \text{ where } \Ts(\pi,a) = 
 \frac{\Bs_a \,\pi}{\sigs(\pi,a)}, \label{eq:piupdate}\eeq
where
$\sigs(\pi,a) = \mathbf{1}_X^\p \Bs_a \tp^\p \pi$ is the normalization factor of the Bayesian update.
In (\ref{eq:piupdate}),  the public belief $\pi_k(i)  = P(x_k = i|a_1,\ldots a_k)$ and $\Bs_a  = \text{diag}(P(a|x=i,\pi),i\in \X ) $ has elements
\begin{align*} 
 & P(a_k = a|x_k=i,\pi_{k-1}=\pi) = \sum_{y\in \Y} P(a|y,\pi)P(y|x_k=i) \\
 &   P(a_k=a|y,\pi) = \begin{cases}  1 \text{ if }  c_a^\p B_y \tp^\p \pi \leq c_{\ta}^\p B_y \tp^\p\pi, \; \ta \in \A \\
 0  \text{ otherwise. }  \end{cases}
  \end{align*}

 The following
result which is well known in the economics literature \cite{BHW92,Cha04} states that if agents follow the above social learning protocol, then   after some finite time $\bar{k}$, an
 {\em information cascade} occurs.\footnote{
  A {\em herd of agents} takes place at time $\bar{k}$, if the actions of all agents after time $\bar{k}$ are identical, i.e., $a_k = a_{\bar{k}}$ for all
time  $k > \bar{k}$. An information cascade implies that a herd of agents occur.
 \cite{TBB10}  quotes the following anecdote of user  influence and herding in a social network: ``... when a popular blogger left his blogging site for a two-week vacation, the site's visitor tally fell, and content produced by three invited substitute bloggers could not stem the decline."}

\begin{theorem}[\cite{BHW92}] 
\label{thm:herd} The social learning protocol  leads to an {\em information cascade}  in finite time
with probability~1. That is, after some finite 
 time $\bar{k}$ social learning ceases and the public belief $\pi_{k+1} = \pi_k$, $k \geq \bar{k}$, and all agents choose the same action  $a_{k+1} = a_k$, $k\geq \bar{k}$.
  \qed\end{theorem}

Instead of reproducing the proof, let us give some insight as to why Theorem \ref{thm:herd} holds. It can be shown using martingale methods
that at some finite time $k=k^*$, the agent's probability $P(a_k|y_k,\pi_{k-1})$ becomes  independent of the private observation $y_k$. Then clearly, $P(a_k=a|x_k=i,\pi_{k-1}) =   P(a_k=a|\pi)$. Substituting this into the social learning filter (\ref{eq:piupdate}), we see
that $\pi_{k} = \pi_{k-1}$.  Thus after some finite time $k^*$, the social learning filter hits a fixed point and social learning stops. 
As a result, all subsequent agents $k> k^*$ completely disregard their private observations and take the same action $a_{k^*}$,
thereby forming
 an information cascade (and therefore a herd).

\subsection{Risk Averse Social Learning and Detecting Market Shocks}  \label{sec:finance}
Here we consider the  statistical signal processing problem involving agent based models of financial markets which at a micro-level are driven by socially aware and risk-averse trading agents. These agents trade (buy or sell) stocks at each trading instant by using the decisions of all previous agents (social learning) in addition to a private (noisy) signal they receive on the value of the stock. We are interested in the following: (1) Modelling the dynamics of these risk averse agents, (2) Sequential detection of a market shock based on the behaviour of these agents. Structural results which characterize social learning under a risk measure, CVaR (Conditional Value-at-risk), are presented and formulation of the Bayesian change point detection problem is provided. The structural results exhibit two interesting properties: (i) Risk averse agents herd more often than risk neutral agents (ii) The stopping set in the sequential detection problem is non-convex.

It is well documented in  behavioural economics  \cite{CLLS75} and  psychology \cite{DS99} that people prefer a certain but possibly less desirable outcome over an uncertain but potentially larger outcome. To model this risk averse behaviour, commonly used risk measures{\footnote{A risk measure $\varrho : \mathcal{L} \rightarrow \mathbb{R}$ is a mapping from the space of measurable functions to the real line which satisfies the following properties: (i) $\varrho(0)=0$. (ii) If $S_{1}, S_{2} \in \mathcal{L}$ and $S_{1} \leq S_{2} ~\text{a.s}$ then $\varrho(S_{1}) \leq \varrho(S_{2})$. (iii) if $a\in\mathbb{R}$ and $S\in\mathcal{L}$, then $\varrho(S+a) = \varrho(S)+a $. The risk measure is coherent if in addition $\varrho$ satisfies: (iv) If $S_{1}, S_{2} \in \mathcal{L}$, then $\varrho(S_{1}+S_{2}) \leq \varrho(S_{1}) + \varrho(S_{1})$. (v) If $a \geq 0$ and $S\in\mathcal{L}$, then $\varrho(aS)=a\varrho(S)$. The expectation operator is a special case where subadditivity is replaced by additivity.}} are Value-at-Risk (VaR), Conditional Value-at-Risk (CVaR), Entropic risk measure and Tail value at risk; see \cite{MJ10}. We consider social learning under CVaR risk measure. CVaR \cite{RU00} is an extension of VaR  that gives the total loss given a loss event and is a coherent risk measure \cite{ADEH02}. Below, we choose CVaR risk measure as it exhibits the following properties \cite{ADEH02}, \cite{RU00}:
(i) It associates higher risk with higher cost.
(ii) It ensures that risk is not a function of the quantity purchased, but arises from the stock.
(iii) It is convex. 
CVaR as a risk measure has been used in solving portfolio optimization problems \cite{PUK99}, \cite{LSU10} and order execution. For an overview of risk measures and their application in finance, see \cite{MJ10}.

\subsubsection*{CVaR Social Learning Model}
 The market micro-structure is modelled as a discrete time dealer market motivated by algorithmic and high-frequency tick-by-tick trading \cite{CJ13}. There is a single traded stock or asset, a market observer and a countable number of trading agents. The asset has an initial true underlying value $x_{0} \in \mathcal{X} = \lbrace 1,2,\hdots, X \rbrace$. The market observer does not receive direct information about $x\in \mathcal{X}$ but only observes the public buy/sell actions of agents, $a_{k} \in \mathcal{A} = \lbrace 1(\text{buy}),2({\text{sell}}) \rbrace$. The agents themselves receive noisy private observations of the underlying value $x$ and consider this in addition to the trading decisions of the other agents visible in the order book. At a random time, $\tau^{0}$ determined by  the transition matrix $P$, the asset experiences a jump change in its value to a new value. The aim of the market observer is to detect the change time (global decision) with minimal cost, having access to only the actions of these socially aware agents. Let $y_{k} \in \mathcal{Y} = \{1,2, \hdots, Y\}$ denote agent $k$'s private observation. The initial distribution is $\pi_{0} = (\pi_{0}(i),i\in \mathcal{X})$ where $\pi_{0}(i) = \Prb(x_{0}=i)$. \\

The  agent based model has the following dynamics:
\begin{compactenum}
\item[1.] \textit{Shock in the asset value}: At time $\tau^{0} > 0$, the asset experiences a jump change (shock) in its value due to exogenous factors. 
The change point  $\tau^0$ is modelled by a   
{\em  phase type (PH) distribution}. 
The family of all PH-distributions forms a dense subset for the set of all distributions
	\cite{Neu89} i.e., for any given distribution function $F$ such that $F(0) = 0$, one can find a sequence of PH-distributions	
$\{F_n , n	\geq	1\}$	 to		approximate	$F$	uniformly over $[0, \infty)$.
The PH-distributed time $\tau^0$ can be constructed via
 a multi-state  Markov chain $x_k$ with state space $\mathcal{X} = \{1,\hdots, X\}$ as follows:
Assume    state `1'   is an absorbing state
and denotes the state after the jump change.    The states $2,\ldots,X$ (corresponding to beliefs $e_2,\ldots,e_X$) can be viewed as  a single composite state that $x$ resides in before the jump. 
 So $\tau^{0} = \text{inf} \lbrace{k:x_{k}=1} \rbrace$ and the transition probability matrix $P$ is of the form
\begin{equation}\label{eq:tr_m}
P = \begin{bmatrix}
       1 & 0        \\
       \underline{P}_{(X-1)\times1}   & {\bar{P}_{(X-1)\times(X-1)}}
     \end{bmatrix}
\end{equation}
 The 
distribution of the absorption time to state 1 is 
\begin{equation} \label{eq:nu}
 \nu_0 = \pi_0(1), \quad \nu_k = \bar{\pi}_0' \bar{P}^{k-1} \underline{P}, \quad k\geq 1, \end{equation}
 where $\bar{\pi}_0 = [\pi_0(2),\ldots,\pi_0(X)]'$.
The key idea  is that by appropriately choosing the pair $(\pi_0,P)$ 
and the associated state space dimension $X$,
 one can approximate any given discrete distribution on $[0, \infty)$ by the distribution $\{\nu_k, k \geq 0\}$; see 
\cite[pp.240-243]{Neu89}.
 The event $\{x_k = 1\}$ means the change point has occurred before time $k$ according
 to PH-distribution (\ref{eq:nu}). In the special case when $x$ is a 2-state Markov chain,
 the change time $\tau^0$  is geometrically distributed.

\item[2.] \textit{Agent's Private Observation}: Agent $k$'s private (local) observation denoted by $y_{k}$ is a noisy measurement of the true value of the asset. It is obtained from the observation likelihood distribution as,
\begin{equation}\label{eq:obs_m}
B_{xy} = \Prb(y_{k}=y|x_{k}=x)
\end{equation} 

\item[3.] \textit{Private Belief update}: Agent $k$ updates its private belief using the observation $y_{k}$ and the prior public belief $\pi_{k-1}(i) = \Prb(X=i|a_{1},\hdots,a_{k-1})$ as the following Hidden Markov Model update
\begin{equation}
\eta_{k} = \frac{B_{y_{k}}P'\pi_{k-1}}{\textbf{1}'B_{y_{k}}P'\pi_{k-1}}
\end{equation}
where $\textbf{1}$ denotes the $X$-dimensional vector of ones.
\item[4.] \textit{Agent's trading decision}: Agent $k$ executes an action $a_{k}\in\mathcal{A}=\lbrace1(\text{buy}),2(\text{sell})\rbrace$ to myopically minimize its cost. Let $c(i,a)$ denote the cost incurred if the agent takes action $a$ when the underlying state is $i$. Let the local cost vector be 
\begin{equation}\label{eq:cst_v}
c_{a} = [c(1,a)~c(2,a) \dots ~c(X,a)]
\end{equation}
The costs for different actions are taken as
\begin{equation}
c(i,j) = p_{j}-\beta_{ij} ~ \text{for} ~ i \in \mathcal{X}, j \in \mathcal{A}
\end{equation}
where  $\beta_{ij}$ corresponds to the agent's demand. Here demand is the agent's desire and willingness to trade at a price $p_{j}$ for the stock. Here $p_{1}$ is the quoted price for purchase and $p_{2}$ is the price demanded in exchange for the stock. We assume that the price is the same during the period in which the value changes. As a result, the willingness of each agent only depends on the degree of uncertainty on the value of the stock.
\begin{remark}
The analysis provided in this paper straightforwardly extends  to the case when different agents are facing different prices like in an order book.  For notational simplicity we assume the cost are time invariant.
\end{remark}

  The agent considers measures of risk in the presence of uncertainty in order to overcome the losses incurred in trading. To illustrate this, let $c(x,a)$ denote the loss incurred with action $a$ while at unknown and random state $x\in{\mathcal{X}}$. When an agent solves an optimization problem involving $c(x,a)$ for selecting the best trading decision, it will take into account not just the expected loss, but also the ``riskiness" associated with the trading decision $a$. 
The agent therefore chooses an action $a_{k}$ to minimize the CVaR   measure\footnote{
For the reader unfamiliar with risk measures, it should be noted that CVaR is one of the `big' developments in risk modelling in finance in the last 15 years.
In comparison, the 
 value at risk (VaR) is the percentile loss namely,  $\text{VaR}_\alpha(x) = \min\{z: F_x(z) \geq \alpha\} $ for cdf $F_x$. While CVaR is a coherent risk measure,
 VaR is not  convex and so not coherent. CVaR has  other remarkable properties \cite{RU00}: it is continuous in $\alpha$
 and jointly convex in $(x,\alpha)$.
For continuous cdf $F_x$, $\text{CVaR}_\alpha(x) = \E\{X | X > \text{VaR}_\alpha(x)\}$. Note that the variance  is
 not  a coherent risk~measure.} of trading as
\begin{align}
a_{k} &=  {\underset{a \in \mathcal{A}}{\text{argmin}}} \{ \cvr \} \\
&=  {\underset{a \in \mathcal{A}}{\text{argmin}}} \{ {\underset{z \in \mathbb{R}}{\text{min}}} ~ \{ z + \frac{1}{\alpha} \mathbb{E}_{y_{k}}[{\max} \{ (c(x_{k},a)-z),0 \rbrace] \} \} \nonumber
\end{align}
Here $\alpha \in (0,1]$ reflects the degree of risk-aversion for the agent (the smaller $\alpha$ is, the more risk-averse the agent is). 
Define 
\begin{equation}
\mathcal{H}_{k} := \sigma \text{- algebra generated by}~ (a_{1},a_{2},\hdots,a_{k-1},y_{k}) 
\end{equation} 
$\mathbb{E}_{y_{k}}$ denotes the expectation with respect to private belief, i.e, $\mathbb{E}_{y_{k}} = \mathbb{E}[.|\mathcal{H}_{k}]$ when the private belief is updated after observation $y_{k}$.
 

\item[5.] \textit{Social Learning and Public belief update}: Agent $k$'s action is recorded in the order book and hence broadcast publicly. Subsequent agents and the market observer update the public belief on the value of the stock according to the social learning Bayesian filter as follows 
\begin{equation}
\pi_{k} = T^{\pi_{k-1}} (\pi_{k-1},a_k)  = \frac{R_{a_{k}}^{\pi_{k-1}}P'\pi_{k-1}}{\textbf{1}'R_{a_{k}}^{\pi_{k-1}}P'\pi_{k-1}}
\end{equation}

Here, $R_{a_{k}}^{\pi_{k-1}} = \text{diag}(\Prb(a_{k}|x=i,\pi_{k-1}),i \in \mathcal{X})$, where $\Prb(a_{k}|x=i,\pi_{k-1}) = {\underset{y \in \mathcal{Y}}{\sum}} \Prb(a_{k}|y,\pi_{k-1})\Prb(y|x_{k}=i)$ and 
\[ \Prb(a_{k}|y,\pi_{k-1}) = \left\{ \begin{array}{ll}
         1 & \mbox{if $ a_{k} = {\underset{a \in \mathcal{A}}{\text{argmin}}}~ \text{CVaR}_{\zeta}(c(x_{k},a))$} ; \\
         0 & \mbox{$\text{otherwise}$}.\end{array} \right. \] 
Note that $\pi_k$ belongs to the unit simplex  $\Pi(X){\overset{\Delta}{=}}\lbrace \pi \in \mathbb{R}^{X} : \textbf{1}_{X}'\pi = 1, 0 \leq \pi \leq 1 ~\text{for all} ~ i \in \mathcal{X}\rbrace $. 
\item[6.]\textit{ Market Observer's Action}: The market observer (securities dealer) seeks to achieve quickest detection by balancing delay with false alarm. At each time $k$, the market observer chooses action\footnote{It is important to distinguish between the ``local'' decisions $a_k$ of the agents and ``global'' decisions $u_k$ of the market observer. Clearly the decisions 
$a_k$ affect the choice of $u_k$ as will be made precise below.} $u_{k}$ as
\begin{equation}
u_{k} \in \mathcal{U} =  \lbrace 1(\text{stop}), 2(\text{continue}) \rbrace
\end{equation}
Here `Stop' indicates that the value has changed and the dealer incorporates this information before selling new issues to investors. The formulation presented considers a general parametrization of the costs associated with detection delay and false alarm costs. 
Define  
\begin{equation}
\mathcal{G}_{k} := \sigma \text{- algebra generated by}~  (a_{1},a_{2},\hdots,a_{k-1},a_{k}).
\end{equation}  
\begin{compactitem}
\item[i)] \textit{Cost of Stopping}: The asset experiences a jump change(shock) in its value at time $\tau^{0}$. If the action $u_{k}=1$ is chosen before the change point, a false alarm penalty is incurred. This corresponds to the event ${\underset{i\geq2}\cup} \lbrace x_{k}=i \rbrace \cap \lbrace u_{k}=1 \rbrace$. Let $\mathcal{I}$ denote the indicator function. The cost of false alarm in state $i,i\in \mathcal{X}$ with $f_{i}\geq0$ is thus given by $f_{i}\mathcal{I}(x_{k}=i,u_{k}=1)$. The expected false alarm penalty is
\begin{align} \label{eq:falc}
C(\pi_{k},u_{k}=1) &= {\underset{i\in\mathcal{X}}{\sum}}f_{i}\mathbb{E}\lbrace\mathcal{I}(x_{k}=i,u_{k}=1)|\mathcal{G}_{k} \rbrace \nonumber \\
 &= \textbf{f}'\pi_{k}
\end{align}
where $\textbf{f} = (f_{1},\hdots,f_{X})$ and it is chosen with increasing elements, so that states further from `$1$' incur higher false alarm penalties. Clearly, $f_{1}=0$.

\item[ii)] \textit{Cost of delay}: A delay cost is incurred when the event $\lbrace x_{k}=1,u_{k}=2 \rbrace$ occurs, i.e, even though the state changed at $k$, the market observer fails to identify the change. The expected delay cost is 
\begin{align} \label{eq:delc}
C(\pi_{k},u_{k}=2) &= d\, \mathbb{E}\lbrace\mathcal{I}(x_{k}=i,u_{k}=1)|\mathcal{G}_{k} \rbrace \nonumber \\
&= de_{1}'\pi_{k}
\end{align}
where $d>0$ is the delay cost and $e_{1}$ denotes the unit vector with 1 in the first position.
\end{compactitem}
\end{compactenum}

\begin{figure}[!t] 
\centering
\includegraphics[scale=0.38]{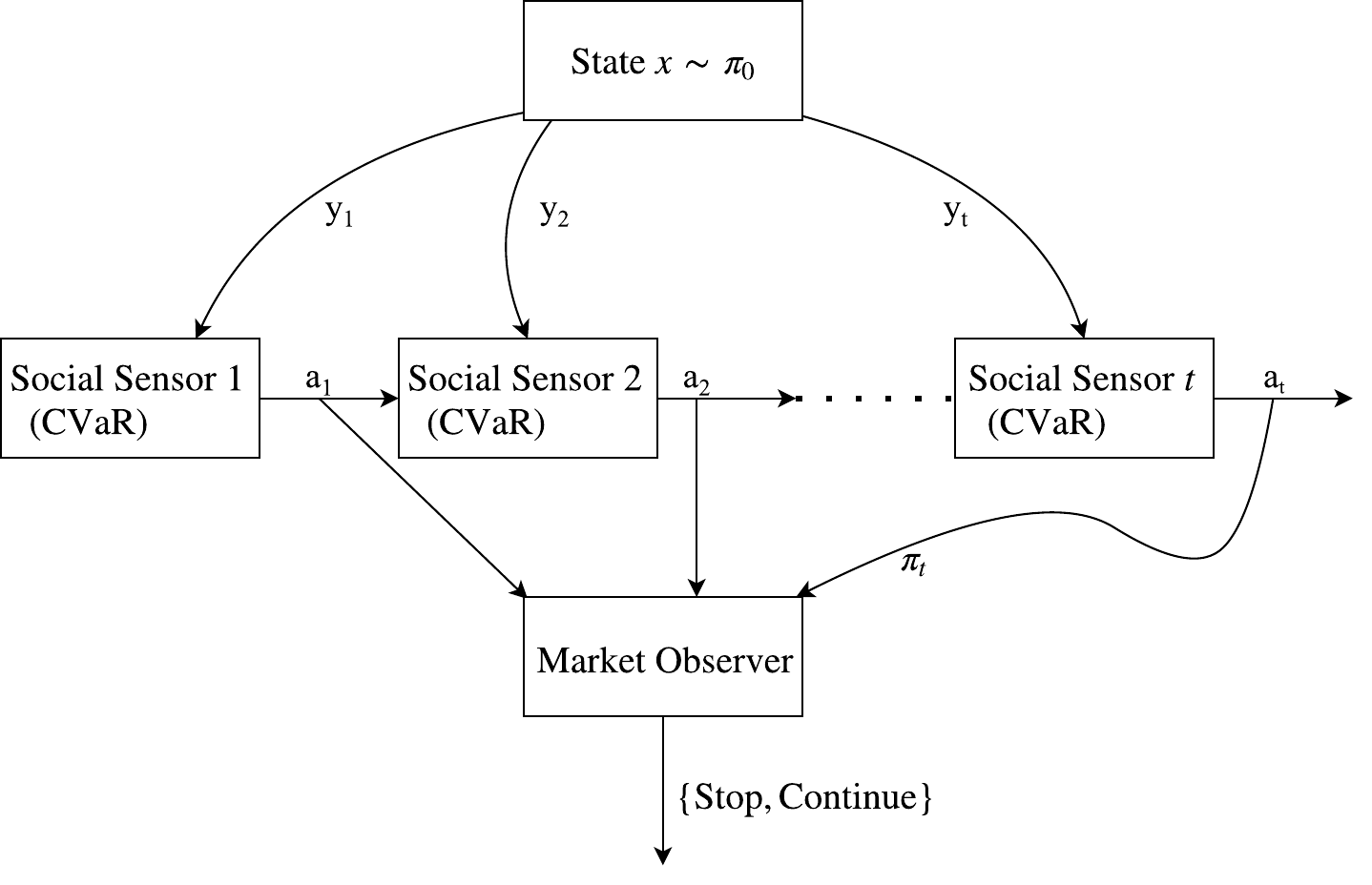}
\caption{Sequential detection with risk-averse social sensors. Each social sensor receives a noisy observation on the state and chooses an action to minimize its CVaR measure of trading. The social sensors communicate their actions to subsequent sensors. The market observer seeks to determine if there is a change in the value of the underlying asset from the actions of the sensors.}
\label{p_m}
\end{figure}

Fig. \ref{p_m} illustrates the above social learning model in which the information exchange between the risk-averse social sensors is sequential. 

\subsubsection*{Market Observer's Quickest Detection Objective}
The market observer chooses its action at each time $k$ as 

\begin{equation} u_k = \mu(\pi_k)  \in  \lbrace 1(\text{stop}), 2(\text{continue}) \rbrace \label{eq:marketaction}
\end{equation}
 where $\mu$ denotes a stationary policy.
For each initial distribution $\pi_{0} \in \Pi(X)$ and policy $\mu$, the following cost is associated
\begin{equation} \label{eq:Obj}
J_{\mu}(\pi_{0}) = \mathbb{E}^{\mu}_{\pi_{0}} \left\{ {\overset{\tau-1} {\underset{k=1} {\sum}}} \rho^{k-1}C(\pi_{k},u_{k}=2)+\rho^{\tau-1}C(\pi_{k},u_{k}=1) \right\}
\end{equation}
Here $\rho \in [0,1]$ is the discount factor which is a measure of the degree of impatience of the market observer. 
(As long as $\textbf{f} $ is non-zero, stopping is guaranteed in finite time and so $\rho = 1$ is allowed.)

Given the cost, the market observer's objective is to determine $\tau^{0}$ with minimum cost by computing an optimal policy $\mu^{*}$ such that
\begin{equation}  \label{eq:pomdp}
J_{\mu^{*}}(\pi_{0}) = {\underset{\mu\in \boldsymbol{\mu}}{\text{inf}}J_{\mu}(\pi_{0})}
\end{equation}
The sequential detection problem 
(\ref{eq:pomdp}) can be viewed as a  partially observed Markov decision process (POMDP) where  the belief update is given by the social
learning filter.

\subsubsection*{Stochastic Dynamic Programming Formulation} \label{sec:sdp}

The optimal policy of the market observer $\mu^{*}:\Pi(X)\rightarrow\{1,2\}$ is the solution of \eqref{eq:Obj} and is given by Bellman's dynamic programming equation as follows: 
\begin{align}
V(\pi) &= \text{min} \left\{ C(\pi,1),C(\pi,2)+\rho{\underset{a\in \mathcal{A}}{\sum}}V(T^{\pi} (\pi,a))\sigma(\pi,a)\right\} \label{eq:val}\\
\mu^{*}(\pi) &= \text{argmin} \left\{ C(\pi,1),C(\pi,2)+\rho{\underset{a\in \mathcal{A}}{\sum}}V(T^{\pi}(\pi,a))\sigma(\pi,a) \nonumber \right\}
\end{align}
where $T^{\pi} (\pi,a) = \frac{R_{a}^{\pi}P'\pi}{\textbf{1}'R_{a}^{\pi}P'\pi}$ is the CVaR-social learning filter and $\sigma(\pi,a) = \textbf{1}'R_{a}^{\pi}P'\pi$ is the normalization factor of the Bayesian update.  $C(\pi,1)$ and $C(\pi,2)$ from \eqref{eq:falc} and \eqref{eq:delc} are the market observer's costs. As $C(\pi,1)$ and $C(\pi,2)$ are non-negative and bounded for $\pi \in \Pi(X)$, the stopping time $\tau$ is finite for all $\rho \in [0,1]$.

The aim of the market observer is then to determine the stopping set $\mathcal{S} = \{ \pi \in \Pi(X) : \mu^{*}(\pi) = 1 \}$ given by:
\begin{equation*}
\mathcal{S} = \left\{ \pi : C(\pi,1) < C(\pi,2) + \rho {\underset{a\in \mathcal{A}}{\sum}}V(T^{\pi}(\pi,a))\sigma(\pi,a) \right\}
\end{equation*}
The dynamic programming equation (\ref{eq:val}) is similar to that for stopping time  POMDP except that the belief update is given by a CVaR
social learning filter. 
As will be shown below, because of the social learning dynamics, quite remarkably, $\mathcal{S}$ is not necessarily a convex set. This is in stark contrast to classical quickest detection where  the stopping region is always convex irrespective of the change time distribution \cite{Kri11}.

 \subsubsection*{Social Learning Behavior of Risk Averse Agents}
\begin{figure}[!t] 
\centering
\includegraphics[scale=0.4]{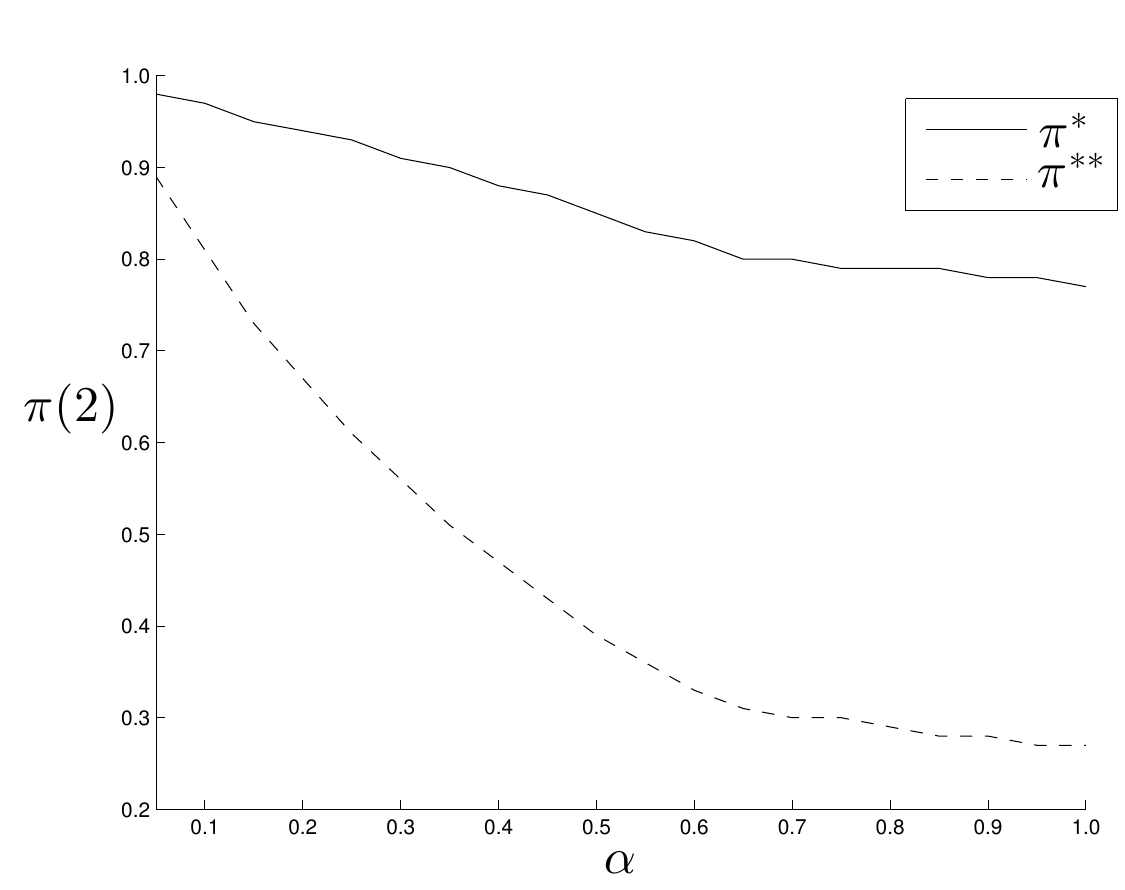}
\caption{The social learning region for the risk-aversion parameter $\alpha \in (0,1]$. It can be seen that the curves corresponding to $\pi^{**}$ and $\pi^{*}$ do not intersect and their separation (social learning region) varies with $\alpha$. Here $P=I$, i.e, the value is a random variable. 
 }
\label{alp_p}
\end{figure}

\begin{figure}[!t] 
\centering
\includegraphics[scale=0.4]{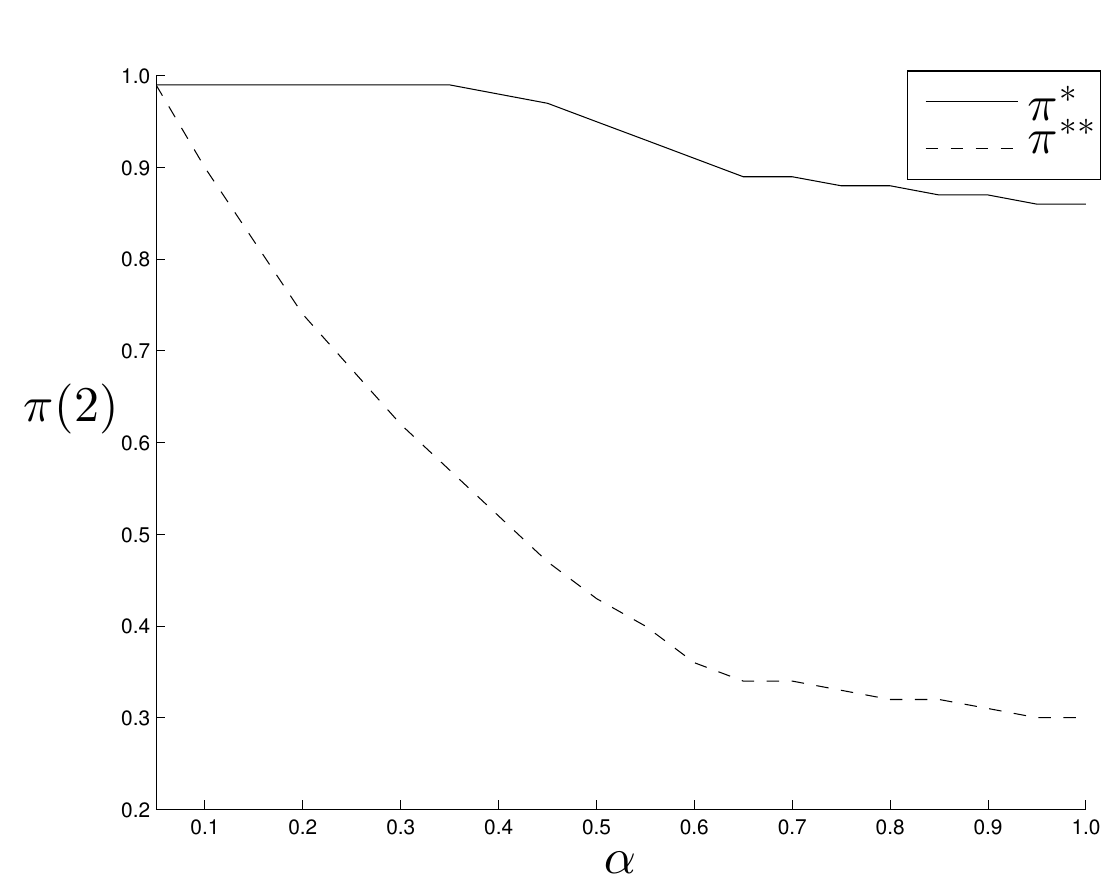}
\caption{The social learning region for the risk-aversion parameter $\alpha \in (0,1]$. It can be seen that the social learning region is absent when agents are sufficiently risk-averse and is larger when the stock value is known to change, i.e, $P\neq I$.
 }
\label{alp_p2}
\end{figure}

The following discussion highlights the relation between risk-aversion factor $\alpha$ and the regions $\mathcal{P}^{\alpha}_{l}$. For a given risk-aversion factor $\alpha$, it can be shown  that there are at most $Y+1$ polytopes on the belief space. It was shown in \cite{Kri12} that for the risk neutral case with $X=2$, and $P = I$ (the value is a random variable) the intervals $\mathcal{P}^{\alpha}_{1}$ and $\mathcal{P}^{\alpha}_{3}$ correspond to the herding region and the interval $\mathcal{P}^{\alpha}_{2}$ corresponds to the social learning region. In the herding region, the agents take the same action as the belief is frozen. In the social learning region there is observational learning. However, when the agents are optimizing a more general risk measure (CVaR), the social learning region is different for different risk-aversion factors. The social learning region for the CVaR risk measure is shown in Fig. \ref{alp_p}. It can be observed from Fig. \ref{alp_p} that $\mathcal{P}^{\alpha}_{1}$ becomes smaller, $\mathcal{P}^{\alpha}_{2}$ becomes smaller and $\mathcal{P}^{\alpha}_{3}$ becomes larger as $\alpha$ decreases.
The following parameters were chosen: 
\[ B = \begin{bmatrix}
0.8 & 0.2 \\
0.3 & 0.7
\end{bmatrix},
P =\begin{bmatrix}
1 & 0 \\
0 & 1
\end{bmatrix},
c =\begin{bmatrix}
1 & 2 \\
3 & 0.5
\end{bmatrix}
\]
This can be interpreted as risk-averse agents showing a larger tendency to go with the crowd rather than ``risk" choosing the other action. With the same $B$ and $c$ parameters, but with 
transition matrix
\[
P = \begin{bmatrix}
1 & 0 \\
0.1 & 0.9
\end{bmatrix}
\]
the social learning region is shown in Fig. \ref{alp_p2}.
From Fig. \ref{alp_p2}, it is observed that when the state is evolving and when the agents are sufficiently risk-averse, social learning region is very small. It can be interpreted as: agents having a strong risk-averse attitude don't prefer to ``learn" from the crowd; but rather face the same consequences, when $P \neq I$.

\subsubsection*{Nonconvex Stopping Set for Market Shock Detection} 
We now illustrate the solution to the Bellman's stochastic dynamic programming equation (\ref{eq:val}), which determines the optimal policy for 
quickest market shock detection, by considering an agent based model with two states. Clearly the agents (local decision makers) and market observer  interact -- the  local decisions $a_k$ taken by the agents determines the public belief $\pi_k$ and hence determines decision $u_{k}$ of the  market observer via (\ref{eq:marketaction}).

\begin{figure}[!t] 
\centering
\includegraphics[scale=0.4]{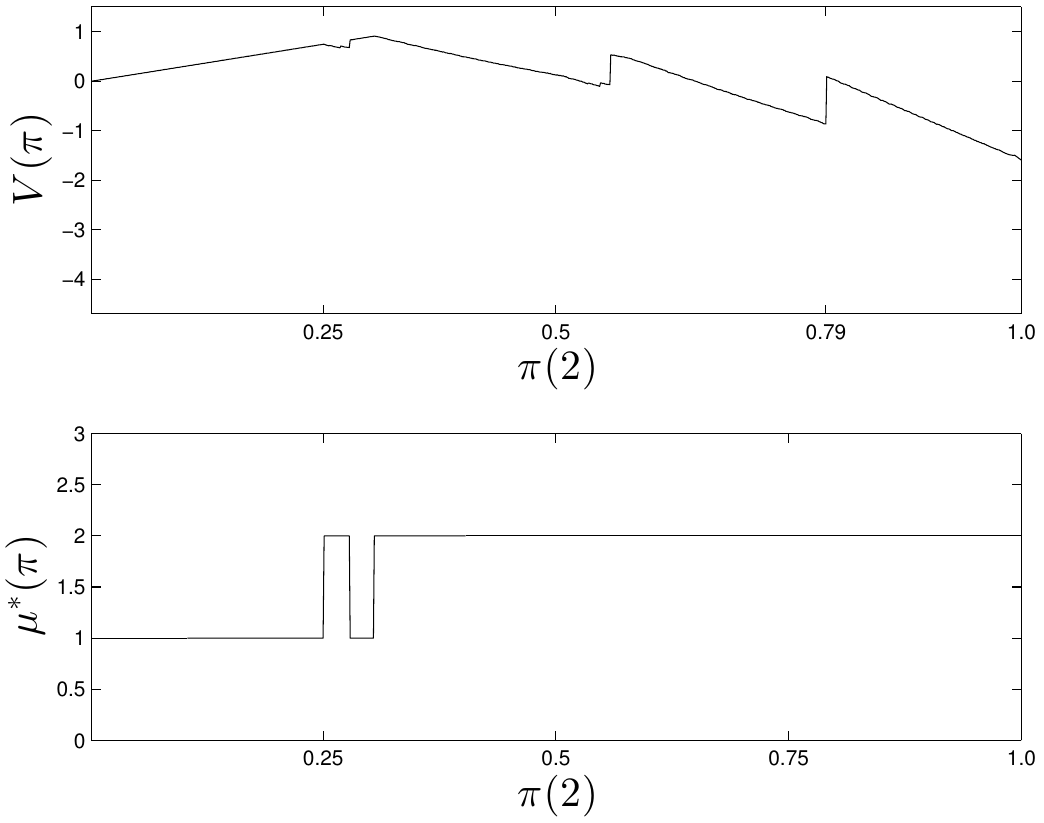}
\caption{The value function $V(\pi)$ and the double threshold optimal policy $\mu^{*}(\pi)$ are plotted over $\pi(2)$. The significance of the double threshold policy is that the stopping regions are non-convex. The implication of non-convex stopping set for the market observer is that - if he believes that it is optimal to stop, it need not be optimal to stop when his belief is larger.}
\label{non_cnv}
\end{figure}

Fig. \ref{non_cnv} displays  the value function and optimal policy for a toy example having the following parameters:
\[ B = \begin{bmatrix}
0.8 & 0.2 \\
0.3 & 0.7
\end{bmatrix}, 
P =\begin{bmatrix}
1 & 0 \\
0.06 & 0.94
\end{bmatrix},
c = \begin{bmatrix}
1 & 2 \\
2.5 & 0.5
\end{bmatrix}
\]

The parameters for the market observer are chosen as:
$d = 1.25$, $\textbf{f}=[0 ~ 3]$, $\alpha = 0.8$ and $\rho = 0.9$. 

From Fig.~\ref{non_cnv} it is clear that the market observer has a double threshold policy and the value function is discontinuous.
The double threshold policy is unusual from a signal processing point of view. Recall that $\pi(2)$ depicts the posterior probability
of no change. The market observer ``changes its mind" - it switches from no change to change as the posterior
probability of change decreases! Thus the global decision (stop or continue) is a non-monotone function of the posterior
probability obtained from local decisions in the agent based model. The example illustrates the unusual behaviour of the social
learning filter.

\subsubsection*{Summary}
In this subsection we  provided a Bayesian formulation of the problem of quickest detection of change in the value of a stock using the decisions of socially aware risk averse agents.  The quickest detection problem was shown to be non-trivial - the stopping region is in general non-convex when the agents' risk attitude was accounted for by considering a coherent risk measure, CVaR. Results which characterize the structural properties of social learning under the CVaR risk measure were provided and the importance of these results in understanding the global behaviour was discussed. It was observed that the behaviour of these risk-averse agents is, as expected, different from risk neutral agents. Risk averse agents herd sooner and do not prefer to ``learn" from the crowd, i.e, social learning region is smaller the more risk-averse the agents are. 
For further structural results on the risk averse social learning filter, please see \cite{KB16}.

 \subsection{Data Incest in Online Reputation Systems} \label{sec:incest}

 In comparison to the previous subsections ,
where social learning model  was formulated on  a line graph,
we now  consider social learning on a family of time dependent directed acyclic graphs  - in such cases, apart from herding, the phenomenon of data incest arises. 






 Consider an online reputation system  comprised of agents $\{1,2,\ldots,S\}$ that aim to estimate an underlying state of nature (a random variable). 
 Let $x \in 
 \X = \{1,2,\ldots,X\}$
 represent the state of nature (such as the quality of a restaurant/hotel) with known prior distribution $\pi_0$. Let $k = 1,2,3,\ldots$ depict epochs at which events occur. These events involve taking observations, evaluating beliefs and choosing actions as  described below. The index $k$ marks the historical order of events. For simplicity, we refer to $k$ as ``time".


 It is convenient also to reduce the coordinates of time $k$ and agent $s$  to a single integer index  $n$:
\begin{equation} \label{reindexing_scheme}
 n \triangleq s+ S(k-1), \quad
s \in \{1,\ldots, S\}, \; k = 1,2,3,\ldots 
\end{equation}
We  refer to $n$ as a ``node" of a time dependent information flow   graph $G_n$ which we now define.
 Let \begin{equation}\label{eq:defG} G_{n} = (V_{n}, E_{n}), \quad n  = 1,2,\ldots \end{equation} denote a sequence of time-dependent {\em directed acyclic graphs} ({\em DAGs})
  \footnote{A DAG is a directed graph with no directed cycles.}
  of information flow in the social network until and including time $k$ where $n = s + S(k-1)$. Each vertex in $V_{n}$ represents an agent $s'$ in the social network at time $k'$ and each edge $(n',n'')$ in $E_{n}\subseteq V_{n} \times V_{n}$ shows that the information (action) of node $n'$ (agent $s'$ at time $k'$) reaches node $n''$ (agent $s''$ at time $k''$).  It is clear that  $G_n$ is  a sub-graph of $G_{n+1}$. 

The Adjacency Matrix $A_n$ of $G_n $ is an $n\times n$ matrix with elements $A_n(i,j)$  given by 
\beq \label{eq:adjacencymatrix}
A_n (i,j)=\begin{cases}
1 &\textrm{ if } (v_j,v_i)\in E \;, \\
0 &\textrm{ otherwise}
\end{cases}\;, \text{  } A_n(i,i)=0.
\eeq

The transitive closure matrix $T_n$ is the  $n\times n$ matrix 
\beq T_n =  \text{sgn}((\mathbf{I}_n-A_n)^{-1}) \label{eq:tc} \eeq
where for  matrix $M$, the matrix $\text{sgn}(M)$ has elements
$$ 
\text{sgn}(M)(i,j) = \begin{cases} 0 & \text{ if } M(i,j)=0\;, \\
1  & \text{ if } M(i,j) \neq 0. \end{cases}
$$
Note that $A_n(i,j) = 1$ if there is a single hop path between nodes $i$ and $j$, In comparison,
$T_n(i,j) = 1$ if there exists a path (possible multi-hop) between  $i$ and $j$.

The information reaching node $n$ depends on the information flow graph  $G_n$.
The following two sets will be used  to specify the incest removal algorithms below:
\begin{align}
\history_n &= \{m : A_n(m,n) = 1 \}  \label{eq:history}  \\
\full_n &= \{m : T_n(m,n) = 1 \} . \label{eq:full}  
\end{align}
Thus $\history_n$ denotes the set of previous nodes $m$ that communicate with node $n$ in a single-hop.
In comparison, $\full_n$
  denotes the set of previous nodes $m$ whose information eventually arrives at node $n$. Thus $\full_n$  contains all possible  multi-hop connections by which information from a node $m$
 eventually reaches node $n$. 

\subsubsection*{Example} 
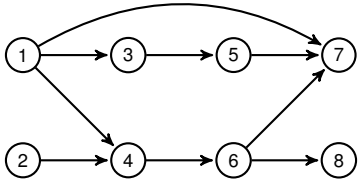
\begin{figure}[h]
\centering
\begin{tikzpicture}[->,>=stealth',shorten >=1pt,auto,node distance=2cm,
  thick,main node/.style={circle,fill=white!12,draw,font=\sffamily},scale=0.7, every node/.style={transform shape}]
 
  \node[main node] (1) {1};
  \node[main node] (2) [below  of=1] {2};
  \node[main node] (3) [right of=1] {3};
  \node[main node] (4) [below  of=3] {4};
  \node[main node] (5) [right of=3] {5};
  \node[main node] (6) [right of=4] {6};
  \node[main node] (7) [right of=5] {7};
  \node[main node] (8) [right of=6] {8};

  \path[every node/.style={font=\sffamily\small}]
    (1) edge node  {} (4)
          edge [bend left] node[left] {} (7)     
         edge node {} (3)
    (2) edge node {} (4)
    (3) edge node {} (5)
    (4) edge node {} (6)
    (5) edge node {} (7)
    (6) edge node {} (7)
         edge node {} (8);
\end{tikzpicture}

\caption{Example of  information flow network with $S=2$  two agents, namely   $s\in \{1,2\}$ and  time points $k=1,2,3,4$.  Circles represent the nodes indexed by $n= s + S(k-1)$
in the social network and each edge depicts a communication link between two nodes.}
\label{sample}
\end{figure}

Consider  $S=2 $ two agents with  information flow graph for three  time points $k=1,2,3$ depicted in Fig.\ref{sample}  characterized by 
the family of DAGs  $\{G_1,\ldots,G_7\}$. 
The adjacency matrices  
$A_1,\ldots,A_7$ are constructed as follows:  $A_n$ is the upper left  $n\times n$ submatrix of $A_{n+1}$ and 
$${\small A_7 = \begin{bmatrix}
0 & 0 & 1 & 1 & 0 & 0 & 1 \\
0 & 0 & 0 & 1 & 0 & 0 & 0 \\
0 & 0 & 0 & 0 & 1 & 0 & 0 \\
0 & 0 & 0 & 0 & 0 & 1 & 0 \\
0 & 0 & 0 & 0 & 0 & 0 & 1 \\
0 & 0 & 0 & 0 & 0 & 0 & 1 \\
0 & 0 & 0 & 0 & 0 & 0 & 0 \\
\end{bmatrix}} .$$

Let us explain these matrices.
Since nodes 1 and 2 do not communicate,
clearly $A_1$ and $A_2$ are zero matrices. Nodes 1 and  3 communicate, hence  $A_3$ has a single
one, etc. Note that if nodes 1,3,4 and 7 are assumed to be the same individual, then at node 7, the individual remembers what happened at node 5 and node 1, but not node 3.
This models the case where the individual has selective memory and remembers certain highlights.
From (\ref{eq:history}) and (\ref{eq:full}),  
$$\history_7=\{1,5,6\}, \quad \full_7 = \{1,2,3,4,5,6\} $$
where $\history_7$ denotes all one hop links to node 7 while $\full_7$ denotes all multihop links to node 7.

\subsection{Data Incest Model and Social Influence Constraint} \label{sec:influence}

Each node $n$ receives recommendations from its immediate friends (one hop neighbors) according to the information flow graph defined above.
 That is, it receives  actions  $\{a_m, m \in \history_n\}$ from  nodes  $m \in \history_n$ and then seeks 
to  compute the associated public beliefs  $\belief_m, m\in \history_n$.
If node $n$ naively (incorrectly)  assumes that the public beliefs $\belief_m, m\in \history_n$  are 
  independent, then
it would    
  fuse  these  as
  \beq  \belief_n{-} = \frac{\prod_{m\in \history_n} \belief_m }{\mathbf{1}_X^\p \prod_{m\in \history_n} \belief_m}.  \text{ (WRONG!)}\label{eq:dataincest}\eeq
  This naive data fusion would result in data incest.
  
\subsubsection{Aim} The aim is to provide each node $n$ the true posterior distribution 
\beq \tbelief_{n-}(i) = P(x = i | \{a_m, m \in \full_n\})   \label{eq:aims}\eeq
subject to the following {\em social influence constraint}:
There exists a fusion algorithm $\mathcal{A}$ such that
  \beq  \tbelief_{n-} = \mathcal{A} (\pi_m, m \in \history_n).   \label{eq:si}\eeq

 \subsubsection{Discussion. Fair Rating and Social Influence}
We briefly pause to discuss 
  (\ref{eq:aims}) and (\ref{eq:si}).\\
(i) We  call $\tbelief_{n-}$ in (\ref{eq:aims})  the {\em true}  or {\em fair online rating} available to
  node $n$
 since 
  $\full_n $ defined in (\ref{eq:full}) denotes all  information (multi-hop links) available  to node $n$. By definition
  $\tbelief_{n-}$ is incest free and is the desired conditional probability that agent $n$ needs.

  Indeed, if node $n$ combines  $\tbelief_{n-}$ together with its own private observation via social learning, then
clearly
\begin{align*}   \eta_n(i) &=  P(x = i | \{ a_m, m \in \full_n\},  y_n ), \quad i \in \X,  \\
\pi_n(i)  &= P(x = i | \{a_m, m \in \full_n\},  a_n ),\quad i \in \X, \end{align*}
are, respectively,   the correct (incest free) private belief for node $n$ and the correct after-action public belief.
If agent $n$ does not use  $\tbelief_{n-}$, then incest can propagate; for example if agent $n$ naively uses (\ref{eq:dataincest}).

Why should an individual $n$ agree to  use  $\tbelief_{n-}$ to combine with its private message? It is here that the social influence constraint (\ref{eq:si}) is important.
   $ \history_n$ can be viewed as the  ``{\em social message}'', i.e., personal friends of node $n$ since they directly communicate to node $n$, while the associated beliefs can be viewed as the
``{\em informational message}''.
As described in the remarkable recent paper  \cite{BFJ12}, the social message  from personal friends exerts a  large social influence\footnote{In a study conducted by social networking site {\em myYearbook}, 81\% of respondents said they had received advice from friends and followers relating to a product purchase through a social site; 74 percent of those who received such advice found it to be influential in their decision. ({\em Click Z}, January 2010).}
 -- it provides significant incentive (peer pressure) for individual $n$ to comply with
the protocol of combining its estimate with  $\tbelief_{n-}$  and thereby prevent incest.
 \cite{BFJ12} shows  that  receiving  messages 
from known friends has significantly more influence on an individual than the information in the messages. This study includes a comparison of information messages and social messages on Facebook and their direct
effect on voting behavior.
 To quote \cite{BFJ12}, ``The effect of social transmission on real-world voting
was greater than the direct effect of the messages themselves..."  In Sec.\ref{sec:expt}, we provide results of an experiment on human subjects that also  illustrates social influence in social learning.  \cite{KKT03} is an influential paper in the area of social influence.

 \subsection{Incest Removal  in  Online Reputation System} \label{sec:removal}

 It is convenient to work with the logarithm of the un-normalized belief\footnote{The un-normalized belief proportional to $\belief_n(i)$ is the numerator of the social learning filter (\ref{eq:piupdate}).
The corresponding un-normalized fair rating corresponding to $\tbelief_{n-}(i) $ is the joint distribution $P(x=i, \{a_m, m \in \full_n\})$.
By taking  the logarithm of the un-normalized belief, Bayes formula merely becomes the sum of the log likelihood and log prior. This allows
 us to devise a data incest  removal algorithm based on linear combinations of the log beliefs.};  accordingly define
$$  \lbelief_n(i) \propto \log \belief_n(i), \quad \lbelief_{n-}(i) \propto  \log \belief_{n-}(i), \quad i \in \X.$$

 The following theorem shows that the logarithm of the fair rating $\tbelief_{n-}$ defined in (\ref{eq:aims}) can be obtained as  a weighted  linear combination of the logarithms of previous public beliefs.

  \begin{theorem}[Fair Rating Algorithm] \label{thm:socialincestfilter} 
Suppose the network administrator runs the following algorithm for an online reputation system:
\begin{align}\label{eq:socialconstraintestimate}
\lbelief_{n-}(i)  &=   w_n^\p \, \lbelief_{1:n-1}(i) 
 \nonumber \\
\text{ where } &\;  w_n =  T_{n-1}^{-1}  t_n.
\end{align}
Then  $\lbelief_{n-}(i) \propto \log \tbelief_{n-}(i)$. That is,  the fair  rating $\log \tbelief_{n-}(i)$ defined in (\ref{eq:aims}) is obtained.
In (\ref{eq:socialconstraintestimate}),  $w_n$ is an  $n-1$ dimensional weight vector.
Recall  that
$t_n$ denotes the first $n-1$ elements of the  $n$th column of the transitive closure matrix $T_n$. \qed
\end{theorem}

Theorem  \ref{thm:socialincestfilter}  says that  the fair rating  $\tbelief_{n-}$ can be expressed as a linear function of the action log-likelihoods
 in terms of the transitive closure matrix $T_n$ of graph $G_n$. This is 
intuitive since $\tbelief_{n-}$ can be viewed as the sum of information collected by the nodes such that there are paths between all these nodes and $n$.

\begin{theorem}[Achievability of Fair Rating]\label{thm:sufficient}
Consider the fair rating algorithm specified by (\ref{eq:socialconstraintestimate}). With available information $(\belief_m, m \in \history_n)$ to achieve the estimates $\lbelief_{n-}$ of algorithm (\ref{eq:socialconstraintestimate}),
a necessary and sufficient condition on the information flow graph $G_n$ is \beq \label{constraintnetwork}
A_n(j,n)=0   \implies w_n(j)= 0.
\eeq
(Recall 
  $w_n$ is specified in (\ref{eq:socialconstraintestimate}).)
\qed
\end{theorem}

Note that the constraint (\ref{constraintnetwork}) is purely in terms of the adjacency matrix $A_n$, since the transitive closure matrix (\ref{eq:tc}) is a function of the adjacency matrix.

 \subsection{Ordinal Decisions  and Bayesian Social Sensors} \label{sec:ordinal}

The social learning protocol 
 assumes that each agent is a Bayesian utility optimizer.
The following discussion puts together ideas from the economics literature
to show that under reasonable conditions, such a Bayesian model is a useful idealization of  agents' behaviors.
This means that the Bayesian social learning follows simple intuitive rules and is therefore, a useful idealization.
(In Sec.\ref{sec:revealed},  we discuss the theory of revealed preferences which yields a non-parametric  test on data to determine
if an agent is a utility maximizer.)

Humans typically make {\em monotone} decisions - the more favorable the private  observation,
the higher the recommendation. Humans   make {\em ordinal} decisions\footnote{Humans typically convert numerical attributes to ordinal scales before making  decisions. For example,
it does not matter if the cost of a meal at a restaurant is \$200 or \$205; an individual would classify this cost as ``high". 
Also credit rating agencies use ordinal symbols such as AAA, AA, A.} since humans tend to think in symbolic ordinal terms.
Under what conditions is the recommendation $a_n$ made by  node $n$ {\em monotone increasing} in its observation
$y_n$ and {\em ordinal}? 
Recall from the social learning protocol (\ref{eq:myopic}) that the actions of agents are 
$$ a_k =   \arg\min_{a\in \A} \{c_a^\p \oprob_{\obs_k} \belief_k\}.  $$
So an equivalent question is: Under what conditions is the $\argmin$ increasing in observation $y_n$?
Note that an increasing argmin is an {\em ordinal} property - that is,
$  \argmin_a c_a ^\p B_{y_n} \tbelief_{n{-}}$ increasing in $y$ implies $\argmin_a \phi(c_a ^\p B_{y_n} \tbelief_{n{-}})$ is also increasing in $y$ for any monotone function $\phi(\cdot)$.

The following result gives sufficient conditions for each agent to give a  recommendation that is monotone and ordinal in its private observation:
\begin{theorem} \label{thm:monotone}
Suppose the observation probabilities and costs satisfy the following conditions:
\begin{compactenum}
\item[(A1)] $B_{iy}$ are TP2 (totally positive of order 2); that is,
$B_{i+1,y}B_{i,y+1} \leq B_{i,y} B_{i+1,y+1}$.
\item[(A2)]  $c(x,a)$ is submodular. That is, $c(x,a+1) - c(x,a) \leq c(x+1,a+1)-c(x+1,a)$.
\end{compactenum}
Then
\begin{compactenum}
\item
Under (A1) and (A2),  the recommendation $a_n(\tbelief_{n-},y_n) $  made by agent $n$ is increasing and hence ordinal in observation $y_n$, for any $\tbelief_{n{-}}$. 
\item Under (A2),     $a_n(\tbelief_{n-},y_n) $ is increasing in belief $\tbelief_{n-}$ with respect to the monotone likelihood ratio (MLR) stochastic order\footnote{ \label{footnotemlr} Given probability mass functions
$\{p_i\}$ and $\{q_i\}$, $i=1,\ldots,X$ then
$p$ MLR dominates $q$ if  $\log p_i - \log p_{i+1} \leq \log q_i - \log q_{i+1}$.}  for any observation~$y_n$.\qed
\end{compactenum} 
\end{theorem}
The proof is in \cite{Kri12}.
We can interpret the above theorem as follows. If agents makes recommendations that are monotone and ordinal in the observations and monotone in the prior, then they mimic
the Bayesian social learning model.  Even if the agent does not exactly follow a Bayesian social learning model,  its monotone ordinal behavior implies that such a Bayesian model is  a useful idealization.

Condition (A1) is widely studied in monotone decision making; see the classical book by  Karlin \cite{Kar68} and  \cite{KR80}; numerous examples of noise distributions are  TP2. Indeed in the highly cited paper
\cite{Mil81} in the economics literature, observation $y+1$ is said to be more ``favorable news''  than observation $y$ if Condition (A1) holds.

Condition (A2) is the well known submodularity condition \cite{Top98,MS92,Ami05}. 
(A2)  makes sense in a reputation system for the costs to be well posed.
 Suppose the recommendations in action set $\A$ are arranged
in increasing order and also  the states in $\X$ for the underlying state are arranged in ascending order. Then (A2) says:  if recommendation $a+1$  is more accurate
than recommendation $a$ for state $x$; 
then recommendation $a+1$ is also more accurate than recommendation $a$ for state $x+1$ (which is a higher quality state than $x$).

In the experiment  results reported in Sec.\ref{sec:expt}, we found that  (A1) and (A2) of Theorem \ref{thm:monotone} are justified.

\subsection{Psychology Experiment Dataset} \label{sec:expt}
To illustrate social learning, data incest and social influence, this section presents an actual 
psychology experiment that was conducted by our colleagues at the Department of Psychology of University of British Columbia in September and  October,  2013. The participants comprised 36 undergraduate students  who participated in the experiment for course credit.

\subsubsection{Experiment Setup}
The experimental study involved 1658 individual trials. Each trial comprised two participants who were asked to
 perform a perceptual task interactively.  
 The perceptual task was as follows:
 Two arrays of circles denoted left and right, were given to each pair of participants. Each participant was asked to judge which array (left or right) had the larger average diameter. The
 participants answer (left of right) constituted their action. So the action space is $\A = \{0 \text{ (left)} ,1 \text{ (right)}\}$.
 
The circles  were prepared for each trial as follows: two $4\times4$ grids of circles were generated by uniformly sampling  from the radii: $\{ 20, 24, 29, 35, 42\}$ (in pixels). The average diameter of each grid was computed, and if the means differed by more than 8\% or less than 4\%, new grids were made. Thus in each trial, the left array and right array
   of circles differed  in the average diameter  by 4-8\% 
   
 For each trial,
one of the two participants was chosen randomly to start the experiment  by choosing an action according to his/her observation. Thereafter, each participant was given access
to their partner's previous response (action) and the participants own previous action prior to making his/her  judgement. This mimics the social learning protocol.
  The participants continued choosing actions according to this procedure until the experiment terminated. The trial terminated when the response of each of the two participants did not change for three successive iterations (the two participants did not necessarily have to agree for the trial to terminate).
 
\begin{figure}[htb]
\centerline{
\includegraphics[width=0.7\textwidth]{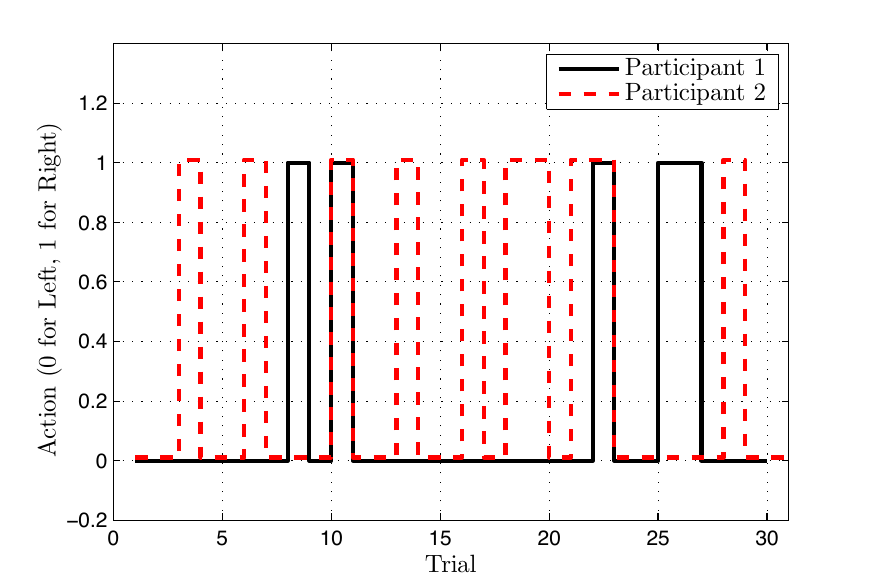}}
\caption{Example of sample  path of actions chosen by  two participants in a single trial of the experiment. In this  trial, both participants eventually chose the correct answer 0 (left).}
\label{Fig:Samplepath}
\end{figure}

In each trial, the actions of participants were recorded along with the  time interval taken to choose their action. As an example, Fig.~\ref{Fig:Samplepath} illustrates the sample path of  decisions made by the two participants
in one of the 1658 trials. In this specific trial, the average diameter of the left array of circles was $32.1875$ and the right array was	$30.5625$ (in pixels); so the ground truth was $0 $ (left).

\begin{figure}[h]
\centerline{
\includegraphics[width=0.7\textwidth]{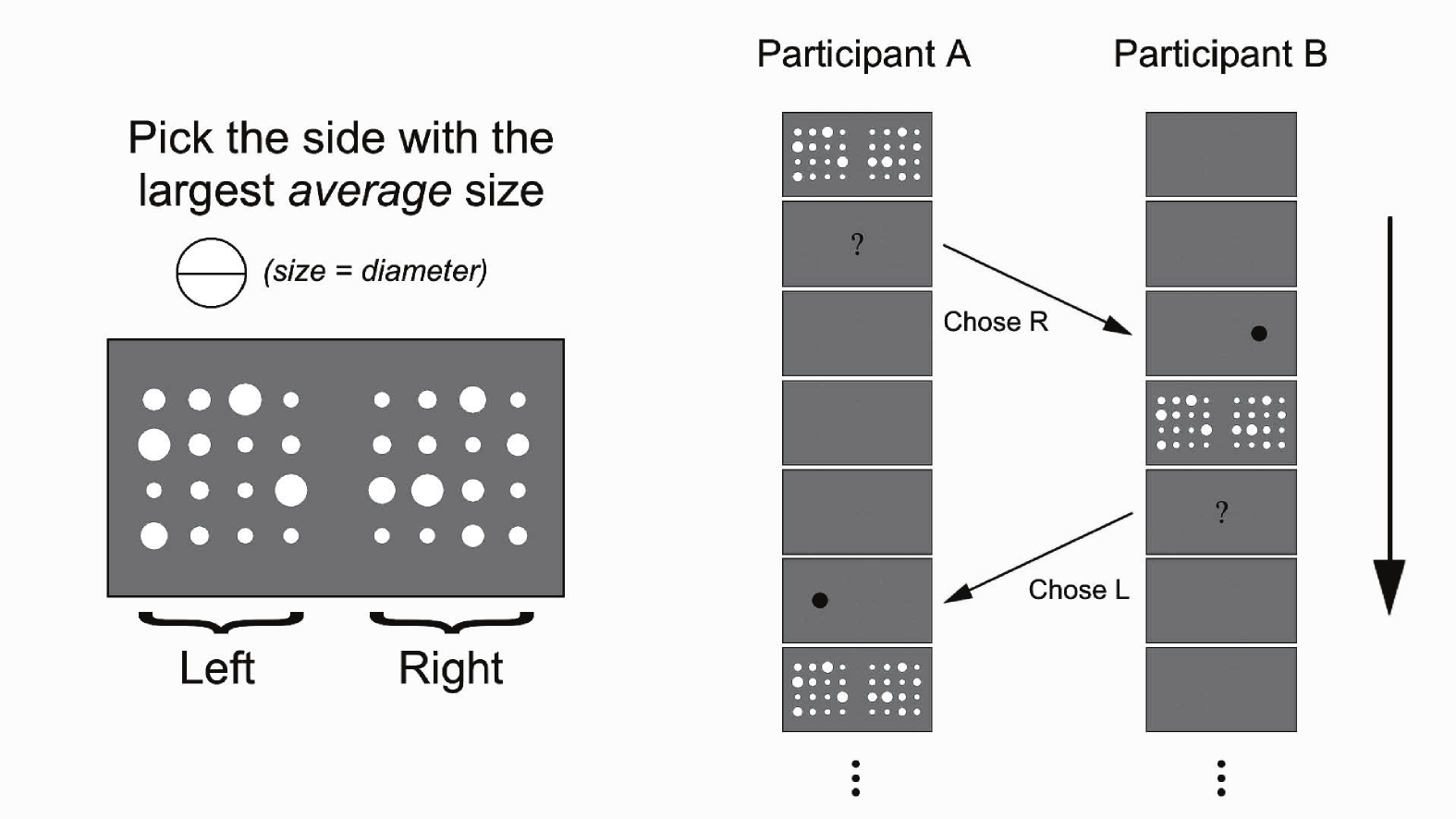}}
\caption{Two arrays of circles were given to each pair of participants on a screen. Their task is to interactively determine which side (either left or right) had the larger average diameter. The partner's previous decision was displayed on screen prior to the stimulus.}
\label{Fig:SocialSensor}
\end{figure}

\subsubsection{Experimental Results}   The results of our experimental study are as follows:

\paragraph{Social learning Model} As mentioned above, the experiment for each pair of participants was continued until both participants' responses stabilized. 
 {\em In what percentage of these experiments, did an agreement occur between the two participants?} 
The answer to this question reveals  whether ``herding" occurred in the experiments
 and whether
the participants performed social learning (influenced by their partners). The experiments show that in  66\% of trials (1102 among 1658), participants reached an agreement; that is herding occurred. Further, in 32\% of the trials, both participants converged to the correct decision after a few interactions. 

To construct  a  social learning model for the experimental data,  we consider  the experiments where both participants reached an agreement.
Define the social learning success rate as
    \begin{equation}\nonumber
\frac{\text{\# expts where both participants chose correct answer}}{\text{\# expts where both participants reached an agreement}}\cdot
    \end{equation}
   In the experimental study, the state space is $\X = \{0,1\}$ where $x = 0$, when the left array of circles has the larger diameter and $x = 1$, when the right array has the larger diameter. The initial belief for both participants is considered to be $\pi_0 = [0.5, 0.5]$. The observation space is assumed to be $\Y = \{0,1\}$. 
   
   To estimate the social learning model parameters (observation probabilities  $B_{iy}$  and  costs  $c(i,a)$), 
we determined the parameters that best fit the learning success rate of the experimental data. The best fit parameters obtained were\footnote{Parameter estimation in social learning is a challenging problem not addressed in this chapter. Due to the formation of  cascades in finite time, construction of an asymptotically  consistent estimator is impossible, since actions after the formation of a cascade contain no information.}
   \begin{align}
   &B_{iy}  = \begin{bmatrix} 0.61 & 0.39 \\ 0.41 & 0.59 \end{bmatrix}, \quad
   c(i,a) = \begin{bmatrix} 0 & 2 \\ 2 & 0 \end{bmatrix}\nonumber.  
   \end{align}
Note that $B_{iy}$ and $c(i,a)$ satisfy both the conditions of the Theorem \ref{thm:monotone}, namely TP2 observation probabilities and single-crossing cost. This implies that the subjects of this experiment made monotone and ordinal decisions.
   
\begin{figure}[h]
\centering
\scalebox{.5}{\includegraphics{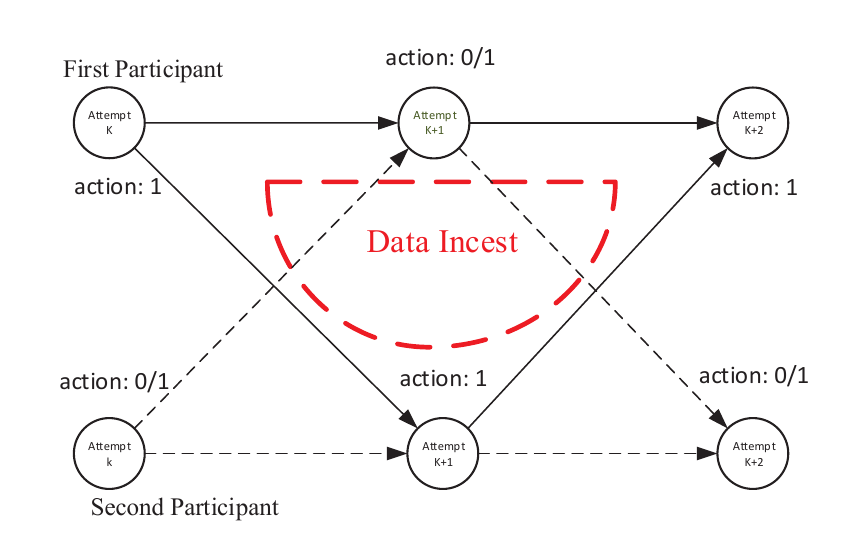}}

\scalebox{.5}{\includegraphics{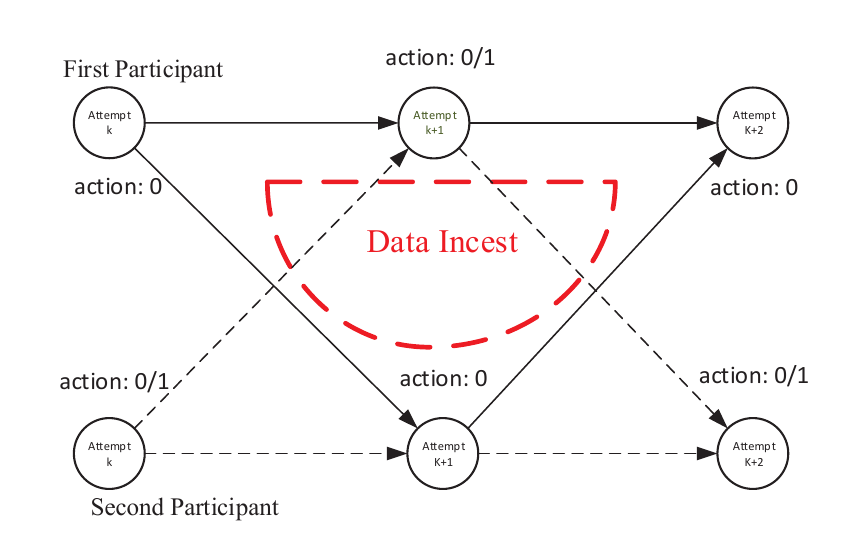}}
\caption{\label{exp:dataincest} Two information  patterns from our experimental studies  which can result in data incest.}
\end{figure}

\paragraph{Data incest} Here, we study the effect of information patterns in  the experimental study that can result in data incest. Since private observations
are highly subjective and participants  did not
document these, we cannot claim with certainty if data incest changed the action of an individual. However, from the experimental data, we can localize specific
information patterns that can result in incest. In particular, we focus on  the two information flow graphs depicted in Fig.\ref{exp:dataincest}.
 In the two graphs of Fig.\ref{exp:dataincest}, the action of the first participant at time $k$ 
 influenced the action of the second participant at time $k+1$, and thus, could have been double
 counted by the first participant at time $k+2$. 
We found that in 
   79\% of experiments, one of the information patterns shown in Fig.\ref{exp:dataincest} occurred (1303 out of  1658 experiments). Further, in 21\% of experiments, 
  the information patterns shown in Fig.\ref{exp:dataincest} occurred and at least one participant
changed his/her decision, i.e., the judgement of participant at time $k+1$ differed from his/her judgements at time $k+2$ and $k$.  These results
show that even for   experiments involving  two participants, data incest information  patterns occur frequently (79\%) and causes individuals to modify their
actions (21\%). It shows that  social learning protocols require careful design to handle and mitigate data incest.


\subsection{Summary and Extensions}
In this section,
we have outlined a controlled sensing problem over a social network in which the administrator controls (removes) data incest
and thereby maintains an unbiased (fair) online reputation system.  The state of nature could be geographical coordinates of an event (in a target localization problem) or quality of a social unit (in an online reputation system).  As discussed above, data incest arises  due to the recursive nature of Bayesian estimation and non-determinism in the timing of 
the  sensing by individuals. Details of proofs, extensions and further numerical studies are presented in~\cite{KH13,HK13}.

We summarize  some  extensions of the social learning framework that are relevant to interactive sensing.


\subsubsection{Wisdom of Crowds} 

Surowiecki's  book \cite{Sur05}  is an excellent popular piece that 
explains the wisdom-of-crowds hypothesis. The wisdom-of-crowds hypothesis predicts that the independent judgments of a crowd of individuals (as measured by any form of central tendency) will be relatively accurate, even when most of the individuals in the crowd are ignorant and error prone. The book also 
studies situations (such as rational bubbles) in which  crowds are not wiser than individuals.  Collect enough people on a street corner staring at the sky, and everyone who walks past will look up. Such herding behavior is typical in social learning.

\subsubsection{In which order should agents act?}
In the social learning protocol, we assumed that the agents act sequentially in a pre-defined order.
However, in many social networking applications, it is important to optimize the order in which agents  act. For example, 
consider an online review site where individual reviewers with different reputations  make their reviews publicly available. 
If a reviewer with high reputation publishes her review first, this review will unduly affect the decision of a reviewer with lower reputation.
In other words, if the most senior agent ``speaks" first it would unduly affect the decisions of more junior  agents. This could lead to an increase in bias of the underlying state estimate.\footnote{To quote  \cite{AK09}: ``In  94\% of cases, groups (of people) used the first answer provided as their final answer... Groups tended to commit to the first answer provided by
any group member.''  People with dominant personalities tend to speak first and most forcefully ``even when they actually lack competence''. } 
On the other hand, if the most junior agent is polled first, then since its variance is large, several agents would need to be polled in order
to reduce the variance. We refer the reader to \cite{OS01} for an interesting description of who should speak first in a public debate.\footnote{As described
in \cite{OS01}, seniority is considered in the rules of debate and voting in the U.S.\ Supreme Court. ``In the past, a vote was taken after the newest
justice to the Court spoke, with the justices voting in order of ascending seniority largely, it was said, to avoid the pressure from long-term members
of the Court on their junior colleagues."}
It turns out that for two agents, the seniority rule is always optimal for any prior -- that is, the  senior agent speaks first followed
by the junior agent; see \cite{OS01} for the proof. However, for more than two agents, the optimal order
depends on the prior, and the observations in general. 

\section{Revealed Preferences: Are social sensors utility maximizers?} \label{sec:revealed}

We now move on to the third main topic of the chapter, namely, the principle of revealed preferences. 
The main question addressed is:
Given a dataset of decisions made by a social sensor, is it possible to determine if the social sensor is a utility maximizer? More generally,  is a dataset from
 a multiagent system consistent with play from a Nash equilibrium? If yes, can the behavior of the social sensors be learned using data from the social network?

These questions are  fundamentally  different to the model-based theme  that is widely used in the signal processing literature in which an 
objective function (typically convex) is proposed and then algorithms are constructed to compute the minimum. In contrast, the revealed preference approach is data-centric--we wish to determine whether the dataset is obtained from an utility maximizer.
%
In simple terms, revealed preference theory seeks to determine if an agent is an utility maximizer subject to budget constraints
based on observing its  choices over time.   
In signal processing terminology, such problems can be viewed as set-valued system identification of an argmax nonlinear system.
The principle of revealed preferences  is widely studied in the micro-economics literature. As mentioned in Sec.\ref{sec:introduction}, Varian  has written several influential papers in this area.
In this section we will use the principle of revealed preferences
on datasets  to determine how social sensors behave as a function of external influence.
 The setup is  depicted in the schematic diagram Fig.\ref{fig:SNrevealedpreference}.

\subsection{Afriat's Theorem for a single agent}
\label{subsec:afriatstheorem}

The theory of revealed preferences was pioneered by Samuelson~\cite{Samuelson38}.
Afriat  published a highly influential paper \cite{Afr67}  in revealed preferences (see also \cite{Afr87}). 
 Given a time-series of data $\dataset=\{(\probe_\tindx,\response_\tindx), \tindx\in\{1,2,\dots,\Tindxter\}\}$ where $\probe_\tindx\in\mathds{R}^m$ denotes the external influence, $\response_\tindx$ denotes the response of agent, and $\tindx$ denotes the time index, is it possible to detect if the
   agent is a {\it utility maximizer}?
An agent is a {\em utility maximizer} if for every external influence $\probe_\tindx$, the chosen response $\response_\tindx$ satisfies
\begin{equation}
\response_\tindx(\probe_\tindx)\in\operatorname*{arg\,max}_{\{\probe_\tindx^\p \response \leq \budget_\tindx\}}\utility(\response)
\label{eqn:singlemaximization}
\end{equation}
with $\utility(\response)$ a non-satiated utility function. Nonsatiated means that an increase in any element of response $\response$ results in the utility function increasing.\footnote{The non-satiated assumption rules out  trivial cases such as a constant utility function which can be  optimized by any response.} As shown by Diewert~\cite{Die73}, without local nonsatiation the maximization problem (\ref{eqn:singlemaximization}) may have no solution. 

 In (\ref{eqn:singlemaximization}) the budget constraint $\probe_\tindx^\p \response \leq \budget_\tindx$ denotes the total amount of resources available to the social sensor for selecting the response $x$  to the external influence $p_t$. For example, if $p_t$ is the electricity price and $x_t$ the associated electricity consumption, then the budget of the social sensor is the available monetary funds for purchasing electricity. In the real-world social sensor datasets provided in this chapter, further insights are provided for the budget constraint.

The celebrated 
``Afriat's theorem"  provides a necessary and sufficient
  condition for a finite dataset $\dataset$ to have originated from an utility maximizer. 
 Afriat's theorem has subsequently been expanded and refined, most notably by Diewert \cite{Die73}, Varian \cite{Var82} and 
 Blundell \cite{Blu05}. 
\begin{theorem}[Afriat's Theorem]{Given a dataset $\mathcal{D}=\{(p_t,x_t):t\in \{1,2,\dots,T\}\}$, the following statements are equivalent:}
\begin{compactenum}
  \item The agent is a utility maximizer and there exists a nonsatiated and concave utility function that satisfies (\ref{eqn:singlemaximization}).
\item For scalars $u_t$ and $\lambda_t>0$ the following set of inequalities has a feasible solution:
\begin{equation}
\utility_\tau-\utility_\tindx-\lambda_\tindx \probe_\tindx^\p (\response_\tau-\response_\tindx) \leq 0 \text{ for } \tindx,\tau\in\{1,2,\dots,\Tindxter\}.\
\label{eqn:AfriatFeasibilityTest}
\end{equation}
\item A nonsatiated and concave utility function that satisfies (\ref{eqn:singlemaximization}) is given by:
\begin{equation}
\utility(\response) = \underset{\tindx\in T}{\operatorname{min}}\{u_\tindx+\lambda_\tindx \probe_\tindx^\p(\response-\response_\tindx)\}
\label{eqn:estutility}
\end{equation}
  \item The dataset $\mathcal{D}$ satisfies the Generalized Axiom of Revealed Preference (GARP), namely for any $k\leq T$, $p_t^\p x_t \geq p_t^\p x_{t+1} \quad \forall t\leq k-1 \implies p_k^\p  x_k \leq p_k^\p  x_{1}.$ \qed
\end{compactenum}
\label{thrm: Afriat's Theorem}
\end{theorem}

As pointed out in  \cite{Var82}, a remarkable feature of Afriat's theorem is that if the dataset can be rationalized by a non-trivial utility function, then it can be rationalized
by a continuous, concave, monotonic utility function. ``Put another way,  violations of continuity, concavity, or monotonicity cannot be detected with only a finite number of demand observations".

Verifying  GARP  (statement 4 of Theorem \ref{thrm: Afriat's Theorem}) on a dataset $\dataset$ comprising $T$ points can be done using Warshall's algorithm with $O(\Tindxter^3)$~\cite{Var82,FST04} computations. Alternatively, determining if Afriat's inequalities (\ref{eqn:AfriatFeasibilityTest}) are feasible can be done via a LP feasibility test (using for example interior point methods \cite{BV04}). Note that the utility (\ref{eqn:estutility}) is not unique and is ordinal by construction. Ordinal means that any monotone increasing  transformation of the utility function will also satisfy Afriat's theorem. Therefore the utility mimics the ordinal behavior of humans, see also Sec.\ref{sec:ordinal}.
Geometrically the estimated utility (\ref{eqn:estutility}) is the lower envelop of a finite number of hyperplanes that is consistent with the dataset~$\dataset$.

Note that GARP is equivalent to the notion of ``cyclical consistency''~\cite{Var83} -- they state that the responses are consistent with utility maximization if no negative cycles are present. As an example, consider a dataset $\dataset$ with $\Tindxter=2$ observations resulting from a utility maximization agent. Then GARP states that $\probe_1^\p\response_1 \geq \probe_1^\p\response_2$ $\implies \probe_2^\p\response_2 \leq \probe_2^\p\response_1$. From (\ref{eqn:singlemaximization}), the underlying utility function must satisfy $\utility(\response_1) \geq \utility(\response_2)$ $\implies \utility(\response_2) \leq \utility(\response_1)$ where the equality results if $\response_1=\response_2$. 

Another remarkable feature of Afriat's Theorem is that no parametric assumptions of the utility function of the agent are necessary. 
 To gain insight into the construction of the inequalities (\ref{eqn:AfriatFeasibilityTest}), let us assume the utility function $\utility(\response)$ in (\ref{eqn:singlemaximization}) is increasing for positive $\response$, concave, and differentiable. If $\response_\tindx$ solves the maximization problem (\ref{eqn:singlemaximization}), then from the Karush-Kuhn-Tucker (KKT) conditions there must exist Lagrange multipliers $\lambda_\tindx$ such that $$\nabla\utility(\response_\tindx)=\lambda_\tindx\nabla(\probe_\tindx^\p\response_\tindx-\budget_\tindx)=\lambda_\tindx\probe_\tindx$$ is satisfied for all $\tindx\in\{1,2,\dots,\Tindxter\}$. Note that since $\utility(\response)$ is increasing, $\nabla\utility(\response_\tindx)=\lambda_\tindx\probe_\tindx > 0$, and since $\probe_\tindx$ is strictly positive, $\lambda_\tindx > 0$. Given that $\utility(\response)$ is a concave differentiable function, it follows  that $$\utility(\response) \leq \utility(\response_\tindx)+\nabla\utility(\response_\tindx)^\p(\response-\response_\tindx)
\quad \forall \tindx\in\{1,2,\dots,\Tindxter\}.$$ Denoting $\utility_\tindx=\utility(\response_\tindx)$ and $\utility_\tau=\utility(\response_\tau)$, and using the KKT conditions and concave differential property, the inequalities (\ref{eqn:AfriatFeasibilityTest}) result. To prove that if the solution of (\ref{eqn:AfriatFeasibilityTest}) is feasible then GARP is satisfied can be performed using the duality theorem of linear programming as illustrated in \cite{FST04}.

\subsection{Revealed Preferences for Multi-agent Social Sensors}

We now consider a 
multi-agent version of Afriat's theorem for deciding if a dataset is generated by  playing from the equilibrium of a potential game\footnote{As in
\cite{Deb08}, we consider potential games since they are sufficiently specialized so that there exist datasets that fail Afriat's test.} An example is the control of power consumption in the electrical grid. Consider a corporate network of financial management operators that select the electricity prices in a set of zones in the power grid. By selecting the prices of electricity the operators are expected to be able to control the power consumption in each zone. The operators wish to supply their consumers with sufficient power however given the finite amount of resources the operators in the corporate network must interact. This behavior can be modelled as a game. Recent analysis of energy use scheduling and demand side management schemes in the energy market have been performed using potential games~\cite{CVH13,ING10,WMHW11}. Another example of potential games are {\it congestion games}~\cite{RR73,HTW06,BKP07,MS96} in which the utility of each player depends on the amount of resource it and other players use.

\begin{figure}[h!]
  \centering
\begin{tikzpicture}[font = \small, scale =0.8,transform shape, american voltages]

\draw [fill= lightgray] (-5.2,-1.5) coordinate (topleft) rectangle (3.4,-0.8) coordinate (bottomright) node[pos=.5] {\large External Influence};
\draw [fill= white] (-4.8,0) coordinate (topleft) rectangle (3,3.3) coordinate (bottomright);
\draw[ultra thick, black, ->] (2,0) -- (2,-0.8);
\draw[ultra thick, black,->] (-4,-0.8) -- (-4,0);
\node at (-3.7,-0.5) {$\probe_\tindx$};
\node at (1.1,-0.4) {$\response_\tindx^1,\cdots,\response_\tindx^\nindx$};

\node[draw,ultra thick,black,circle,minimum width=0.8cm] (C1) at (-2,2.8) {1};
\node[draw,ultra thick,black,circle,minimum width=0.8cm] (C2) at (0,2.8) {2};
\node[draw,ultra thick,black,circle,minimum width=0.8cm] (C3) at (2,1.7) {3};
\node[draw,ultra thick,black,circle,minimum width=0.8cm] (C4) at (0,0.5) {4};
\node[draw,ultra thick,black,circle,minimum width=0.8cm] (Cn1) at (-2,0.5) {n-1};
\node[draw,ultra thick,black,circle,minimum width=0.8cm] (Cn) at (-4,1.7) {n};
\draw[thick, black, <->] (Cn.east) -- (C3.west);
\draw[thick, black, <->] (C1.east) -- (C2.west);
\draw[thick, black, <->] (C2.east) -- (C3.north west);
\draw[thick, black, <->] (C3.south west) -- (C4.east);
\draw[thick, black, <->] (C2.south) -- (C4.north);
\draw[thick, black, <->] (Cn1.north) -- (C1.south);
\draw[thick, black, <->] (Cn.45) -- (C1.180);
\draw[thick, black, <->] (Cn.320) -- (Cn1.180);
\draw[thick, black, dotted] (C4.west) -- (Cn1.east);
\draw[thick, black, <->] (C1.south east) -- (C4.north west);
\draw[thick, black, <->] (C2.south west) -- (Cn1.north east);
\draw[thick, black, <->] (Cn.22) -- (C2.205);
\draw[thick, black, <->] (Cn.-21) -- (C4.155);
\draw[thick, black, <->] (C3.155) -- (C1.-25);
\draw[thick, black, <->] (C3.-155) -- (Cn1.25);

\node at (-3.6,3.1) {Social Network};
\end{tikzpicture}
  \caption{Schematic of a social network containing $\nindx$ agents where $\probe_\tindx\in\mathds{R}^m$  denotes the external influence, and $\response_\tindx^i\in\mathds{R}^m$ the response of agent $i$ in response to the external influence and other agents at time $\tindx$. Note that dotted line denotes consumers $4,\dots,n-1$.  The aim is to determine
  if the  dataset $\dataset$ 
defined in (\ref{eqn:ResponseData}),  is consistent with play from a Nash equilibrium.} 
\label{fig:SNrevealedpreference}
\end{figure}
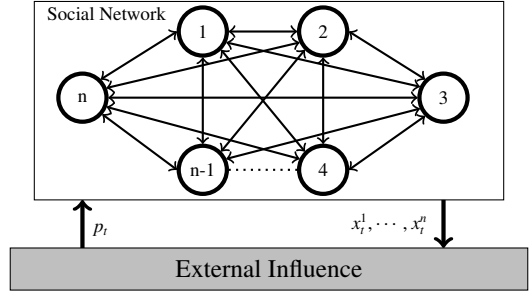

Consider the social network presented in Fig.\ref{fig:SNrevealedpreference}, given a time-series of data from $N$ agents $\dataset=\{(\probe_\tindx,\response_\tindx^1,\dots,\response_\tindx^n): \tindx\in\{1,2,\dots,\Tindxter\}\}$ with $\probe_\tindx\in\mathds{R}^m$ the external influence, $\response_\tindx^i$ the response of agent $i$, and $\tindx$ the time index, is it possible to detect if the dataset originated from agents that play
a potential game?

The characterization of how agents behave as a function of external influence, for example price of using a resource, and the responses of other agents in a social network, is key for analysis. Consider the social network illustrated in Fig.\ref{fig:SNrevealedpreference}. There are a total of $\nindx$ interacting agents in the network and each can produce a response $\response_\tindx^i$ in response to the other agents and an external influence $\probe_\tindx$. Without any {\it a priori} assumptions about the agents, how can the behaviour of the agents in the social network be learned? In the engineering literature the behaviour of agents is typically defined {\it a priori} using a {\it utility function}, however our focus here is on learning the behaviour of agents. The {\it utility function} captures the satisfaction or payoff an agent receives from a set of possible responses, denoted by $\setresponse$. Formally, a utility function $\utility:\setresponse\rightarrow\mathds{R}$ represents a preference relation between responses $\response_1$ and $\response_2$ if and only if for every $\response_1,\response_2\in\setresponse$, $\utility(\response_1) \leq \utility(\response_2)$ implies $\response_2$ is preferred to $\response_1$. Given a time-series of data $\dataset=\{(\probe_\tindx,\response_\tindx^1,\dots,\response_\tindx^n): \tindx\in\{1,2,\dots,\Tindxter\}\}$ with $\probe_\tindx\in\mathds{R}^m$ the external influence, $\response_\tindx^i$ the response of agent $i$, and $\tindx$ the time index, is it possible to detect if the series originated from an agent that is a {\it utility maximizer}?

In a  network of social sensors (Fig.\ref{fig:SNrevealedpreference}), the responses of agents may be dependent on both the external influence $\probe_\tindx$ and the responses of the other agents in the network, denoted by $\response_\tindx^{-i}$. The utility function of the agent must now include the responses of other agents--formally if there are $\nindx$ agents, each has a utility function $\utility^i(\response^i,\response_\tindx^{-i})$ with $\response^i$ denoting the response of agent $i$, $\response_\tindx^{-i}$ the responses of the other $\nindx-1$ agents, and $\utility^i(\cdot)$ the utility of agent $i$. Given a dataset $\dataset$, is it possible to detect if the data is consistent with agents playing a game and maximizing their individual utilities? Deb, following Varian's and Afriat's work, shows that refutable restrictions exist for the dataset $\dataset$, given by (\ref{eqn:ResponseData}), to satisfy Nash equilibrium (\ref{eqn:NashEquation})~\cite{Deb08,Afr67,Var06}. These refutable restrictions are however, satisfied by most $\dataset$~\cite{Deb08}. The detection of agents engaged in a concave potential game, and generating responses that satisfy Nash equilibrium, provide stronger restrictions on the dataset $\dataset$~\cite{Deb08,Deb09}. We denote this behaviour as {\em Nash rationality}, defined as follows:
\begin{definition}[\cite{Deb09,Ui08,Ney97}]\label{eq:NashEquilibrium}
Given a dataset
\begin{equation}
 \mathcal{D}=\{(\probe_\tindx,\response_\tindx^1,\response_\tindx^2,\dots,\response_\tindx^\nindx): \tindx\in\{1,2,\dots,\Tindxter\}\},
\label{eqn:ResponseData}
\end{equation}
$\dataset$ is consistent with {\it Nash equilibrium} play if there exist utility functions $\utility^i(\response^i,\response^{-i})$ such that 
\begin{equation}
\response_\tindx^i=\response_\tindx^{i*}(\probe_t)\in\operatorname*{arg\,max}_{\{\probe_\tindx^\p \response^i \leq \budget_\tindx^i\}}\utility^i(\response^i,\response^{-i}). 
\label{eqn:NashEquation}
\end{equation}
In (\ref{eqn:NashEquation}), $\utility^i(\response,\response^{-i})$ is a nonsatiated utility function in $\response$, $\response^{-i} = \{\response^j\}_{j\neq i}$ for $i,j\in\{1,2,\dots,\nindx\}$, and the elements of $\probe_\tindx$ are strictly positive. Nonsatiated means that for any $\epsilon>0$, there exists a $\response^i$ with $\norm{\response^i-\response^i_\tindx}_2<\epsilon$ such that $\utility^i(\response^i,\response^{-i}) > \utility^i(\response_\tindx^i,\response_\tindx^{-i})$. If  for all $\response^i, \response^j\in \setresponse^i$, there exists a concave potential function $\potfun$ that satisfies
\begin{equation}
\utility^i(\response^i,\response^{-i})   -\utility^i(\response^j,\response^{-i}) > 0 \text{ iff } \potfun(\response^i,\response^{-i})-\potfun(\response^j,\response^{-i}) > 0 
\label{eqn:cordpotfun}
\end{equation} 
for all the utility functions $\utility^i(\cdot)$ with $i\in\{1,2,\dots,\nindx\}$, then the dataset $\dataset$ satisfies {\it Nash rationality}. \qed
 \end{definition}
Just as with the utility maximization budget constraint in (\ref{eqn:singlemaximization}), the budget constraint $\probe_\tindx^\p \response^i \leq \budget_\tindx^i$ in (\ref{eqn:NashEquation}) models the total amount of resources available to the social sensor for selecting the response $\response^i_\tindx$ to the external influence $\probe_\tindx$.

The detection test for Nash rationality (Definition~\ref{eq:NashEquilibrium}) has been used in \cite{CDFQ13} to detect if oil producing countries are collusive, and in \cite{Deb08} for the analysis of household consumption behaviour. 

In the following sections, decision tests for utility maximization, and a non-parametric learning algorithm  for
predicting agent responses are presented. Three real world datasets are analyzed using the non-parametric decision tests and learning algorithm. The datasets are comprised of bidders auctioning behaviour, electrical consumption in the power grid, and on the tweeting dynamics of agents in the social network Twitter illustrated in Fig.\ref{fig:SNrevealedpreference}. 

\subsection{Decision Test for Nash Rationality}
\label{sec:detectiontests}

This section presents a non-parametric test for Nash rationality given the dataset $\mathcal{D}$ defined in (\ref{eqn:ResponseData}). If the dataset $\mathcal{D}$ passes the test, then it is consistent with play according to a Nash equilibrium of a concave potential game. In Sec.\ref{sec:Non-parametricLearningAlgorithmforResponseForecasting}, a learning algorithm is provided that can be used to predict the response of agents in the social network provided in Fig.\ref{fig:SNrevealedpreference}.

The following theorem provides necessary and sufficient conditions for a dataset $\dataset$ to be consistent with Nash rationality (Definition~\ref{eq:NashEquilibrium}). The proof is analogous to Afriat's Theorem when the concave potential function of the game is differentiable~\cite{Deb08,Deb09,HK14}. 

\begin{theorem}[Multiagent Afriat's Theorem] Given a dataset $\dataset$ (\ref{eqn:ResponseData}), the following statements are equivalent:
\begin{compactenum}
\item $\dataset$ is consistent with Nash rationality (Definition~\ref{eq:NashEquilibrium}) for an $n$-player concave potential game.
\item Given scalars $v_t$ and $\lambda_t^i>0$ the following set of inequalities have a feasible solution for $t,\tau \in\{1,\dots,T\},$
\begin{equation}
v_{\tau}-v_t-\sum\limits_{i=1}^n\lambda_t^i\probe_t^\p(\response_{\tau}^i-\response_t^i) \leq 0. \quad \label{eqn:NashRationFesTest}
\end{equation}
\item A concave potential function that satisfies (\ref{eqn:NashEquation}) is given by:
\begin{equation}
\hat{V}(x^1,x^2,\dots,x^n) = \underset{t\in T}{\operatorname{min}}\{v_t+\sum_{i=1}^n\lambda_t^ip_t^\p(x^i-x^i_t)\}.   \label{egn:est_potential}
\end{equation}
\item The dataset $\mathcal{D}$ satisfies the Potential Generalized Axiom of Revealed Preference (PGARP) if the following two conditions are satisfied. 
\begin{compactenum}
\item For every dataset $\dataset_\tau^i=\{(\probe_\tindx,\response_\tindx^i): \tindx\in\{1,2,\dots,\tau\}\}$ for all $i\in\{1,\dots,n\}$ and all $\tau\in\{1,\dots,T\},$ $\dataset_\tau^i$ satisfies GARP. 
\item The responses $\response_\tindx^i$ originated from players in a concave potential game. \qed
\end{compactenum}
\end{compactenum}
\label{thrm:NashRationFeasibility}
\end{theorem}
Note that if only a single agent (i.e. $n=1$) is considered, then Theorem~\ref{thrm:NashRationFeasibility} is identical to Afriat's Theorem. Similar to Afriat's Theorem, the constructed concave potential function (\ref{egn:est_potential}) is ordinal--that is, unique up to positive monotone transformations. Therefore several possible options for $\estpotfun(\cdot)$ exist that would produce identical preference relations to the actual potential function $\potfun(\cdot)$. In 4) of Theorem~\ref{thrm:NashRationFeasibility}, the first condition only provides necessary and sufficient conditions for the dataset $\dataset$ to be consistent with a Nash equilibrium of a game, therefore the second condition is required to ensure consistency with the other statements in the Multiagent Afriat's Theorem. The intuition that connects statements 1 and 3 in Theorem~\ref{thrm:NashRationFeasibility} is provided by the following result from~\cite{Ney97}; for any smooth potential game that admits a concave potential function $V$, a sequence of responses $\{\response^i\}_{i\in \{1,2,\dots,n\}}$ are generated by a pure-strategy Nash equilibrium if and only if it is a maximizer of the potential function,
\begin{align}
&x_t=\{\response_t^1,\response_t^2,\dots,\response_t^n\}\in\operatorname*{arg\,max}V(\{\response^i\}_{i\in \{1,2,\dots,n\}}) \nonumber\\
&\text{s.t. } \quad\probe_t^\p\response^i \leq  I_t^i \quad\quad\forall i\in \{1,2,\dots,n\}
\label{eqn:PotMax}
\end{align}
for each probe vector $\probe_t\in\mathds{R}_+^m$.

The non-parametric test for Nash rationality involves determining if (\ref{eqn:NashRationFesTest}) has a feasible solution. Computing parameters $v_t$ and $\lambda^i_t>0$ in (\ref{eqn:NashRationFesTest}) involves solving a linear program with $T^2$ linear constraints in $(n+1)T$ variables, which has polynomial time complexity~\cite{BV04}. In the special case of one agent, the constraint set in (\ref{eqn:NashRationFesTest}) is the dual of the {\it shortest path problem} in network flows. Using the graph theoretic algorithm presented in \cite{BV06}, the solution of the parameters $u_t$ and $\lambda_t$ in (\ref{eqn:AfriatFeasibilityTest}) can be computed with time complexity $O(T^3)$.

\subsection{Learning Algorithm for Response Prediction} 
\label{sec:Non-parametricLearningAlgorithmforResponseForecasting}

In the previous section a non-parametric tests to detect if a dataset $\dataset$ is consistent with Nash rationality was provided. If the $\dataset$ satisfies Nash rationality, then the Multiagent Afriat's Theorem can be used to construct the concave potential function of the game for agents in the social network illustrated in Fig.\ref{fig:SNrevealedpreference}. In this section we provide a non-parametric learning algorithm that can be used to predict the responses of agents using the constructed concave potential function (\ref{egn:est_potential}). 

To predict the response of agent $i$, denoted by $\hat{x}_\tau^i$, for probe $p_\tau$ and budget $I_\tau^i$, the optimization problem (\ref{eqn:PotMax}) is solved using the estimated potential function $\hat{V}$ (\ref{egn:est_potential}), $p_\tau$, and $I_\tau^i$. Computing $\hat{x}_\tau^i$ requires solving an optimization problem with linear constraints and concave piecewise linear objective. This can be solved using the interior point algorithm~\cite{BV04}. The algorithm used to predict the response $\hat{x}_\tau = (\hat{x}_\tau^1,\hat{x}_\tau^2, \cdots,\hat{x}_\tau^n)$ is given below: 
\begin{addmargin}[0mm]{0mm}
\begin{compactenum}
  \item[\bf Step 1:] Select a probe vector $p_\tau\in\mathds{R}^{m}_+$, and response budget $I_\tau^i$ for the estimation of optimal response $\hat{x}_\tau\in\mathds{R}_+^{m\times n}$.
  \item[\bf Step 2:] For dataset $\mathcal{D}$, compute the parameters $v_t$ and $\lambda_t^i$ using (\ref{eqn:NashRationFesTest}). 
\item[\bf Step 3:] The response $\hat{x}_\tau$ is computed by solving the following linear program given $\{\mathcal{D},p_\tau,I_\tau^i\}$, and $\{v_t,\lambda_t^i\}$ from Step~2:
\begin{equation}
\begin{array}{rl}
\max & z \\
\mbox{s.t.} & z \leq  v_t+\sum\limits_{i=1}^n\lambda_t^ip_t^\p(\hat{x}^i_\tau-x^i_t) \text{ for } t=1,\dots,T \quad \\
&p_\tau^\p \hat{x}^i_\tau \leq I^i_\tau \quad \forall i\in \{1,2,\dots,n\}
\label{eqn:forcalgr}
\end{array}
\end{equation}
\end{compactenum}
\end{addmargin}


\subsection{Dataset 1: Online Multiwinner Auction}


\label{subsec:MultiwinnerAuction}

This  section illustrates how Afriat's Theorem from Sec.\ref{subsec:afriatstheorem} can be used to determine if bidders in an online multiwinner auction are 
utility optimizers.
Online auctions are rapidly gaining popularity since bidders do not have to gather at the same geographical location.
Several researchers have focused on the timing of bids and multiple bidding behavior  in Amazon and eBay auctions~\cite{BBK06,OR02,PW13,Del09}. The analysis of the bidding behavior can be exploited by auctioneers to target suitable bidders and thereby increase  profits. 

The multiwinner auction dataset was obtained from an experimental study conducted amongst undergraduate students in Electrical Engineering at Princeton University in March $25^\text{th}$ 2011\footnote{The experimental data is provided by Leberknight {\it et al.}~\cite{LICP12}.}.
The multiwinner auction consists of bidders competing for questions that will aid them for an upcoming midterm exam. 
The social network is composed of $\nindx=12$ bidders where the bidders do not interact with other bidders, they only interact with the external influence, refer to Fig.\ref{fig:SNrevealedpreference}. 
Each bidder is endowed with 500 tokens prior to starting the multiwinner auction. The number of questions being auctioned is not known to the bidders, this prevents the bidders from immediately submitting their entire budget in the final auction. Each auction consists of auctioning a single question at an initial price of 10 tokens and has a duration of 30 min. At the beginning of each auction the bidders are provided with the number of winners, denoted $k$, that auction will have, and the budget of each bidder. The bids are private information with each bidder only informed when their bid has been outbid. The bidders do not communicate with each other during the auction. At the end of each auction, the first $k$ highest bidders are selected, denoted by $\zeta_1, \zeta_2,\dots,\zeta_k$. The bidders $\zeta_1, \zeta_2,\dots,\zeta_k$ are awarded with the question, and pay the second largest bid amount (i.e. bidder $\zeta_k$ pays $\zeta_{k-1}$'s bid amount). In the multiwinner auction it is in the self interest of bidders to force other bidders to spend too much eliminating them from competing in successive auctions.

If the bidding behaviour of agents satisfies Afriat's test (\ref{eqn:AfriatFeasibilityTest}) for utility maximization, the next goal is to classify the behaviour of bidders into two categories: strategic and frantic. If a bidder fails Afriat's test then they are classified as irrational. Strategic bidders will typically submit a large number of bids and a smaller bid amount when compared to frantic bidders. With this bidding behaviour, strategic bidders force the other bidders to spend too much eliminating them from competing in future auctions. Frantic bidders are however only interested in winning the current auction. 

To apply Afriat's test  (\ref{eqn:AfriatFeasibilityTest}), the external influence $\probe_\tindx$, and bidder responses $\response_\tindx$ must be defined. The external influence for each bidder is defined by $\probe_\tindx^i = [\probe_\tindx^i(1), \probe_\tindx^i(2)]$ with $\probe^i_\tindx(1) = $ {\it initial bid amount} representing the bidders interest level for winning, and $\probe^i_\tindx(2) = $ {\it \# of winners} representing the perception of winning where $i$ is the bidder and $t$ the auction. Two datasets are considered for analysis denoted by $\dataset_1$ and $\dataset_2$. An identical external influence is used to construct both $\dataset_1$ and $\dataset_2$. The responses in $\dataset_1$ are given by $\response_\tindx^i=[\response_\tindx^i(1), \response_\tindx^i(2)]$ where $\response_\tindx^i(1)=$ {\it \# of bids} and $\response_\tindx^i(2) = $ {\it mean bid amount}; and for $\dataset_2$ the inputs of $\response_\tindx^i$ are given by $\response_\tindx^i(1) = $ {\it \# of bids} and $\response_\tindx^i(2) = $ {\it mean change in bid amount}. The response $\response_\tindx^i(2)$ in $\dataset_1$ provides the expected bid amount, and $\response_\tindx^i(2)$ in $\mathcal{D}_2$ a measure of the statistical dispersion of the bids. The budget $I_t^i$ of each bidder has units of {\it tokens} multiplied by {\it \# of bids}, and is constrained as the number of tokens and auction duration are finite. The datasets $\dataset_1$ and $\dataset_2$ are constructed from $T=6$ auctions. The non-parametric test (\ref{eqn:AfriatFeasibilityTest}) is applied to each dataset $\dataset_1$ and $\dataset_2$ to detect irrational bidders. For dataset $\dataset_1$ bidder 4 is irrational, and for $\dataset_2$ bidder 11 is irrational. Note that the classification of irrational behaviour is dependent on the choice of response signals used for analysis by the experimentalist.  

\begin{figure}[h]
  \centering
  \subfigure[Estimated utility function $u(x_t)$ using dataset $\bar{\mathcal{D}}_1$ defined in Sec.\ref{subsec:MultiwinnerAuction}.]{\label{fig:estutilauctD1}\includegraphics[angle=0,width=2.1in]{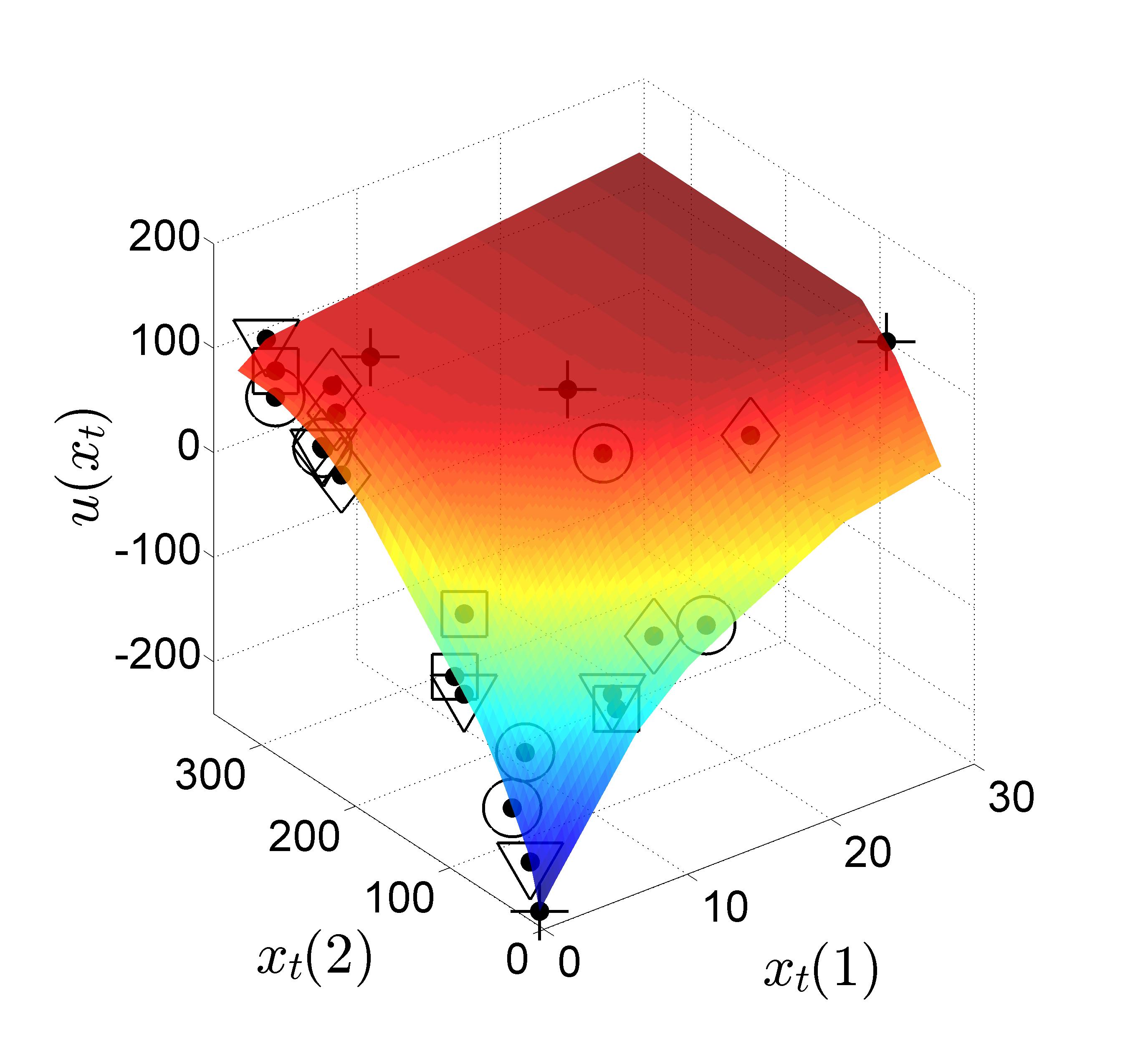} } 
  \hspace{0.00in}
  \subfigure[Estimated utility function $u(x_t)$ using dataset $\bar{\mathcal{D}}_2$ defined in Sec.\ref{subsec:MultiwinnerAuction}.]{\label{fig:estutilauctD2}\includegraphics[angle=0,width=2.1in]{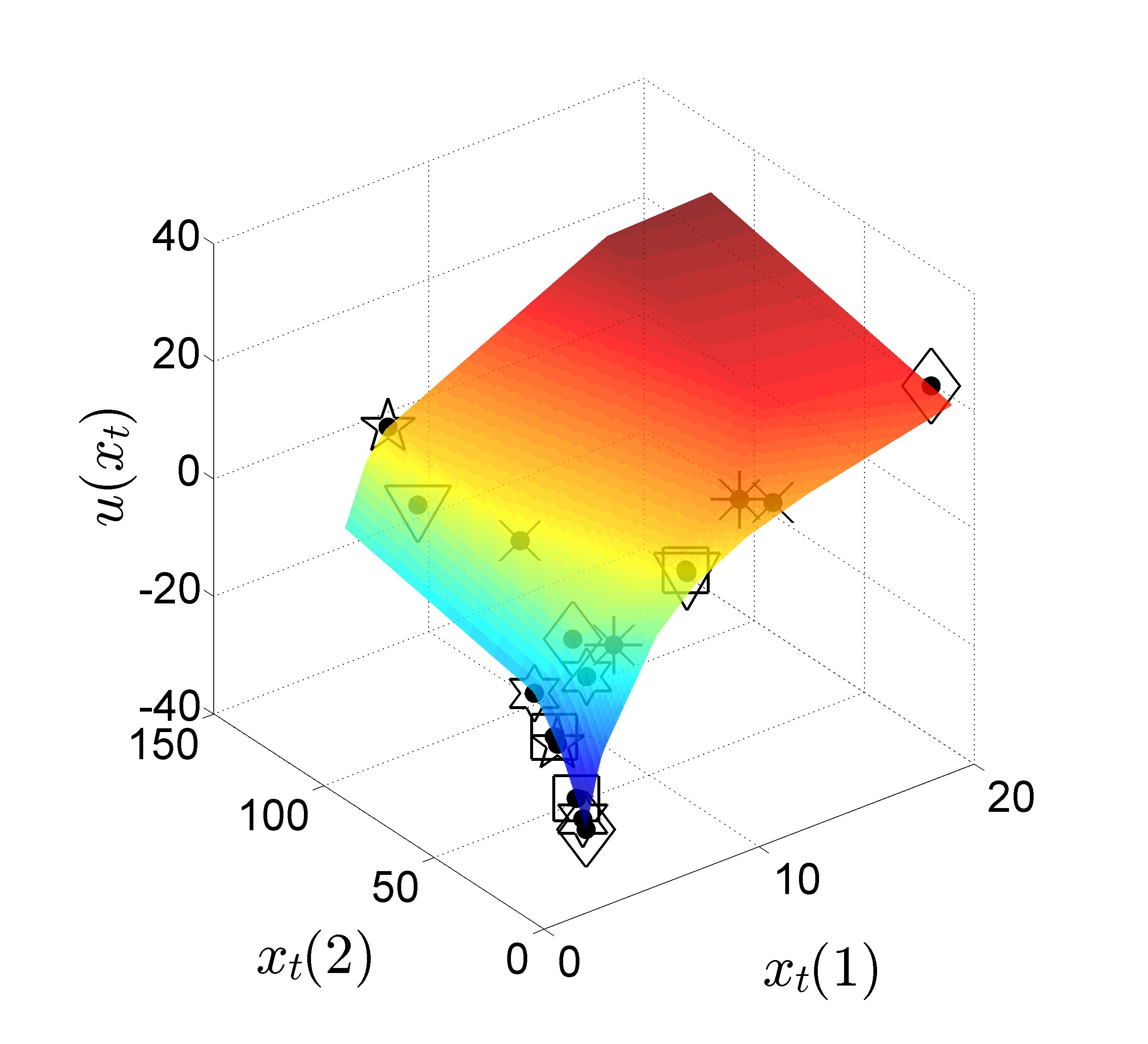} }
\vspace{-5pt}
\caption{Estimated utility function of bidders is constructed using (\ref{eqn:estutility}) using the datasets $\bar{\mathcal{D}}_1$ and $\bar{\mathcal{D}}_2$ defined in Sec.\ref{subsec:MultiwinnerAuction}. The black dots represent the utility associated with experimentally measured responses, and the colour (blue to red) indicates the utility level. The black dots indicate the observed demands, and the shape (i.e. circle, diamond etc.) denotes the respective bidder.}
\label{fig:estutilauct}
\vspace{-5pt}
\end{figure}

For utility maximization bidders, an estimate of the utility function of each bidder is required to classify them as strategic or frantic. To estimate the utility function of the bidders, a subset of data from $\mathcal{D}_1$, denoted as $\bar{\mathcal{D}}_1$, is selected such that the preferences of all agents $i$ in $\bar{\mathcal{D}}_1$ are identical. It was determined that $\bar{\mathcal{D}}_1 =  \{(p_t^i,x_t^i): i\in\{1,3,5,7,12\}\}$. Since the preferences of these bidders are identical, we can consider all the data in $\bar{\mathcal{D}}_1$ as originating from a single representative bidder. This allows an improved estimate of the utility function of these bidders as compared to learning the utility function for each bidder separately. An analogous explanation is used for the construction of $\bar{\mathcal{D}}_2 =  \{(p_t^i,x_t^i): i\in\{1,2,3,4,7,8,9\}\}$ from the dataset $\mathcal{D}_2$. The estimated utility function for $\bar{\mathcal{D}}_1$ is given in Fig.\ref{fig:estutilauctD1} and for $\bar{\mathcal{D}}_2$ in Fig.\ref{fig:estutilauctD2}. As seen from Fig.\ref{fig:estutilauctD1} and Fig.\ref{fig:estutilauctD2}, bidders have a preference to increase the number of bids compared with increasing the mean bid amount or the difference in mean bid amount. This follows logically as $x_t^i(2)$ increases, the bidder will have to pay more tokens to win the question limiting their ability to bid in future auctions. Interestingly, the bidders show strategic and frantic behaviour in both datasets $\mathcal{\bar{D}}_1$ and $\mathcal{\bar{D}}_2$, as seen in Fig.\ref{fig:estutilauctD1} and Fig.\ref{fig:estutilauctD2}. This is consistent with the results in \cite{PW13} which show that bidders change their bidding behavior between successive auctions. 

The analysis shows that auctioneers should target bidders that show frantic bidding behavior as they are likely to overspend on items increasing the revenue of the auctioneer. Such behavior can be detected using utility maximization test and constructed utility function from Afriat's Theorem. 

\addtolength{\subfigcapskip}{-8pt}

\subsection{Dataset 2: Ontario Electrical Energy Market Dataset}

In this section we consider the aggregate power consumption of different zones in the Ontario power grid.  A sampling period of $\Tindxter=79$ days starting from January 2013 is used to generate the dataset $\dataset$ for the analysis. All price and power consumption data is available from the {\it Independent Electricity System Operator}\footnote{http://ieso-public.sharepoint.com/} (IESO) website. Each zone is considered as an agent in the corporate network illustrated in Fig.\ref{fig:SNrevealedpreference}. The study of corporate  social networks was pioneered by Granovetter~\cite{Granovetter85,Granovetter05} which shows that the social structure of the network can have important economic outcomes. Examples include agents choice of alliance partners, assumption of rational behavior, self interest behavior, and the learning of other agents behavior. Here we test for rational behavior (i.e. utility maximization and Nash rationality), and if true then learn the associated behavior of the zones. This analysis provides useful information for constructing demand side management (DSM) strategies for controlling power consumption in the electricity market. For example, if a utility function exists it can be used in the DSM strategy presented in~\cite{DSZ11,NSH12}. 

The zones power consumption is regulated by the associated price of electricity set by the senior management officer in each respective zone. Since there is a finite amount of power in the grid, each officer must communicate with other officers in the network to set the price of electricity. Here we utilize the aggregate power consumption from each of the $\nindx=10$ zones in the Ontario power grid and apply the non-parametric tests for utility maximization (\ref{eqn:AfriatFeasibilityTest}) and Nash rationality (\ref{eqn:NashRationFesTest}) to detect if the zones are demand responsive. If the utility maximization or Nash rationality tests are satisfied, then the power consumption behaviour is modelled by constructing the associated utility function (\ref{eqn:estutility}) or concave potential function of the game (\ref{egn:est_potential}).

To perform the analysis the external influence $\probe_\tindx$ and response of agents $\response_\tindx$ must be defined. In the Ontario power grid the wholesale price of electricity is dependent on several factors such as consumer behaviour, weather, and economic conditions. Therefore the external influence is defined as $\probe_\tindx=[\probe_\tindx(1),\probe_\tindx(2)]$  with $\probe_\tindx(1)$ the average electricity price between midnight and noon, and $\probe_\tindx(2)$ as the average between noon and midnight with $\tindx$ denoting day. The response of each zone correspond to the total aggregate power consumption in each respective tie associated with $\probe_\tindx(1)$ and $\probe_\tindx(2)$ and is given by $\response_\tindx^i=[\response_\tindx^i(1), \response_\tindx^i(2)]$ with $i\in\{1,2,\dots,\nindx\}$. The budget $I_t^i$ of each zone has units of dollars as $\probe_\tindx$ has units of \$/kWh and $\response_\tindx^i$ units of kWh.

We found that 
the aggregate consumption data of each zone does not satisfy Afriat's utility maximization test (\ref{eqn:AfriatFeasibilityTest}). This points to the possibility that  the zones are engaged in a concave potential game--this would not be a surprising result as network congestion games have been shown to reduce peak power demand in distributed demand management schemes~\cite{ING10}. To test if the dataset $\dataset$ is consistent with Nash rationality the detection test (\ref{eqn:NashRationFesTest}) is applied. The dataset for the power consumption in the Ontario power gird is consistent with Nash rationality. Using (\ref{eqn:forcalgr}), a concave potential function for the game is constructed. Using the constructed potential function, when do agents prefer to consume power? The {\it marginal rate of substitution}\footnote{The amount of one good that an agent is willing to give up in exchange for another good while maintaining the same level of utility.} (MRS) can be used to determine the preferred time for power usage. Formally, the MRS of $\response^i(1)$ for $\response^i(2)$ is given by
\begin{equation*}
 \operatorname{MRS}_{12}=\frac{\partial\hat{V}/\partial \response^i(1)}{\partial\hat{V}/\partial \response^i(2)}.
 \end{equation*} 
From the constructed potential function we find that $\operatorname{MRS}_{12} > 1$ suggesting that the agents prefer to use power in the time period associated with $\response_\tindx(1)$--that is, the agents are willing to give up $\operatorname{MRS}_{12}$ kWh of power in the time period associated with $\response^i(2)$ for 1 additional kWh of power in time period associated with $\response^i(1)$.  

The analysis in this section suggests that the power consumption behavior of agents is consistent with players engaged in a concave potential game. Using the Multiagent Afriat's Theorem the agents preference for using power was estimated. This information can be used to improve the DSM strategies presented in~\cite{DSZ11,NSH12} to control power consumption in the electricity market. 


\subsection{Dataset 3: Twitter Data}
\label{subsec:twitterdata}
Does the tweeting behaviour of Twitter agents satisfy a utility maximization process? The goal is to investigate how tweets and trend indices\footnote{Here we define the {\it trend index} as the frequency of tweets containing a particular word~\cite{GOT13}.} impact the tweets of agents in the Twitter social network. The information provided by this analysis can be used in social media marketing strategies to improve a brand and for brand awareness. As discussed in \cite{GOT13}, Twitter may relay on a huge amount of agent-generated data which can be analyzed to provide novel personal advertising to agents. 

\begin{figure}[h!]
\centering
\includegraphics[angle=0,width=3in]{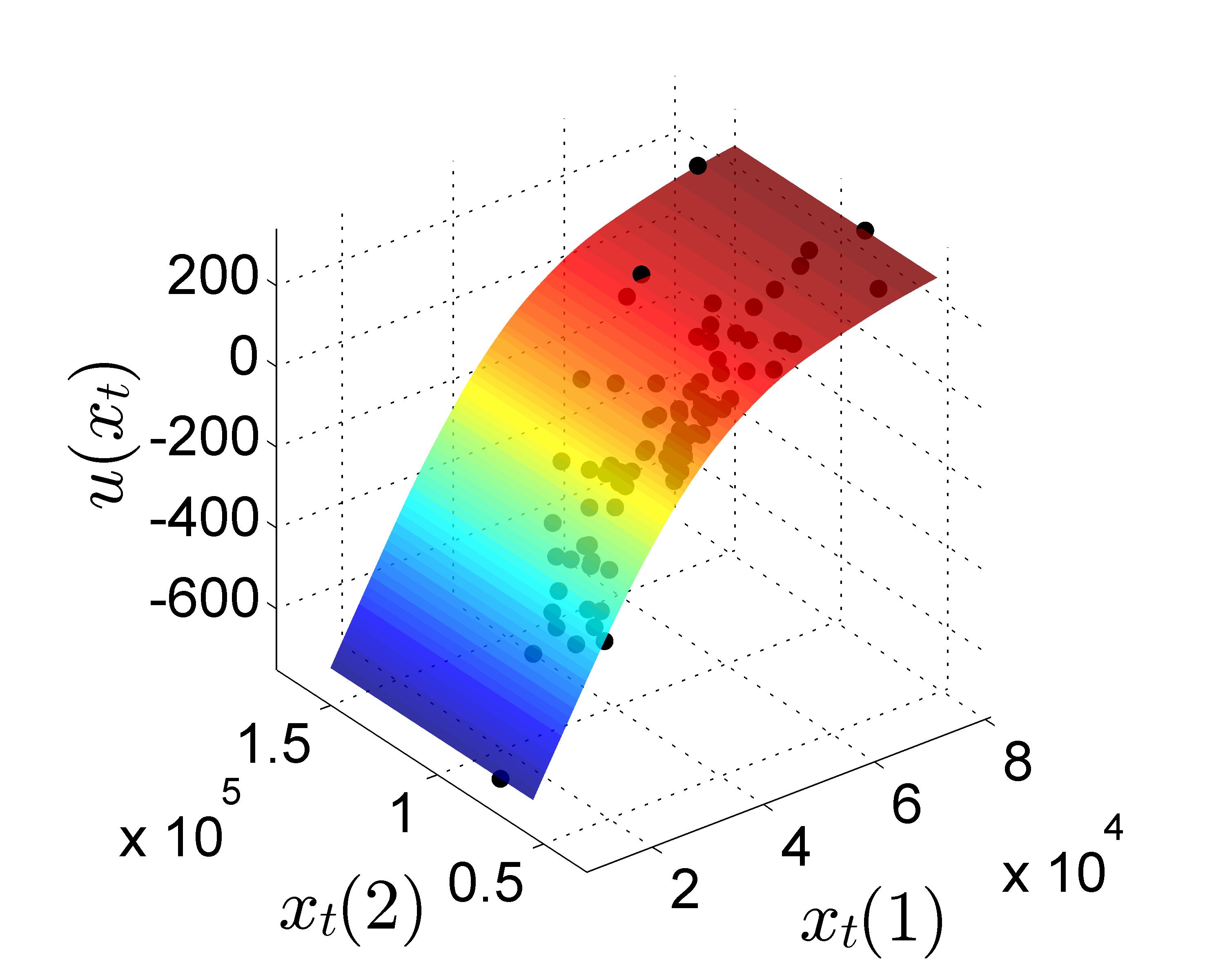} 
\caption{Estimated utility function $\utility(\response_\tindx)$ using dataset $\dataset$ defined in Sec.\ref{subsec:twitterdata} and constructed using the non-parametric learning algorithm (\ref{eqn:estutility}) from Afriat's Theorem.}
\label{fig:twitterutility}
\end{figure}

To apply Afriat's utility maximization test (\ref{eqn:AfriatFeasibilityTest}), we choose the external influence and response as follows. External influence  $\probe_\tindx=[\text{\it \#Sony}, 1/\text{\it \#Playstation}]$ for each day $\tindx$. The associated response taken by the agents in the network is given by $\response_\tindx=[\text{\it \#Microsoft}, \text{\it \#Xbox}]$. Notice that the probe $\probe_\tindx(2)$ can be interpreted as the frequency of tweets with the word {\it Playstation} (i.e. the trending index). The dataset $\dataset$ of external influence and responses is constructed from $T=80$ days of Twitter data starting from January 1$^\text{st}$ 2013. The dataset $\dataset$ satisfies the utility maximization test (\ref{eqn:AfriatFeasibilityTest}). This establishes that utility function exists for agents that is dependent on the number of tweets containing the words {\it Microsoft} and {\it Xbox}. The data shows that tweets containing the word {\it Microsoft} and {\it Xbox} are dependent on the number of tweets containing {\it Sony} and trending index of {\it Playstation}. This dependency is expected as Microsoft produces the game console Xbox, and Sony produces the game console Playstation both which have a large number of brand followers (e.g. Xbox has over 3 million, and Playstation over 4 million). To gain further insight into the behaviour of the agents, (\ref{eqn:estutility}) from Afriat's Theorem is used to construct a utility function for the agents. Fig.\ref{fig:twitterutility} shows the constructed utility function of the agents. As seen, agents have a higher utility for using the word {\it Microsoft} as compared to {\it Xbox}--that is, agents prefer to use the word {\it Microsoft} to that of {\it Xbox}. Interestingly, if we define the response to be $\response_\tindx=[\text{\it \#Microsoft}, 1/\text{\it \#Xbox}]$, then the dataset satisfies utility maximization. From the constructed utility function, not shown, the agents prefer to increase the tweets containing the word {\it Microsoft} compared to increasing the trend index of {\it Xbox}. If instead $\response_\tindx=[1/\text{\it \#Microsoft}, 1/\text{\it \#Xbox}]$, then the dataset satisfies utility maximization and agents prefer to increase the trend index of {\it Microsoft} compared to that of {\it Xbox}. 

To summarize, the above  analysis suggests the following interesting fact: {\it Xbox} has a lower utility than {\it Microsoft} in terms of Twitter sentiment.
 Therefore, online marketing strategies should target the brandname {\it Microsoft} instead of {\it Xbox}.


\subsection{Summary and Extensions}

The principle of revealed preferences is an active research  area with numerous recent papers. 
We have already mentioned the papers \cite{Die73,Var82,Blu05}. Below we summarize some related literature that extends
the basic framework of Afriat's theorem.

Afriat's theorem holds for finite datasets and gives an explicit construction of a class of concave utility functions that  rationalize the dataset.
Mas Colell \cite{Col78} has given sufficient conditions under which as the  data set size $\Tindxter$  grows to infinity,
the underlying utility function of the consumer can be fully identified.

Though the classical Afriat's theorem holds for linear budget constraints $\probe_\tindx^\p \response \leq \budget_\tindx$ in (\ref{eqn:singlemaximization}), an identical formulation holds for certain non-linear budget constraints as illustrated in~\cite{FM09}. The budget constraints considered in
\cite{FM09} are of the form $\{ x \in \reals^m_+ \vert  g(x) \leq 0\} $ where $g: \reals^m_+ \rightarrow \reals$ is an increasing continuous function and
$\reals^m_+$ denotes the positive orthant. Also \cite{FM09} shows how the results in \cite{Col78} on recoverability of the utility
function can be extended to such nonlinear budget
constraints. However, learning the utility function from a finite dataset in the case of a non-linear budget constraint requires sophisticated machine learning algorithms~\cite{Lahaie10}. The machine learning algorithms can only guarantee that the estimated utility function is approximately consistent with the dataset $\dataset$--that is, the estimated utility is not guaranteed to contain all the preference relations consistent with the dataset $\dataset$.

In \cite{BV06},  results in statistical learning theory are applied to  the principle of revealed preferences to address the question:
Is the class of demand functions derived from monotone concave utilities efficiently probably approximately correct (PAC) learnable?
It is shown that  Lipschitz utility functions are efficiently PAC learnable. In \cite{HK14}, the authors extend the results of \cite{BV06} and show that for agents engaged in a concave potential game that satisfy Nash rationality, if the underlying potential function satisfies the Lipschitz condition then the potential function of the game is PAC learnable. 

In many cases, the responses of agents are observed in noise. Then determining if an agent is a utility maximizer (or a multiagent
system's response is consistent with play from a Nash) becomes a statistical decision test. In \cite{KH12} it is shown how stochastic optimization
algorithms can be devised to optimize the probe signals to minimize the type II errors of the decision test subject to a fixed type I error.

\subsubsection*{Change Point Detection in Utility Functions}
 \cite{AK17} 
extends the classical revealed preference framework  to agents with a ``dynamic utility function''. The utility function jump changes at an unknown time instant by a linear perturbation. Given the dataset of probe and responses of an agent, the objective in \cite{AK17} is to develop a nonparametric test to detect the change point and the utility functions before and after the change, which is henceforth referred to as the change point detection problem. 

Such change point detection problems arise in online search in social media. 
Online search is currently the most popular method for information retrieval~\cite{SEW04} and can be viewed as an agent maximizing the information utility, i.e.\ the amount of information consumed by an online agent given the limited resource on time and attention. 
There has been a gamut of research which links internet search behaviour to ground truths such as symptoms of illness, political election, or major sporting events~\cite{ZYF11}. 
Detection of utility change in online search, therefore, is helpful to identify changes in ground truth. 
Also, the intrinsic nature of the online search utility function motivates such a study under a revealed preference setting. 

The problem of detecting a sudden linear perturbation change in the utility function is motivated by several reasons. 
First, the linear perturbation assumption provides sufficient selectivity such that the non-parametric test is not trivially satisfied by all datasets but still provides enough degrees of freedom. 
Second, the linear perturbation can be interpreted as the change in the marginal rate of utility relative to a ``base'' utility function. 
In online social media, the linear perturbation coefficients measure the impact of marketing or the measure of severity of the change in ground truth on the utility of the agent. 
This is similar to the linear perturbation models used to model taste changes~\cite{ABBC15,MF12,FIS15} in microeconomics. 
Finally, in social networks, linear change in the utility is often used to model the change in utility of an agent based on the interaction with the agent's neighbours~\cite{CS10}. 
Compared to the taste change model, our model is unique in that we allow the linear perturbation to be introduced at an unknown time.

A related important practical issue that we also consider in this paper is the application of revealed preference framework to high dimensional data (``big-data''). 
As an example of high dimensional data arising in online social media, we investigate the detection of the utility maximization process inherent in video sharing via YouTube. 
Detecting utility maximization behaviour with such high dimensional data is computationally demanding. \cite{AK17} uses  dimensionality reduction to overcome the computational cost associated with high dimensional data.  The high dimensional data is projected into a lower dimensional subspace using the Johnson-Lindenstrauss (JL) transform.

\section{Social Interaction of Channel Owners and YouTube Consumers}
\label{sec:SocialInteractionofChannelOwnersandYouTubeConsumers}
In this section time-series analysis methods are applied to real-world YouTube data to determine how social sensors interact with YouTube channel owners. Several key results are presented that elucidate the dynamics of social sensors in the YouTube social network. This section contains five main results. 
\begin{compactenum}
\item Sec.\ref{subsec:SocialSensorEngagementSensitivitytoMeta-LevelOptimization} illustrates the sensitivity of social sensor engagement to changes in meta-level (title, thumbnail, tags) features of YouTube videos. It is found that meta-level feature optimization causes an increase in user engagement for approximately 50\% of videos. Optimization of the title of the video causes a significant improvement of users finding the video from YouTube search results. Additionally, optimization of the thumbnail causes an increase in users accessing the video from the related video list\footnote{In YouTube, the suggested videos refers to the overall list to the right of the video player on the watch page which is populated with suggestions for what to watch next. A subset of these suggested videos, known as related videos, can also be displayed at the end of a YouTube video.}. 
\item In Sec.\ref{subsec:causalrelationYouTube} Granger causality is used to show that a causal relationship exists between a channels subscriber count and the social sensor engagement of videos on the channel. However, this causal relationship is dependent on the category. For example, 80\% of the ``Entertainment'' channels satisfy the Granger causality test while only 40\% of the ``Food'' channels satisfy the test. 
\item In Sec.\ref{subsec:schedulingYouTube} it is determined that for popular gaming YouTube channels with a dominant (constant) upload schedule, deviating from the schedule increases the views and the comment counts of the channel (e.g. increases user engagement). Specifically, when the channel goes off schedule the channel gains views $97\%$ of the time and the channel gains comments $68\%$ of the time. 
\item In Sec.\ref{subsec:SocialSensorEngagementDynamicswithYouTubeVideos} we illustrate that the social sensor engagement dynamics with YouTube videos can be modeled using a generalized Gompertz model. The generalized Gompertz model accounts for the initial viral increase in views from subscribers, the subsequent linear growth that results from non-subscribers, and views from exogenous events such as promotion on other popular social media platforms. It is important to account for exogenous events when estimating the efficiency of meta-level optimization procedures.
\item In Sec.\ref{subsec:SocialSensorEngagementforChannelPlaythroughs} the generalized Gombertz model is used to study the dynamics of social sensors to video playthroughs (squence of videos on the same topic). It is illustrated that the early view count dynamics are highly correlated with the view count dynamics of future videos. Both the short term view count and long term migration of users to future videos in the playthrough decrease after the initial video in the playthrough is posted. This results even when the channels subscriber count increases. A possible reason for this decrease is that subsequent videos in the playthrough become repetitive and hence decrease user engagement. 
\end{compactenum}

The results in this section are based on the extensive BroadBandTV\footnote{BroadBandTV is one of the largest YouTube video partners in the world.
 \url{http://bbtv.com/press/broadbandtv-now-the-largest-multi-platform-network-worldwide}}  (BBTV) dataset.
Extrapolating these results to other YouTube datasets is an important problem worth addressing by the reader. For example, an extension of this work could involve
studying the effect of video characteristics on different traffic sources, for example the affect of tweets or posts of videos on Twitter or Facebook.

\subsection{YouTube Dataset}
\label{subsec:YouTubeDataset}

All the results in this section are constructed using the extensive YouTube dataset provided by BBTV\footnote{http://bbtv.com/}. The dataset contains daily samples of meta-level features of YouTube videos and channels on the BBTV platform from April, 2007 to May, 2015, and has a size of several terabytes. The meta-level features include: views, comments, likes, dislikes, shares, and subscribers which are recorded each day since the video was published. The dataset contains information for over $6$ million videos spread over $25$ thousand channels. Table~\ref{tab:dataset:summary} shows the statistics summary of the videos present in the dataset. 
\begin{table}[h!]
	\centering
	\caption{Dataset summary}
	\begin{tabular}{c|c}
		\toprule
		Videos &  $6$ million\\
		Channels &  $26$ thousand\\
		Average number of videos (per channel)& 250 \\
		Average age of videos & 275 days\\
		Average number of views  (per video) & $10$ thousand \\
		\bottomrule
	\end{tabular}
	\label{tab:dataset:summary}
\end{table}

\begin{figure}[h]
	\centering
	\includegraphics[scale=0.6]{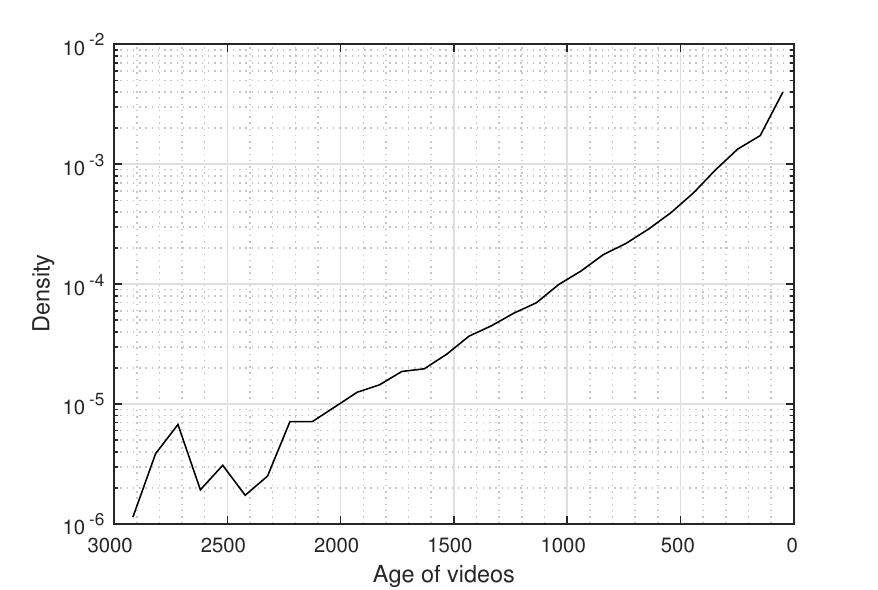}
	\caption{The fraction of videos in the dataset as a function of the age of the videos. There is a significant percentage of newer videos (videos with less age) compared to older videos. Hence, the dataset capture the exponential growth of the number of videos uploaded to YouTube. }
	\label{fig:hist:age}
\end{figure}

Table~\ref{tab:category:dataset:summary}, shows the summary of the various category of the videos present in the dataset. The dataset contains a large percentage of gaming videos.    
\begin{table}[!h]
	\centering
	\caption{YouTube dataset categories (out of $6$ million videos) }
	\begin{tabular}{c|c}
		\toprule
		Category & Fraction \\
		\midrule
		Gaming & 0.69 \\
		Entertainment & 0.07 \\
		Food & 0.07 \\
		Music & 0.035 \\
		Sports & 0.017\\
		\bottomrule
	\end{tabular}
	\label{tab:category:dataset:summary}
\end{table}
Fig.~\ref{fig:hist:age} shows the fraction of videos as a function of the age of the videos. There is a large fraction of videos uploaded within a year. 
Also, the dataset captures the exponential growth in the number of videos uploaded to YouTube. 
Similar to~\cite{RAEJLP14}, we define three categories of videos based on their popularity: Highly popular, popular, and unpopular. 
Table~\ref{tab:popularity:dataset:summary} gives a summary of the fraction of videos in the dataset belonging to each category. 
As can be seen from Table~\ref{tab:popularity:dataset:summary}, the majority of the videos in the dataset belong to the popular category.  
\begin{table}[!h]
	\centering
	\caption{Popularity distribution of videos in the dataset}
	\begin{tabular}{c|c}
		\toprule
		Criteria & Fraction \\
		\midrule
		Highly Popular (Total Views $> 10^4$)  &  0.12\\
		Popular ($150 < $  Total Views $< 10^4$)&  0.67\\
		Unpopular (Total Views $< 150$) &  0.21\\
		\bottomrule
	\end{tabular}
	\label{tab:popularity:dataset:summary}
\end{table}

A unique feature of the dataset is that it contains information about the ``\metalevel optimization'' for videos.  The \metalevel optimization is a change in the title, tags or thumbnail, of an existing video in order to increase the popularity.  BBTV markets a product that intelligently automates the \metalevel optimization. Table~\ref{tab:optimization:dataset:summary} gives a summary of the statistics of the various \metalevel optimization present in the dataset. 
\begin{table}[!h]
	\centering
	\caption{Optimization summary statistics}
	\begin{tabular}{c|c}
		\toprule
		Optimization & \# Videos \\
		\midrule
		Title change &  $21$ thousand \\
		Thumbnail change &  $13$ thousand \\
		Keyword change &  $21$ thousand\\
		\bottomrule
	\end{tabular}
	\label{tab:optimization:dataset:summary}
\end{table}

\subsection{Social Sensor Engagement Sensitivity to Meta-Level Optimization}
\label{subsec:SocialSensorEngagementSensitivitytoMeta-LevelOptimization}
Here we analyze how changing \metalevel features, after a video is posted, impacts the user engagement of the video. Meta-level data plays a significant role in the discovery of content, through YouTube search, and in video recommendation, through the YouTube related videos\footnote{Related and suggested videos appear surrounding the current video being viewed by the user.}. Hence, ``optimizing'' the \metalevel data to enhance the discoverability and user engagement of videos is of significant importance to content providers. Therefore, in this section, we study how optimizing the title, thumbnail or keywords affect the view count of YouTube videos. 

To perform the sensitivity analysis we utilize the dataset presented in Sec.\ref{subsec:YouTubeDataset}, and remove any time-sensitive videos. Time-sensitive videos are those videos that are relevant for a short period of time and the popularity of such videos cannot be improved by optimization. We removed the following two time-sensitive categories of videos:  ``politics'' and ``movies and trailers''. In addition, we removed videos (from other categories) which contained the following keywords in their video meta-data: ``holiday'', ``movie'', or ``trailers''. For example, holiday videos are not watched frequently during off-holiday times. 

Let $\hat{\tau}_i$ be the time at which the \metalevel optimization was performed on video $i$ and let $s_i$, denote the corresponding sensitivity. 
We characterize the sensitivity to \metalevel optimization as follows: 
\begin{equation}
	s_i = \frac{
		\left(
		\sum_{t =\hat{\tau}_i}^{\hat{\tau}_i+6}
			v_i(t)
			\right)/7
		}
		{
		\left(
		\sum_{t =
			\hat{\tau}_i-6}^{\hat{\tau}_i}
			v_i(t) 
			\right)/7}
	\label{eqn:sensitivity:optimization:characterize}
\end{equation}
The numerator of~\eqref{eqn:sensitivity:optimization:characterize} is the mean value of the view count $7$ days after optimization. Similarly, the denominator of~\eqref{eqn:sensitivity:optimization:characterize} is the mean value of the view count $7$ days before optimization. The results are provided in Table~\ref{tab:sensitivity:optimization} for optimization of the title, thumbnail, and keywords. 
\begin{table}[!h]
	\centering
	\begin{tabular}{c|c}
		\toprule
		Optimization & Fraction of Videos with increased popularity \\
		\midrule
		Title change &  0.52 \\
		Thumbnail change &  0.533 \\
		Keyword change &  0.50 \\
		No change\footnotemark &  0.35 \\
		\bottomrule
	\end{tabular}
	\caption{Sensitivity to Meta-Level Optimization. The table shows than in more than $50\%$ the videos, \metalevel optimization resulted in an increase in the popularity of the video. \label{tab:sensitivity:optimization}}
\end{table} \footnotetext{``No change'' was obtained by randomly selecting $10^4$ videos which performed no optimization and evaluating $s_i$ $3$ months from the date of posting the video. }

As shown in Table~\ref{tab:sensitivity:optimization}, at least half of the optimizations resulted in an increase in the popularity of the video. In addition, compared to videos with no optimization, the \metalevel optimization improves the probability of increased popularity by $45\%$. This is consistent with YouTube and BBTV recommendation to optimize \metalevel features to increase user engagement.  
However, some class of videos benefit from optimizing meta-data much more than others. The effect may be due to small user channels, which have limited number of videos and subscribers, gain by optimizing the meta-level data of the video compared to hugely popular channels such as Sony or CNN.  
The highly popular channel (e.g.\ Sony or CNN) upload videos frequently (even multiple times daily), so video content becomes irrelevant quickly. 
The question of which class of users gain by optimizing the meta level features of the video is part of our ongoing research.

Table~\ref{tab:sensitivity:traffic} summarizes the impact of various meta-level changes on the three major sources of YouTube traffic, i.e. YouTube search\footnote{Video views that resulted users selecting the video from the YouTube search results.}, YouTube promoted\footnote{Video views that result from channels paying YouTube to increase their probability of being included at the top of search result lists.} and traffic from related videos\footnote{Video views that resulted from users clicking on a thumbnail that was listed on the page of another video they were viewing.}. For those videos where \metalevel optimization increased the popularity (the ratio of the mean value of the views after and before optimization is higher than one), we computed the sensitivity for various traffic sources as in~\eqref{eqn:sensitivity:optimization:characterize}. Table~\ref{tab:sensitivity:traffic} summarizes the median statistics of the ratio of the traffic sources before and after optimization. 
\begin{table}[!h]
\begin{minipage}{\columnwidth}
		\begin{tabular*}{\textwidth}{@{\hspace{\tabcolsep}\extracolsep{\fill}}cccc}
		\toprule
		Optimization & Related &  Promoted & Search\\
		\midrule
		Title change &   $1.13$ & NA\footnote{Not enough data available: A binomial test to check for the true hypothesis with $95\%$ confidence interval requires that the sample size, $n$, should be at least $\left(\frac{1.96}{0.04}\right)^2 p (1-p)$. With $p = 0.5$, $n > 600$. \label{foot:data:na}} & $\mathbf{1.24}$ \\
		Thumbnail change &  $\mathbf{1.20}$ & NA\footref{foot:data:na} & $1.125$\\
		Keyword change & $\mathbf{1.10}$ & $\mathbf{1.16}$ & $1$  \\
		\bottomrule
		\end{tabular*}
\end{minipage}%
	\caption{Sensitivity of various traffic sources to \metalevel optimization, for videos with increased popularity. The title optimization resulted in significant improvement (approximately $25\%$) from the YouTube search. Similarly, thumbnail optimization improved traffic from the related videos and keyword optimization resulted in increased traffic from related and promoted videos. }
	\label{tab:sensitivity:traffic}
\end{table}
The title optimization resulted in significant improvement (approximately $25\%$) from the YouTube search. Similarly, thumbnail optimization improved traffic from the related videos and keyword optimization resulted in increased traffic from related and promoted videos.

{\em Summary}: This section studied the sensitivity of \viewcount with respect to \metalevel optimization. 
The main finding is that \metalevel optimization increased the popularity of video in the majority of cases. 
In addition, we found that optimizing the title improved traffic from YouTube search. Similarly, thumbnail optimization improved traffic from the related videos and keyword optimization resulted in increased traffic from related and promoted videos.

\subsection{Causal Relationship Between Channel Subscribers and Social Sensor Engagement}
\label{subsec:causalrelationYouTube}
\begin{figure}[h]
	\centering
	\input{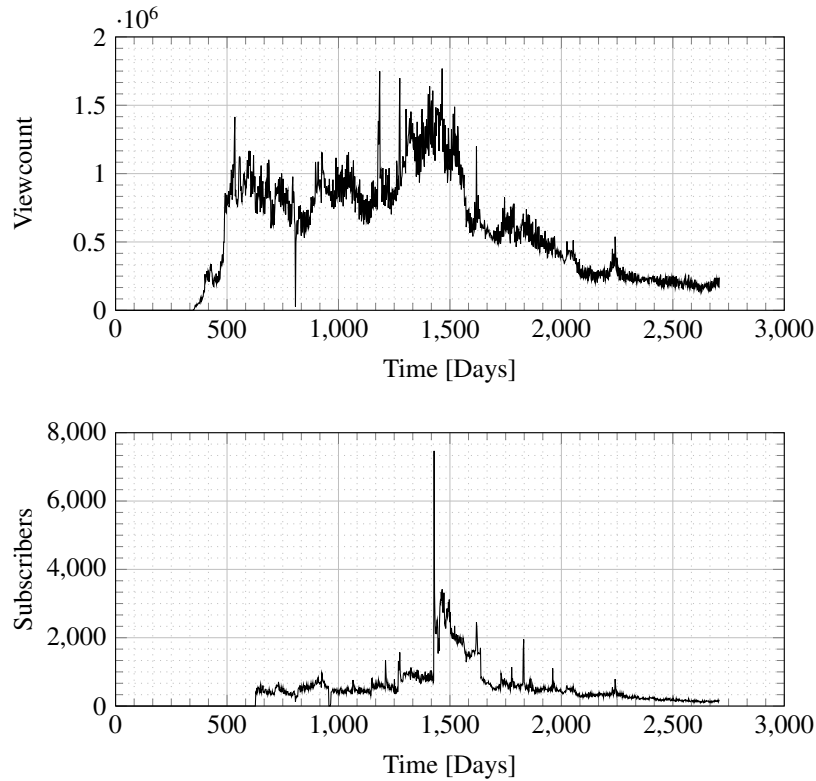}
	\caption{Viewcount and subscribers for the popular movie trailer channel: VISOTrailers. The Granger causality test for view counts ``Granger causes'' subscriber count is true with a p-value of $5\times 10^{-8}$. }
	\label{fig:Channel6:sub:view}
\end{figure}

In this section the goal is to detect if a causal relationship exists between subscriber and viewer counts and how it can be used to estimate the next day subscriber count of a channel. The results are of interest for measuring the popularity of a YouTube channel. Fig.~\ref{fig:Channel6:sub:view} displays the subscriber and view count dynamics of a popular movie trailer channel in YouTube.  It is clear from Fig.~\ref{fig:Channel6:sub:view} that the subscribers ``spike'' with a corresponding ``spike'' in the view count. In this section we model this causal relationship of the subscribers and \viewcount using the Granger causality test from the econometric literature~\cite{granger1969investigating}. 

The main idea of Granger causality is that if the value(s) of a lagged time-series can be used to predict another time-series, then the lagged time-series is said to ``Granger cause'' the predicted time-series. 
%
To formalize the Granger causality model, let $s^j(t)$ denote the number of subscribers to a channel $j$ on day $t$, and $v_i^j(t)$ the corresponding view count for a video $i$ on channel $j$ on day $t$. The total number of videos in a channel on day $t$ is denoted by $\mathcal{I}(t)$. Define, 
\begin{equation}
	\hat{v}^j(t) = \sum_{i=1}^{\mathcal{I}(t)}v_i^j(t),
\end{equation}
as the total view count of channel $j$ at time $t$. 
The Granger causality test involves testing if the coefficients $b_i$ are non-zero in the following equation which models the relationship between subscribers and view counts: 
\begin{equation}
	s^j(t) = \sum_{k = 1}^{n_s}a_k^j s^j(t-k) + \sum_{i=k}^{n_v}b_k^j \hat{v}^j(t-k) + \varepsilon^j(t),
	\label{eqn:arma:model}
\end{equation}
where $\varepsilon^j(t)$ represents normal white noise for channel $j$ at time $t$. The parameters $\{a_i^j\}_{\{i=1,\dots,n_s\}}$ and $\{b_i^j\}_{\{i=1,\dots,n_v\}}$ are the coefficients of the AR model in~\eqref{eqn:arma:model} for channel $j$, with $n_s$ and $n_v$ denoting the lags for the subscriber and view counts time series respectively. 
If the time-series $\mathcal{D}^j=\{s^j(t),\hat{v}^j(t)\}_{t\in\{1,\dots,T\}}$ of a channel $j$ fits the model~\eqref{eqn:arma:model}, then we can test for a causal relationship between subscribers and view count. 
In equation~\eqref{eqn:arma:model}, it is assumed that $|a_i| < 1$, $|b_i| < 1$ for stationarity. The causal relationship can be formulated as a hypothesis testing problem as follows:  
\begin{equation}
	H_0: b_1 = \dots = b_{n_v}=0 \text{ vs. } H_1: \text{Atleast one } b_i \neq 0. 
	\label{eqn:hypothesis}
\end{equation}
The rejection of the null hypothesis, $H_0$, implies that there is a causal relationship between subscriber and view counts.

First, we use Box-Ljung test~\cite{LB78} is to evaluate the quality of the model~\eqref{eqn:arma:model} for the given dataset $\mathcal{D}^j$. If satisfied, then the Granger causality hypothesis~\eqref{eqn:hypothesis} is evaluated using the Wald test~\cite{Wald73}. If both hypothesis tests pass then we can conclude that the time series $\mathcal{D}^j$ satisfies Granger causality--that is, the previous day subscriber and view count have a causal relationship with the current subscriber count. 

A key question prior to performing the Granger causality test is what percentage of videos in the YouTube dataset (Appendix) satisfy the AR model in~\eqref{eqn:arma:model}. To perform this analysis we apply the Box-Ljung test with a confidence of 0.95 (p-value = $0.05$). First, we need to select $n_s$ and $n_v$, the number of lags for the subscribers and view count time series. For $n_s=n_v = 1$, we found that only $20\%$ of the channels satisfy the model~\eqref{eqn:arma:model}. When $n_s$ and $n_v$ are increased to $2$, the number of channels satisfying the model increases to $63\%$.  For $n_s=n_v = 3$, we found that $91\%$ of the channels satisfy the model~\eqref{eqn:arma:model}, with a confidence of $0.95$ (p-value = $0.05$). Hence, in the below analysis we select $n_s=n_v = 3$. It is intersting to note that the mean value of coefficients $b_i$ decrease as $i$ increases indicating that older view counts have less influence on the subscriber count. Similar results also hold for the coefficients $a_i$. Hence, as expected, the previous day subscriber count and the previous day view count most influence the current subscriber count.  

The next key question is does their exist a causal relationship between the subscriber dynamics and the view count dynamics. This is modeled using the hypothesis in~\eqref{eqn:hypothesis}. To test~\eqref{eqn:hypothesis} we use the Wald test with a confidence of $0.95$ (p-value = $0.05$) and found that approximately $55\%$ of the channels satisfy the hypothesis. For approximately $55\%$ of the channels that satisfy the AR model~\eqref{eqn:arma:model}, the view count ``Granger causes'' the current subscriber count. Interestingly, if different channel categories are accounted for then the percentage of channels that satisfy Granger causality vary widely as illustrated in Table~\ref{tab:category:causality}. For example, $80\%$ of the Entertainment channels satisfy Granger causality while only $40\%$ of the Food channels satisfy Granger causality. These results illustrate the importance of channel owners to not only maximize their subscriber count, but to also upload new videos or increase the views of old videos to increase their channels popularity (i.e. via increasing their subscriber count). Additionally, from our analysis which illustrates that the view count of a posted video is sensitive to the number of subscribers of the channel, increasing the number of subscribers will also increase the view count of videos that are uploaded by the channel owners.

\begin{table}[!h]
\begin{minipage}{\columnwidth} 
	\begin{center}
	\begin{tabular}{c|c}
		\toprule
		Category\footnote{YouTube assigns a category to videos, rather than channels. The category of the channel was obtained as the majority of the category of all the videos uploaded by the channel. } & Fraction \\
		\midrule
		Gaming & 0.60 \\
		Entertainment & 0.80 \\
		Food & 0.40 \\
		Sports & 0.67\\
		\bottomrule
	\end{tabular}
\end{center}
\end{minipage}%
	\caption{Fraction of channels satisfying the hypothesis: View count ``Granger causes'' subscriber count, split according to category.}
	\label{tab:category:causality}
\end{table}

\subsection{Video Upload Scheduling and Social Sensor Engagement}
\label{subsec:schedulingYouTube}
Here we investigate how the video upload scheduling dynamics of YouTube channels impacts social sensor engagement. We find the interesting property that for popular gaming YouTube channels with a dominant (constant) upload schedule, deviating from the schedule increases the views and the comment counts of the channel (e.g. increases user engagement). 

Creator Academy\footnote{YouTube website for helping with channels} in their best practice section recommends to upload videos on a regular schedule to get repeat views. The reason for a regular upload schedule is to increase the user engagement and to rank higher in the YouTube recommendation list. However, we show in this section that going ``off the schedule'' can be beneficial for a gaming YouTube channel, with a regular upload schedule, in terms of the number of views and the number of comments.  

From the dataset, we ``filtered out'' video channels with a \emph{dominant} upload schedules, as follows: The dominant upload schedule was identified by taking the periodogram of the upload times of the channel and then comparing the highest value to the next highest value. 
If the ratio defined above is greater than $2$, we say that the channel has a dominant upload schedule. 
From the dataset containing $25$ thousand channels, only $6500$ channels contain a dominant upload schedule. 
Some channels, particularly those that contain high amounts of copied videos such as trailers, movie/TV snippets upload videos on a daily basis. 
These have been removed from the above analysis. The expectation is that by doing so we concentrate on those channels that contain only user generated content. 

We found that channels with gaming content account for $75\%$ of the $6500$ channels with a dominant upload schedule\footnote{This could also be due to the fact gaming videos account for 70\% of the videos in the dataset.} and the main tags associated with the videos were: ``game'', ``gameplay'' and ``videogame''\footnote{We used a topic model to obtain the main tags.}.  We computed the average views when the channel goes off the schedule and found that on an average when the channel goes off schedule the channel gains views $97\%$ of the time and the channel gains comments $68\%$ of the time.  This suggests that channels with ``gameplay'' content have periodic upload schedule and benefit from going off the schedule.

\subsection{Social Sensor Engagement Dynamics with YouTube Videos}
\label{subsec:SocialSensorEngagementDynamicswithYouTubeVideos}
Several time-series analysis methods have been employed in the literature to model the \viewcount dynamics of YouTube videos. These include ARMA time series models~\cite{GCM11}, multivariate linear regression models~\cite{PAG13}, hidden Markov models~\cite{JMYLH14}, normal distribution fitting~\cite{FBA11}, and parametric model fitting~\cite{RAEJLP14,REJAL15}. Though all these models provide an estimate of the \viewcount dynamics of videos, we are interested in segmenting \viewcount dynamics of a video resulting from subscribers, non-subscribers and exogenous events.  Exogenous events are due to video promotion on other social networking platform such as Facebook or the video being referenced by a popular news organization or celebrity on Twitter. Detecting and accounting for exogenous events is motivated by the need for extracting accurate view counts resulting from exogenous events that provide an estimate of the efficiency of video promotion methods and meta-level feature optimizations.

The view count dynamics of popular videos in YouTube typically show an initial viral behaviour, due to subscribers watching the content, and then a linear growth resulting from non-subscribers. The linear growth is due to new users migrating from other channels or due to interested users discovering the content either through search or recommendations (we call this phenomenon \emph{migration} similar to~\cite{RAEJLP14}). Hence, without exogenous events, the \viewcount dynamics of a video due to subscribers and non-subscribers can be estimated using piecewise linear and non-linear segments.  In~\cite{RAEJLP14}, it is shown that a Gompertz time series model can be modeled the \viewcount dynamics from subscribers and non-subscribers, if no exogenous events are present. In this chapter, we generalize the model in~\cite{RAEJLP14} to account for views from exogenous events. It should be noted that classical change-point detection methods~\cite{TNB14} cannot be used here as the underlying distribution generating the \viewcount is unknown.

To account for the \viewcount dynamics introduced from exogenous events we use the generalized Gompertz model given by: 
\begin{align}
\begin{aligned}
\bar{v}_i(t) &= \sum_{k=0}^{K_\text{max}} w^k_i(t)u(t-t_k), \\
w_i^k(t) &= M_k\left(1-e^{-\eta_k\left(e^{b_k\left(t-t_k\right)}-1\right)}\right)+c_k(t-t_k),
\end{aligned}
\label{eqn:gompertz}
\end{align}
where $\bar{v}_i(t)$ is the total \viewcount for video $i$ at time $t$, $u(\cdot)$ is the unit step function, $t_0$ is the time the video was uploaded, $t_k$ with $k\in\{1,\dots,K_\text{max}\}$ are the times associated with the $K_\text{max}$ exogenous events, and $w^k_i(t)$ are Gompertz models which account for the \viewcount dynamics from uploading the video and from the exogenous events. In total there are $K_\text{max}+1$ Gompertz models with each having parameters $t_k, M_k, \eta_k, b_k$. $M_k$ is the maximum number of requests not including migration for an exogenous event at $t_k$, $\eta_k$ and $b_k$ model the initial growth dynamics from event $t_k$, and $c_k$ accounts for the migration of other users to the video. In (\ref{eqn:gompertz}) the parameters $\{M_k,\eta_k,b_k\}_{k=0}$ are associated with the subscriber views when the video is initially posted, the parameters $\{t_k,M_k,\eta_k,b_k\}_{k=1}^{K_\text{max}}$ are associated with views introduced from exogenous events, and the views introduced from migration are given by $\{c_k\}_{k=0}^{K_\text{max}}$. Each Gompertz model (\ref{eqn:gompertz}) captures the initial viral growth when the video is initially available to users, followed by a linearly increasing growth resulting from user migration to the video. 

\begin{figure}[h]
	\centering
	\includegraphics[angle=0,width=2.5in]{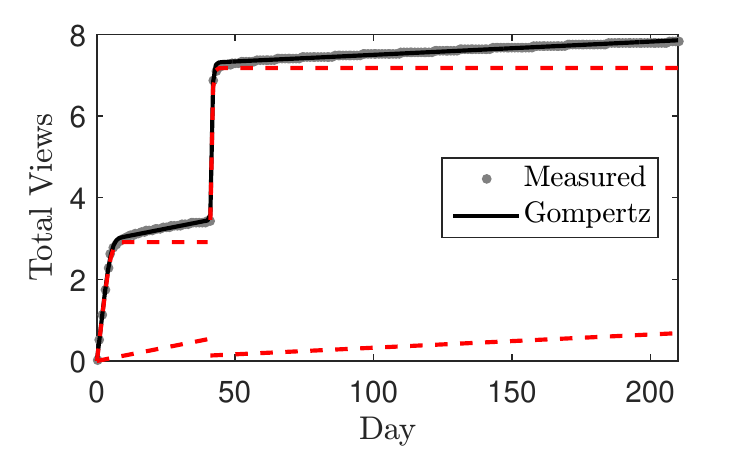}
	\caption{Due to an exogenous event on day $41$, there is a sudden increase in the number of views. The total view count fitted by the Gompertz model $\bar{v}_i(t)$ in~\eqref{eqn:gompertz} is shown in black with the virality (exponential) and migration (linear) illustrated by the dotted red. }
	\label{fig:ExoFitExoGompertz}
\end{figure}

The parameters $\theta_i=\{a_k,t_k,M_k,\eta_k,b_k,c_k\}_{k=0}^{K_\text{max}}$ in (\ref{eqn:gompertz}) can be estimated by solving the following mixed-integer non-linear program:
\begin{align}
\theta_i&\in\operatorname{arg\ min}\Big\{\sum_{t=0}^{T_i}\big(\bar{v}_i(t)-v_i(t)\big)^2+\lambda K\Big\} \nonumber\\
&K=\sum_{k=0}^{K_\text{max}}a_k, \quad\quad a_k\in\{0,1\} \quad\quad k\in\{0,\dots,K_\text{max}\},
\label{eqn:gompertzMINLP}
\end{align}
with $T_i$ the time index of the last recorded views of video $v_i$, and $a_k$ a binary variable equal to $1$ if an exogenous event is present at $t_k$. Note that (\ref{eqn:gompertzMINLP}) is a difficult optimization problem as the objective is non-convex as a result of the binary variables $a_k$~\cite{BL12}. In the YouTube social network when an exogenous event occurs this causes a large and sudden increase in the number of views, however as seen in Fig.~\ref{fig:ExoFitExoGompertz}, a few days after the exogenous event occurs the views only result from migration (i.e. linear increase in total views). Assuming that each exogenous event is followed by a linear increase in views we can estimate the total number of exogenous events $K_\text{max}$ present in a given time-series by first using a segmented linear regression method, and then counting the number of segments of connected linear segments with a slope less then $c_\text{max}$. The parameter $c_\text{max}$ is the maximum slope for the views to be considered to result from viewer migration. Plugging $K_\text{max}$ into (\ref{eqn:gompertzMINLP}) results in the optimization of a non-linear program for the unknowns $\{t_k,M_k,\eta_k,b_k,c_k\}_{k=0}^{K_\text{max}}$. This optimization problem can be solved using sequential quadratic programming techniques~\cite{Ber99}. 

To illustrate how the Gompertz model~\eqref{eqn:gompertz} can be used to detect for exogenous events, we apply (\ref{eqn:gompertz}) to the \viewcount dynamics of a video that only contains a single exogenous event. Fig.~\ref{fig:ExoFitExoGompertz} displays the total \viewcounts of a video where an exogenous event occurs at time $t=41$ (i.e.\ $t_1 = 41$ in~\eqref{eqn:gompertz}) days after the video is posted\footnote{Due to privacy reasons, we cannot detail the specific event. Some of the reasons for the sudden increase in the popularity of the video include: Another user on YouTube mentioning the video, this will encourage viewers from that channel to view the video, resulting in a sudden increase in the number of views. Another possibility is that the channel owner or a YouTube Partner like BBTV did significant promotional initiatives on other social media sites such as Twitter, Facebook, etc. to promote the channel or video.}. The initial increase in views for the video for $t\leq7$ days results from the $2910$ subscribers of the channel viewing the video. For $7\leq t \leq 41$, other users that are not subscribed to the channel migrate to view the video at an approximately constant rate of $13$ views/day. At $t=41$, an exogenous event occurs causing an increase in the views per day. The difference in viewers, resulting from the exogenous event, is $7174$. For $t\geq 43$, the views result primarily from the migration of users to approximately $2$ views/day. Hence, using the generalized Gompertz model (\ref{eqn:gompertz}) we can differentiate between subscriber views, views caused by exogenous events, and views caused by migration.

\subsection{Social Sensor Engagement for Channel Playthroughs}
\label{subsec:SocialSensorEngagementforChannelPlaythroughs}

\begin{figure}[h!]
  \centering
  \subfigure[Actual and predicted \viewcount of playthrough. We plot the \nth{1}, \nth{5}, \nth{10}, \nth{15}, \nth{20} and \nth{25} video from the playlist containing $25$ videos. In the legend, \emph{Exp} and \emph{Pred} corresponds to the actual and the predicted value using~\eqref{eqn:gompertz}, respectively. Figure shows that the view counts decreases for subsequent videos in the playlist. ]{\label{fig:numpred}\includegraphics[scale=0.5]{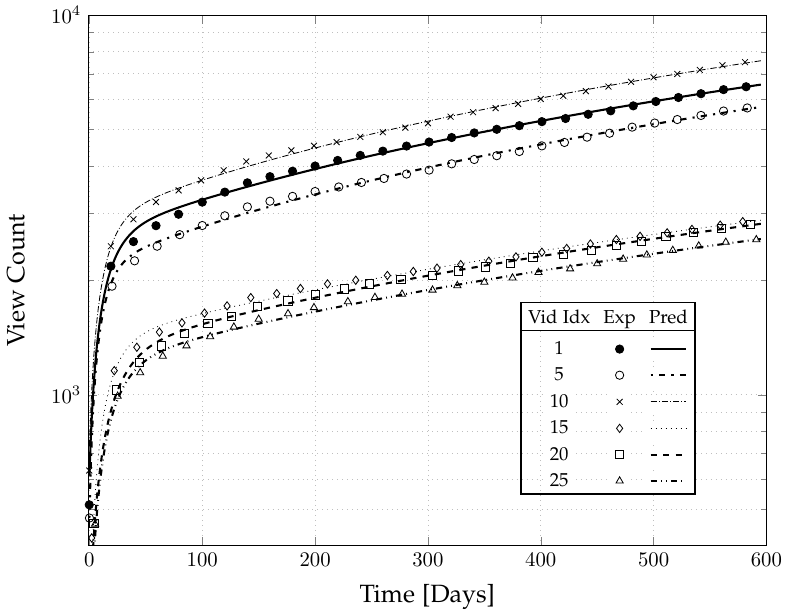}} \\
  \subfigure[The virality rate specifies the early views due to subscribers, and the migration rate (in units of views/1000 days) specifies the subsequent linear growth due to non-subscribers. ]{\label{fig:earlylaterelation}\includegraphics[scale=0.5]{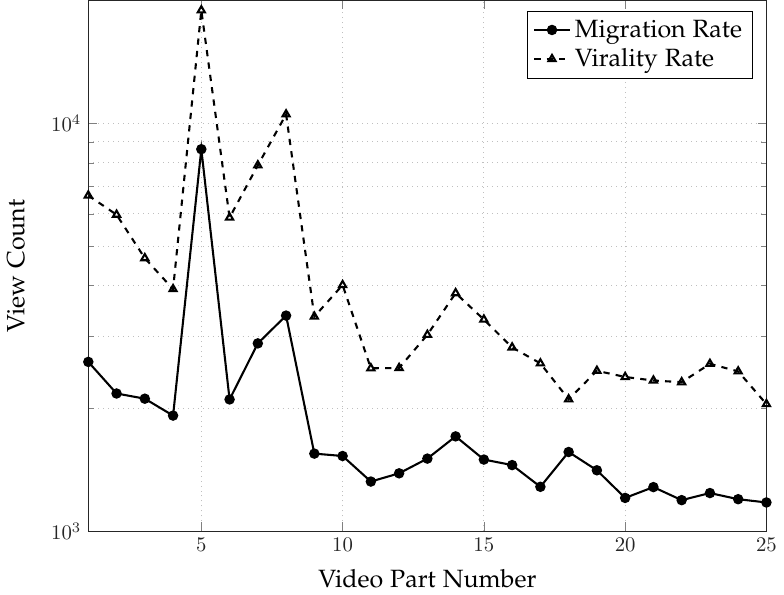}}
\vspace{-0pt}
\caption{Actual and predicted \viewcount of a playthrough containing 25 YouTube videos for the game ``BioShock Infinite''. The predictions are computed by fitting a modified Gompertz model (\ref{eqn:gompertz}) to the measured \viewcounts for each video in the playthrough.}
\label{fig:BioInfPlaythrough}
\vspace{-0pt}
\end{figure}

One of the most popular sequences of YouTube videos is the video game ``playthrough''. A video game playthrough is a set of videos for which each video has a relaxed and casual focus on the game that is being played and typically contains commentary from the user presenting the playthrough. Unlike YouTube channels such as CNN, BBC, and CBC in which each new video can be considered independent from the others, in a video playthrough the future \viewcount of videos are influenced by the previously posted videos in the playthrough. 
To illustrate this effect we consider a video playthrough for the game ``BioShock Infinite''--a popular video game released in 2013. 
The channel, popular for hosting such video playthroughs, contains close to $4500$ videos and $180$ video playthroughs. 
The channel is highly popular and has garnered a combined view count close to $100$ million views with $150$ thousand subscribers over a period of $3$ years. 
Fig.~\ref{fig:BioInfPlaythrough} illustrates that the early view count dynamics are highly correlated with the \viewcount dynamics of future videos. Both the short term \viewcount and long term migration of future videos in the playthrough decrease after the initial video in the playthrough is posted. This results for two reasons, either the viewers purchase the game, or the viewers leave as the subsequent playthroughs become repetitive as a result of game quality or video commentary quality. A unique effect with video playthroughs is that although the number of subscribers to the channel hosting the videos in Fig.~\ref{fig:BioInfPlaythrough} increases over the 600 day period, the linear migration is still maintained after the initial 50 days after the playthrough is published. Additionally, the slope of the migration is related to the early total \viewcounts as illustrated in Fig.~\ref{fig:earlylaterelation}. 

\subsection{Summary and Extensions}
The application of time-series analysis and machine learning methods to gain insight into the social sensor dynamics on YouTube is an active area of research with several promising outcomes. First, they can be used to reduce the operating cost of content distribution networks. In~\cite{HGKDZ15} a two time-scale game-theoretic learning algorithm is constructed to optimally cache videos in the future 5G mobile network based on the dynamics of the social sensors. In~\cite{SHV17} the optimal caching decision is formulated as a mixed-integer linear program that accounts for the dynamics of the social sensors. Second, knowledge of the user dynamics can be used to optimize the meta-level features of videos to maximize user engagement as illustrated in this section. 

\begin{figure}
\vspace{-10pt}
  \centering
    \subfigure[Initial: 239 users, 3,112 edges.]{\label{fig:youtubesocialnetworkA}\includegraphics[angle=0,width=2.2in,trim={0 0 0 0},clip]{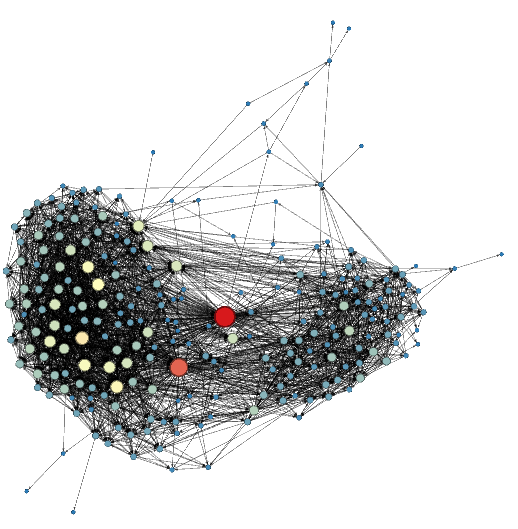}}
    \hspace{10pt}
 \subfigure[After 3 months: 503 users, 8,558 edges.]{\label{fig:youtubesocialnetworkB}\includegraphics[angle=0,width=2.2in,trim={0 0 40 0},clip]{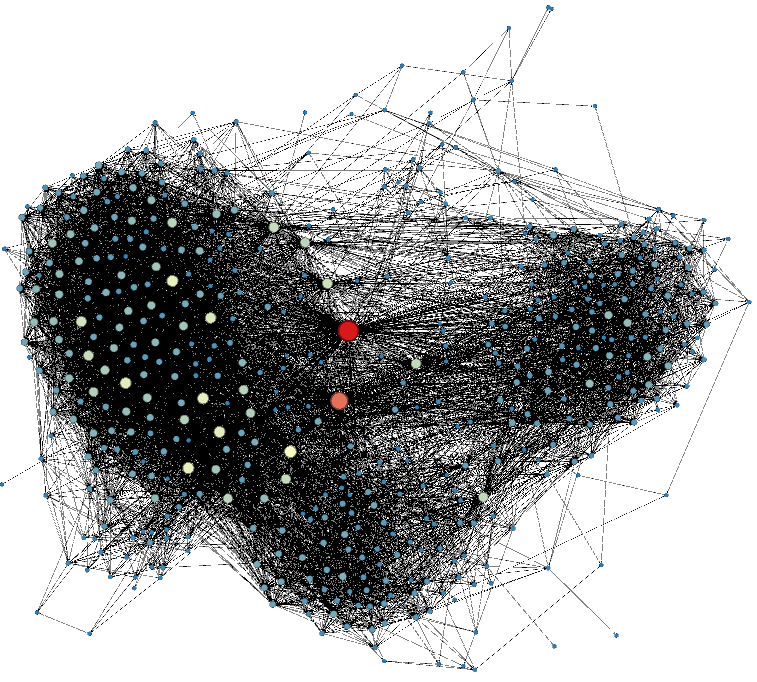}}
    \hspace{5pt}
    \subfigure[After 6 months: 1,391 users, 31,701 edges.]{\label{fig:youtubesocialnetworkC}\includegraphics[angle=0,width=4.0in,trim={0 0 0 0},clip]{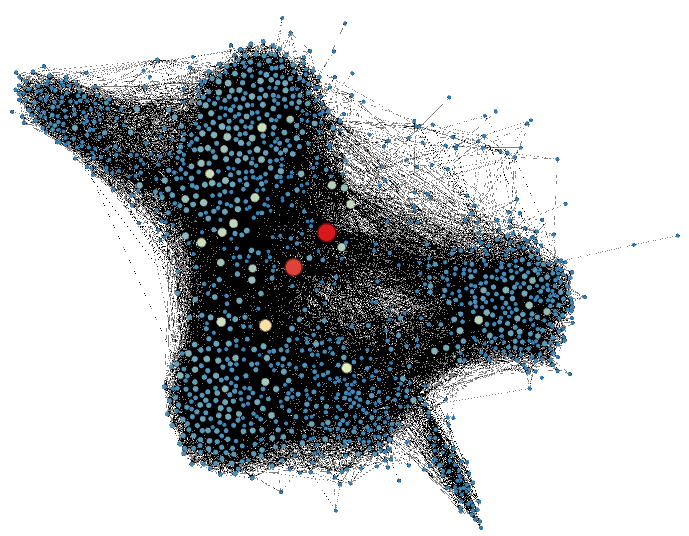}}
\vspace{-10pt}
\caption{Snapshots of the YouTube social network at 0 months, 3 months, and 6 months constructed using the dataset discussed in Sec.\ref{subsec:YouTubeDataset}. Node sizes and color indicate their degree (red is higher, blue is lower). The complete social network is composed of over 1.13 million users, and contains over 2.98 million edges. Only users with at least 1000 edges are displayed.}
\label{fig:youtubesocialnetwork}
\vspace{-5pt}
\end{figure}

Significant work remains on the analysis of user dynamics in the YouTube social network. Recall that in the YouTube social network interaction between users and channel owners include:
\begin{compactenum}
\item Commenting on users videos. Commenting is YouTube's version of engagement, and it has some of the most involved, engaged and dedicated users. Additionally, users can comment on other users comments which is very similar to users interaction on blog posting sites except related to the uploaded videos.
\item Subscribing to YouTube channels provides a method of forming relationships between users. 
\item Users can directly comment on a YouTube channel without the need to only interact when a video is posted. 
\item Users can also interact by embedding videos from another users channel directly into their own channel to promote exposure or form communities of users. 
\end{compactenum}
In addition to the social incentives, YouTube provides monetary incentives to promote users increasing their popularity and engagement. As more users view and interact with a users video or channel, YouTube will pay the user proportional to the advertisement exposure on the users channel. Therefore, users not only maximize exposure to increase their social popularity, but also for monetary gain which introduces unique dynamics in the formation of edges in the YouTube social network. Using the dataset discussed in Sec.\ref{subsec:YouTubeDataset}, Fig.\ref{fig:youtubesocialnetwork} plots the communication network where an edge indicates comments and responses between users that have interacted at least 1000 times. From Fig.\ref{fig:youtubesocialnetworkA}, initially there appears to be two clusters of users that have strong interactions indicating that user preferences play a significant role in forming the edges in these clusters. After a period of 3 months, Fig.\ref{fig:youtubesocialnetworkB} illustrates that more users have entered the network however there still appears to be two primary clusters of interacting users. At 6 months, Fig.\ref{fig:youtubesocialnetworkC} shows a dense interaction between several users in the social network. The dynamics of these interaction links are governed by both the users preferences and the video content that is uploaded by the users. Prior to edge formation, these clustered communities can detected by applying the homophilic community detection tests introduced in~\cite{GHK16}. These tests are designed to cluster users based on their content preferences. 

The dynamics of edge formation/destruction and user popularity in the social network (illustrated in Fig.\ref{fig:youtubesocialnetwork}) are governed by the user-user interaction and the user-content-user interaction. Two key question to address in the YouTube social network is: How do the social dynamics (subscribing, commenting, video content quality, video category, etc.) impact the popularity of videos and the dynamics of the communication network between users? Answers to this question provides valuable insight into the evolving dynamics of the social network illustrated in Fig.\ref{fig:youtubesocialnetwork}.

\section{Closing Remarks} \label{sec:close}
This chapter has discussed four important and inter-related themes regarding the dynamics of social sensors, namely,
diffusion models for  information in social networks,  Bayesian social learning, revealed preferences and how social sensors interact
over YouTube channels. In each case, examples involving
 real datasets were 
given to illustrate the various concepts.
The unifying theme  behind these  three topics stems from predicting global behavior given local behavior:  individual social sensors make decisions and learn from
other social sensors and we are interested in understanding the behavior of the entire network. In Sec.\ref{sec:diffusion} we showed that the global degree of infected nodes
can be determined by mean field dynamics. In Sec.\ref{sec:socialmain}, it was shown that despite the apparent simplicity in  information flows
between social sensors, the global system can  exhibit unusual behavior such as herding and data incest.
Finally, in Sec.\ref{sec:revealed} a non-parametric method was used to determine the utility functions of a multiagent system - this can  be used
to predict the response of the system.

This chapter has  dealt with social sensing issues of relevance to a signal processing audience.
There are several 
topics of relevance to social sensors that are omitted due to space constraints, including:
\begin{compactitem}
\item Coordination of decisions via game-theoretic learning  \cite{HM00,HMB13,NKY13} or Bayesian game models such as global games \cite{AHP07}
\item Consensus formation over social networks and cooperative models of network formation  \cite{KB08,TJ10}
  \item Controlled information fusion in social learning \cite{BK17}.
\item Small world models \cite{Wat99,Kle00}
\item Peer to peer media sharing \cite{HK07,ZLL11}
\item Privacy and security modelling \cite{Lan07,LYM03}
\item Influence Maximization; see \cite{NK18} are references therein.
\item Polling using friendship paradox; see \cite{NK18b} are references therein.
\item The mobile edge cloud and popularity prediction for YouTube
  \cite{HTK18}.
\end{compactitem}

\bibliographystyle{abbrv}
\bibliography{extracted}

\end{document}